%% file: main.tex
\documentclass[5p]{elsarticle}%[review]

\usepackage{amsmath,amssymb}
\usepackage{mathpazo}
\usepackage{enumitem}
\usepackage{times}
\usepackage{graphicx}
\usepackage{subfigure}
\usepackage{epsfig}
\usepackage{epstopdf}
\usepackage{multirow}
\usepackage{esvect}
\usepackage{pifont}
\usepackage{units}
\usepackage{url}
\usepackage{xcolor}
\usepackage{tikz}
\usepackage{arydshln}
\usepackage[flushleft]{threeparttable}
\usepackage{widetext}
\usepackage{amsmath}
\usepackage{amssymb}        % for \varnothing
\usepackage{colortbl}% for color cell
\usepackage{mathtools}
\usepackage{pifont}
\usepackage[linesnumbered,ruled]{algorithm2e}
\usepackage{rotating} % for vertical text in table cells
\usepackage[colorlinks=true]{hyperref}
\usepackage{todonotes}
\usepackage{fancyhdr}

\usepackage{cleveref}

\usepackage[pagewise]{lineno}
\modulolinenumbers[250]
\newcommand{\lref}[1]{\label{#1}\linelabel{#1L}}

\newenvironment{brown}{\color{black}}{}

\newcommand{\cmark}{\ding{52}}
\newcommand{\xmark}{\ding{54}}

\def\checkmark{\tikz\fill[scale=0.3](0,.35) -- (.25,0) -- (1,.7) -- (.25,.15) -- cycle;}
\newcommand{\secref}[1]{\S\ref{#1}}

\emergencystretch=\maxdimen
\usetikzlibrary{trees}
\usetikzlibrary{shadings}
\tikzstyle{every node}=[draw=black,thin,anchor=west, minimum height=2.5em]

\biboptions{authoryear}

\begin{document}
%=================================
% \input{response_letter.tex}
%=================================
\linenumbers

\begin{frontmatter}

% \title{Medical Image Segmentation: A Review of Modern Solutions for Meager Datasets}
% \title{Medical Image Segmentation with Imperfect Datasets: A Review}
% \title{Medical Image Segmentation with Imperfect Datasets: A Review}
\title{ Embracing Imperfect Datasets: \\ A Review of Deep Learning Solutions for Medical Image Segmentation}
\author[mymainaddress]{Nima Tajbakhsh, Laura Jeyaseelan, Qian Li, Jeffrey N. Chiang, Zhihao Wu, and Xiaowei Ding}
\address[mymainaddress]{VoxelCloud, Inc.}
% \address[mysecondaryaddress]{}

\begin{abstract}
The medical imaging literature has witnessed remarkable progress in high-performing segmentation models based on convolutional neural networks. Despite the new performance highs, the recent advanced segmentation models still require large, representative, and high quality annotated datasets. However, rarely do we have a perfect training dataset, particularly in the field of medical imaging, where data and annotations are both expensive to acquire. Recently, a large body of research has studied the problem of medical image segmentation with imperfect datasets, tackling two major dataset limitations:  scarce annotations where only limited annotated data is available for training, and weak annotations where the training data has only sparse annotations, noisy annotations, or image-level annotations. In this article, we provide a detailed review of the solutions above, summarizing both the technical novelties and empirical results. We further  compare the benefits and requirements of the surveyed methodologies and provide our recommended solutions\iffalse to the problems of scarce and weak annotations\fi. We hope this survey article increases the community awareness of the techniques that are available to handle imperfect medical image segmentation datasets.% and inspire further research for more effective solutions. 

\end{abstract}


\end{frontmatter}

% to include header on page 1 only
% \thispagestyle{fancy}

% \todo[inline, color=green]{change post-segmentation to segmentation}
% \todo[inline, color=red]{section-wise summary}
\section{Introduction}
\label{sec:intro}
\input{chapters/introduction.tex}

\section{Related works}
\label{sec:related}
\input{chapters/related_work.tex}

\section{Organization of survey}
\label{sec:organization}
\figurename~\ref{fig:problem_tree} shows the organization of this survey.  We have broadly  grouped  limitations associated with medical image segmentation datasets into two sections: 1) scarce annotations (Section~\ref{sec:scarce}), which covers methodologies that can handle datasets where only a small fraction of images are densely annotated; and 2) weak annotations (Section~\ref{sec:weak}), which covers methodologies that leverage datasets with sparse, noisy, or only image-level annotations. 

The methodologies presented for scarce annotations in Section~\ref{sec:scarce} are further grouped into three categories according to their methodology principles. The first category consists of the methods that aim 
to enlarge the training set (\cref{sec:aug,sec:dataset,sec:cost,sec:unlabeled}) through data augmentation, leveraging external but related labeled datasets, cost effective annotation, and leveraging unlabeled data. Although these methods share the same philosophy, they differ in the required data resources and whether or not they require the expert in the loop. The second category (\cref{sec:regularized}) consists of methods that strengthen regularization during model training, where the regularization can be applied to the input space by changing the image representation, to the output space by constraining the segmentation results with shape priors, or directly to the gradients by leveraging additional supervision signals through multi-task learning. Except for multi-task learning, regularization-based methods do not require any further data or annotations. The third category consists of methods that aim to refine the segmentation mask (\cref{sec:post}) using different variants of conditional random fields (CRFs) either as a post-processing or during model training. These methods require no further data other than the currently available segmentation dataset.  

% \begin{enumerate}
%     \item dataset expansion (\cref{sec:aug,sec:dataset,sec:cost,sec:unlabeled}) where the goal is to enlarge the training set through data augmentation, leveraging external labeled datasets, cost effective annotation, and leveraging unlabeled data.
    
%     \item regularized training (Section~\ref{sec:regularized}) where the regularization can be applied to the input space by changing the image representation, to the output space by constraining the segmentation results with shape priors, or directly to the gradients by leveraging additional supervision signals through multi-task learning.
    
%     \item segmentation refinement where the goal is to improve the segmentation results using different variants of conditional random fields (CRFs) as a post-processing or during training.
% \end{enumerate}. 
 The methodologies presented in Section~\ref{sec:weak} for the problem of weak annotations are further grouped by the type of annotation weakness into 3 sections: 1) methods that tackle image-level annotations (\cref{sec:weak_l}), which use different variants of class activation maps to leverage weak image-level labels for medical image segmentation; 2) methods that leverage sparse annotations (\cref{sec:sparse_l}), which are typically based on some variants of selective loss where only sparsely labeled pixels contribute to the segmentation loss; 3) methods that handle noisy annotations (\cref{sec:noisy_l}), which typically use noise-resilient loss functions to learn from noisy annotations. 
 
 In section~\ref{sec:discussion}, we summarize this survey by comparing the methodologies under review from the perspectives of performance gain, implementation difficulty, and required data resources. We further provide our recommended solutions based on a cost-gain trade-off. 
%  Finally, this survey is concluded in Section \ref{sec:conclusion}.

%\figurename~\ref{fig:problem_tree} shows the common limitations of medical image segmentation datasets and the suggested solutions accordingly. We have broadly categorized dataset limitations into two categories: 1) scarce annotations (Section~\ref{sec:scarce}) where only a small fraction of images in the dataset are densely annotated; 2) weak annotations (Section~\ref{sec:weak}) where the provided annotations are sparse, noisy, or only at image-level. We have further grouped the solutions for scarce annotations by their methodological principles, covering methods based on data augmentation in Section \ref{sec:aug}, leveraging external labeled datasets in Section \ref{sec:dataset}, cost effective annotation in Section \ref{sec:cost}, leveraging unlabeled data in Section \ref{sec:unlabeled}, regularized training in Section \ref{sec:regularized}, and finally post segmentation refinement in Section \ref{sec:post}. The solutions suggested for weak annotations are grouped by the type of weakness these methodologies tackle. Specifically, we  cover methods for image-level annotations in Section \ref{sec:weak_l}, sparse annotations in Section \ref{sec:sparse_l}, and noisy annotations in Section \ref{sec:noisy_l}. We compare the solutions under review in Section~\ref{sec:discussion}, and provide our recommended solutions based on the underlying advantages and required resources. Finally, this survey is concluded in Section \ref{sec:conclusion}.

\section{Problem I: Scarce annotation}
\label{sec:scarce}

Scarce annotation is a common problem when using supervised deep learning methods for medical image segmentation. Traditional solutions to this problem are data augmentation, transfer learning from natural images, and weight regularization. However, these techniques can only partially address the problem of limited annotation. For example, traditional data augmentation is handicapped by the large correlation between the original training set and the augmented examples. Transfer learning from natural images only benefits 2D medical image segmentation models, with no benefits to the common 3D medical image segmentation models. 

The limited capability of the traditional methods in handling the problem of scarce annotations has led to the development of modern reactive and proactive approaches.
The reactive methods tackle the problem of scarce annotation through a post segmentation refinement using variants of conditional random fields. The proactive approaches, on the other hand, actively enlarge the training set through cost-effective annotation and synthetic data generation or change the training paradigm by leveraging unlabeled data and using additional model regularization during training. In the following, we provide a comprehensive summary of such modern solutions to the ubiquitous problem of scarce annotations in medical image segmentation.

\subsection{Data augmentation}
\label{sec:aug}
Data augmentation has served as an effective solution to the problem of over-fitting, particularly in the absence of large labeled training sets. {\brown In this section, we cover the  data augmentation methods based on traditional spatial and intensity transforms, data augmentation by mixing images, and modern image synthesis methods based on adversarial networks. While the scope of this section is limited to medical images, the readers can refer to the survey by \cite{shorten2019survey} for a comprehensive review of data augmentation methods for both natural and medical image domains.}

% Data augmentation has served as an effective solution to the problem of over-fitting, particularly in the absence of large labeled training sets.  However, traditional data augmentation methods result in images that are highly correlated; as such, their impact can be limited. Apart from intrinsic high correlation between augmented and original images, rare conditions may also not be properly enhanced during typical data augmentations. Synthetic data augmentation, on the other hand, samples more diverse  examples, with sufficiently different visual appearance, from the same manifold as the original training set. In this section, we review both traditional and synthetic data augmentation methods as they can offer complimentary values.

\subsubsection{Traditional Augmentation}
\label{sec:trad}

\input{chapters/traditional_augmentation.tex}

\subsubsection{Mixing Augmentation}
\label{sec:mixing}
\input{chapters/mixing.tex}

\subsubsection{Synthetic Augmentation}
\label{sec:syn}
\input{chapters/data_synthesis.tex}

\subsubsection{Summary}
In this section, we first reviewed the traditional data augmentation methods, which manipulate image appearance, quality, or layout to generate new training examples. Although simple to implement, these methods result in augmented images that are typically correlated with the original images; and thus, their impact may be limited. {\brown We then reviewed data augmentation methods that generate new images and masks through linear combination of existing labeled images. We also reviewed data augmentation methods based on image synthesis,}  which generate images with larger appearance variability than those generated by the traditional data augmentation. Image synthesis methods achieve this by sampling from the manifold on which the original training set reside. Although these methods are more effective for handling data scarcity as well as rare conditions, they are more demanding to implement, because their training schemes typically require adversarial networks and additional labeled or unlabeled datasets. 

\subsection{Leveraging External Labeled Datasets}
\label{sec:dataset}
\input{chapters/multiple_datasets.tex}

\subsection{Cost-effective  Annotation}
\label{sec:cost}
Perhaps, the most reliable approach to the scarce annotation problem is to obtain additional labeled examples. This approach requires  the availability of unlabeled medical images, access to a pool of expert annotators, and more importantly additional annotation budget. However, to fully utilize the annotation budget, one must decide how to choose examples for annotation from a large set of unlabeled images and how to accelerate the annotation process given the limited availability of medical experts. The former question is addressed by active learning, which determines the next batch of samples for annotation so as to maximize model's performance, and the latter is addressed by interactive segmentation, which assists the expert annotators by propagating their modifications through the entire segmentation mask.

\subsubsection{Active Learning}
\label{sec:active}
\input{chapters/active_learning.tex}

\subsubsection{Interactive Segmentation}
\label{sec:interactive}
\input{chapters/interactive_segmentation.tex}

\subsubsection{Summary}
In this section, we covered the methodologies available for cost-effective image annotation. We first reviewed the active learning approach, which enables informed decisions as to which unlabeled images should be annotated first. While active learning methods are typically iterative and slow, recent works \cite{mahapatra2018efficient, zheng2019biomedical} present a 1-shot, fast approach to active learning with comparable performance to the iterative counterparts, making active learning an even more attractive methodology. Nearly all active learning works reviewed in this survey base sample selection on informativeness and diversity of samples, neglecting the cost associated with annotating a sample. The exception is the work of \cite{kuo2018cost} where the authors included annotation cost in their sample selection method. We believe that respecting annotation cost is fundamental to realistic active learning. We further reviewed interactive segmentation as a means of accelerating the annotation process. In particular, the work by \cite{sakinis2019interactive} was quite effective, where 1 user click led to a Dice of 0.64 for segmenting colon cancer, outperforming the best automated model with a Dice of 0.56. However, the efficacy of interactive segmentation methods in reducing annotation cost should be corroborated through more systematic user studies.

\subsection{Leveraging Unlabeled Data}
\label{sec:unlabeled}
Unlabeled medical images, although lack annotations, can still be used in conjunction with labeled data to train higher-performing segmentation models. We have identified three scenarios  wherein unlabeled medical images have aided medical image segmentation: 1) self-supervised pre-training where unlabeled  images are used to pre-train a segmentation network; {\brown 2) semi-supervised learning with pseudo labels where unlabeled  images are labeled by a segmentation model and then used as new examples during training; and 3) semi-supervised learning without pseudo labels} where both labeled and unlabeled images are used jointly to train a segmentation model.

\subsubsection{Self-supervised Pre-training}
\label{sec:ss}
\input{chapters/self_supervised.tex}

%\subsubsection{Self Learning}
\subsubsection{Semi-supervised learning with pseudo annotations}
\label{sec:sl}
\input{chapters/self_learning.tex}

\subsubsection{Semi-supervised learning without pseudo annotations}
\label{sec:ssl}

\input{chapters/semi_supervised_learning.tex}

\subsubsection{Summary}
In this section, we reviewed three strategies that enable training a segmentation model with both labeled and unlabeled data. We started with self-supervised learning, which consists of pre-training the segmentation network  using unlabeled medical images followed by fine-tuning the pre-trained model using the target labeled dataset. Self-supervised learning offers several advantages: 1) demonstrated performance gains over the counterpart models trained from scratch; 2) ease of implementation owing to the intuitive underlying proxy tasks; and 3) ease of use, because no or only minor architectural changes to the segmentation network is required. 

We further reviewed semi-supervised learning with pseudo annotations where the model learns from its own prediction on unlabeled images. The suggested methods often show only moderate performance gains over the counterparts trained using only labeled images. This is because model-generated annotations can be noisy, which has detrimental effects to the subsequent segmentation model. This limitations has been recently addressed in  \cite{min2018two, nie2018asdnet}, but at the expense of relatively complex segmentation architectures.

Semi-supervised learning without pseudo annotations was the third strategy we reviewed, which uses  unlabeled data along with the labeled data during training. Since these methods accommodate the segmentation task for the labeled data and an unsupervised task for the unlabeled data, they  typically consist of complex neural architectures.
{\brown Nevertheless, these semi-supervised schemes are effective in coping with limited labeled training sets, improving the Dice score by a couple of points in most cases. One caveat, however, is that these methods are not as effective when the training set grows in size, suggesting that one should not be overoptimistic about the capabilities of these methods.} Compared to the previous strategy, these methods do not attempt to generate annotations for the unlabeled images; as such, they are not vulnerable to annotation noise.  

% Semi-supervised learning differs from self-supervised learning in that the latter uses the unlabeled data only in the pre-training phase whereas the former uses the unlabeled data during training the segmentation model. Semi-supervised learning also differs from self learning in that the former does not attempt to label the unlabeled images whereas the latter does; as such, semi-supervised learning is not vulnerable to annotation noise. 

\subsection{Regularized Training}
\label{sec:regularized}
Having a large number of parameters, deep supervised models are prone to over-fitting, particularly in the absence of large training sets. The traditional regularization to the problem of over-fitting is weight regularization whereby  the network is encouraged to keep the weights small, resulting in a simpler and more robust model. While effective, weight regularization is only one form of model regularization. In this section, we cover other forms of regularization: altered image representation, multi-task learning, and shape regularization.

\subsubsection{Altered Image Representation}
\label{sec:altered}
\input{chapters/altered_img_rep.tex}

\subsubsection{Multi-task Learning}
\label{sec:multi}

\input{chapters/multi_task_learning.tex}

\subsubsection{Shape Regularization}
\label{sec:shape}
\input{chapters/shape_prior.tex}

\subsubsection{Summary}
In this section, we covered three forms of regularization. We first reviewed altered image representations, where regularization is applied to the input space. This form of data regularization results in a lower dimensional input space or an altered input space that facilitates the task of representation learning for the model particularly in the absence of large labeled datasets. We then reviewed multi-task learning for medical image segmentation, which leads to consistent improvement over the single-task segmentation models and further enables a consolidated framework for multi-modality multi-condition medical image segmentation. Lastly, we reviewed shape regularization, which imposes a shallow or deep shape prior on the predicted segmentation results. While no prior works have considered the combination of the above three forms of regularization, these approaches are independent and can potentially offer complementary advantages. Also, except for multi-task learning which requires additional annotations, input space regularization and shape regularization require no further annotations; and thus, can be taken advantage of at the cost of changing the data pipeline and minor architectural changes, respectively.

\subsection{Post segmentation refinement}
\label{sec:post}

\input{chapters/post-processing.tex}

\subsubsection{summary}
In this section, we reviewed different variants of CRF for the task of post-segmentation refinement. We started with locally connected CRF, which encourages smoothness constraints on local regions of segmentation maps. We then reviewed fully connected CRF, which solves a global optimization problem over the entire segmentation mask. While both locally- and fully-connected CRF have proved effective in 2D segmentation applications, their extension to 3D applications has shown only mixed results. These methods further require extensive parameter tuning and may be susceptible to image noise as they operate directly on pixel information. Finally, we reviewed RNN-CRF, which operates on CNN feature maps and is trained end-to-end along with the segmentation model, thereby addressing the limitations of parameter tuning and susceptibility to image noise. 

\section{Problem II: Weak Annotations}
\label{sec:weak}
\label{sec:imperfect_annot}

\input{chapters/imperfect_annotations.tex}

\section{Discussion}
\label{sec:discussion}
\input{chapters/discussion.tex}

\section{Conclusion}
\label{sec:conclusion}

In this survey, we covered data limitations associated with medical image segmentation datasets, namely, scarce annotations and weak annotations. For the problem of scarce annotations, we reviewed a diverse set of solutions, ranging from semi-automated solutions that require human experts in the loop such as active learning and interactive segmentation,  to fully-automated solutions that leverage unlabeled and synthetic data from the same domain or labeled data from similar domains. For the problem of weak annotations, we studied solutions with the capability of handling sparse, noisy, or only image-level annotations. We further compared the suggested methodologies in terms of required data resources, difficulty of implementation, and performance gains, highlighting methodologies with the best cost-gain trade-off. We hope this survey increases the community awareness of the strategies for handling scarce and weak annotations in medical image segmentation datasets, and further inspires efforts in this impactful area of research.

\section{Acknowledgment}
We would like to thank Ju Hu for helping us with compiling the list of related works in the initial stage of this research.

% \section*{References}

% \bibliographystyle{model2-names}
\bibliography{refs/refs,refs/refs_al,refs/refs_is,refs/refs_imgrep,refs/refs_sp,refs/refs_hp,refs/refs_sl,refs/refs_ssl,refs/refs_self,refs/refs_tda,refs/refs_ds,refs/refs_wst,refs/refs_domada,refs/refs_intro,refs/refs_related,refs/refs_mixing} 

% ,refs/refs_sp
% ,refs/refs_hp
% ,refs/refs_he
% ,refs/refs_imb

\end{document}

%% file: chapters/introduction.tex
%http://jmai.amegroups.com/article/view/4659/html
Medical imaging literature has witnessed great progress in the designs and performance of deep convolutional models for medical image segmentation. Since the introduction of UNet \cite{ronneberger_u-net:_2015}, neural architectures for medical image segmentation have transformed markedly. State-of-the-art architectures now benefit from re-designed skip connections \cite{zhou2018unet++}, residual convolution blocks \cite{alom2018recurrent}, dense convolution blocks \cite{li2018h}, attention mechanisms \cite{oktay_attention_2018}, hybrid squeeze-excitation modules \cite{roy2018concurrent}, to name a few. Although the architectural advancements have enabled new performance highs, they still require large, high-quality annotated datasets---more so than before. 

However, rarely do we have a perfectly-sized and carefully-labeled dataset to train an image segmentation model, particularly for medical imaging applications, where both data and annotations are expensive to acquire. The common limitations of medical image segmentation datasets include  scarce annotations where only
limited annotated data is available for training, and weak annotations where the training data has only sparse annotations, noisy annotations, or image-level annotations. In the presence of these dataset shortcomings,  even the most advanced segmentation models may fail to generalize to datasets from real-world clinical settings. In response to this challenge,
researchers from the medical imaging community have actively sought solutions, resulting in a diverse and effective set of techniques with demonstrated capabilities in handling scarce and weak annotations for the task of medical image segmentation. In this article, we have reviewed these solutions in depth, summarizing both the technical novelties and empirical results. We hope this review increases the community awareness of the existing solutions for the common limitations of medical image segmentation datasets,  and further inspires the research community to  explore solutions for the less explored dataset problems.

\input{chapters/data_lim_flowchart.tex}

%% file: chapters/data_lim_flowchart.tex
\begin{figure*}[!htb]
\resizebox{\linewidth}{!}{

\begin{tikzpicture}[
    root/.style={%
        edge from parent fork down,
        level distance=2.5cm,
        text centered, text width=5cm},
    problem/.style={%
        edge from parent fork down,
        level distance=2.5cm,
        text centered, text width=3.5cm},
    idea/.style={%
        text centered, text width=2.3cm,
        level distance=2.5cm,
        font=\small,
        fill=gray!10},
    solution/.style={%
        grow=down, xshift=-1cm, % Horizontal position of the child node
        text centered, text width=2cm,
        font=\footnotesize,
        edge from parent path={(\tikzparentnode.200) |- (\tikzchildnode.west)}},
    level1/.style ={level distance=2cm},
    level2/.style ={level distance=4cm},
    level3/.style ={level distance=6cm},
    level 1/.style={sibling distance=12.5cm},
    level 1/.append style={level distance=2.5cm},
    level 2/.style={sibling distance=2.8cm},
    % level4/.style ={sibling distance=4.5cm},  
]
%   \draw[help lines] (0,0) grid (4,3);

    % Title / Root
    \node[anchor=south,root]{Dataset Limitations}
    [edge from parent fork down]

    % Problems, High Level Ideas and Specific Solutions
    child{node [problem] {Scarce Annotations \secref{sec:scarce}}
        child{
            node [idea] {Data\\Augmentation\\\secref{sec:aug}}
            child[solution,level1] {node {Traditional\\Augmentation\\\secref{sec:trad}}}
            child[solution,level2] {node {Mixing\\Augmentation\\\secref{sec:mixing}}}
            child[solution,level3] {node {Synthetic\\Augmentation\\\secref{sec:syn}}}
            }
        child{node [idea] {Leveraging External Labeled Dataets\\\secref{sec:dataset}}
            child[solution,level1] {node {Transfer learning\\\secref{sec:transfer_learning}}}
            child[solution,level2] {node {Domain Adaptation\\\secref{sec:domain_ada}}}
            child[solution,level3] {node {Dataset Fusion\\\secref{sec:fusion}}}}
        child{node [idea] {Cost Effective\\ Annotation\\\secref{sec:cost}}
            child[solution,level1] {node {Active\\Learning\\\secref{sec:active}}}
            child[solution,level2] {node {Interactive\\Segmentation\\\secref{sec:interactive}}}}
        child{node [idea] {Leveraging Unlabeled \\Data\\\secref{sec:unlabeled}}
            child[solution,level1] {node {Self-supervised\\Learning\\\secref{sec:ss}}}
            child[solution,level2] {node {Semi-supervised\\Learning w/ pseudo labels\\\secref{sec:sl}}}
            child[solution,level3] {node {Semi-supervised\\Learning w/o pseudo labels\\\secref{sec:ssl}}}}
                    child{node [idea] {Regularized Training\\\secref{sec:regularized}}
            child[solution,level1] {node {Altered Image Representation\\\secref{sec:altered}}}
            child[solution,level2] {node {Multi-task Learning\\\secref{sec:multi}}}
            child[solution,level3] {node {Shape Regularization\\\secref{sec:shape}}}}
        child{node [idea] {Post Segmentation\\Refinement\\\secref{sec:post}}
            child[solution,level1] {node {Locally Connected CRF\\\secref{sec:local_crf}}}
            child[solution,level2] {node {Full Connected CRF\\\secref{sec:fc_crf}}}
            child[solution,level3] {node {CRF as RNNs\\\secref{sec:e2e_crf}}}}
}
    child{node [problem] {Weak Annotations \secref{sec:weak}}
        child{node [idea] {Learning w/\\Image\\Labels\\\secref{sec:weak_l}}
            child[solution,level1] {node {Class Activation Maps\\\secref{sec:activation}}}
            child[solution,level2] {node {Multiple Instance Learning\\\secref{sec:mil}}}}
        child{node [idea] {Learning w/\\Sparse Labels\\\secref{sec:sparse_l}}
            child[solution,level1] {node {Selective Loss w/ Mask Completion\\\secref{sec:selective_w}}}
            child[solution,level2] {node {Selective Loss w/o Mask Completion\\\secref{sec:selective_wo}}}}
        child{node [idea] {Learning w/\\Noisy Labels\\\secref{sec:noisy_l}}
            child[solution,level1] {node {Robust Loss w/o Mask Refinement\\\secref{sec:robust}}}
            child[solution,level2] {node {Robust Loss w/ Iterative Mask Refinement\\\secref{sec:iterative}}}}};
 
\end{tikzpicture}}
\caption{Organization of this review paper. We broadly categorize the  limitations of medical image segmentation datasets into scarce annotations and weak annotations. For each problem, we then present the  strategies (highlighted in grey) followed by the suggested solutions.% grouped by data or the type of methodology they leverage. %Click on the box to redirect to the corresponding section.
}
\label{fig:problem_tree}
\end{figure*}
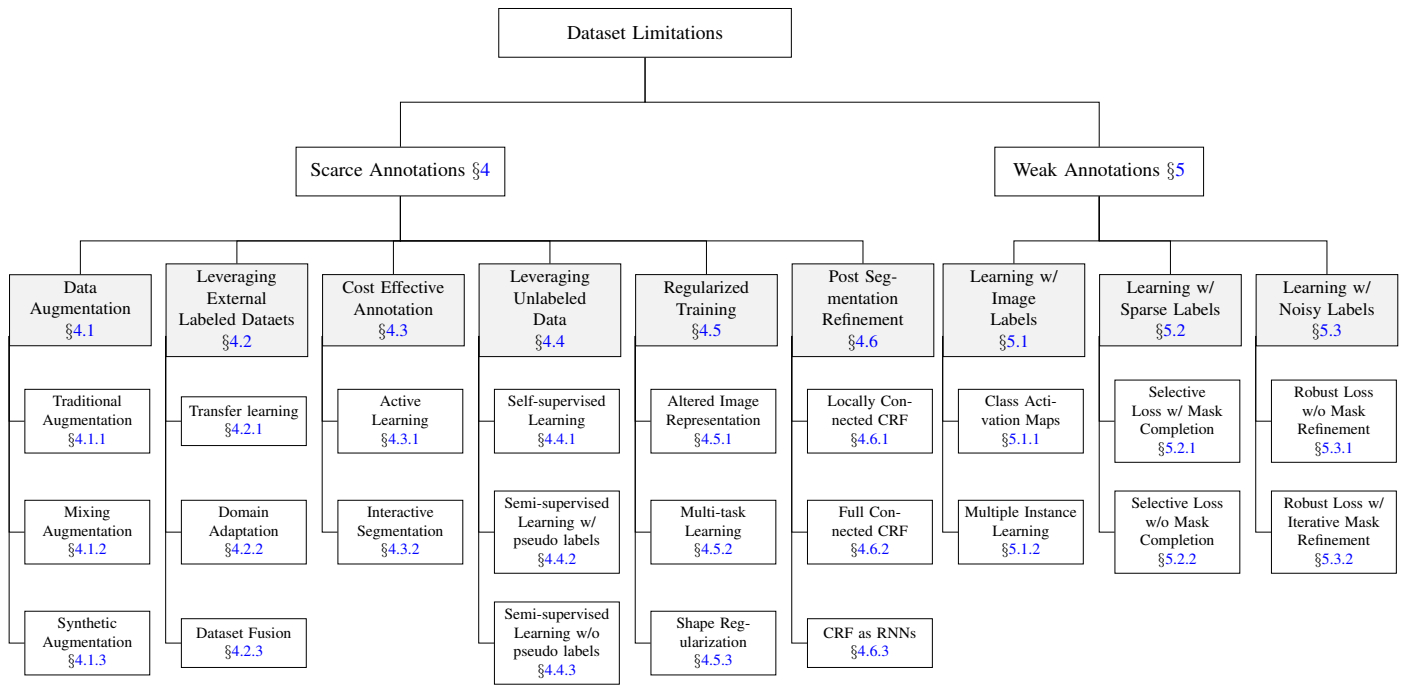

%% file: chapters/related_work.tex
\lref{nt:missed_survey}{\brown \cite{litjens2017survey} surveyed the early deep learning solutions for various medical imaging applications including image classification, object detection, and object segmentation. Following this seminal survey,} \cite{yi2018generative} broadly investigated the use of generative adversarial networks (GANs) in medical imaging. \cite{cheplygina2019not} reviewed semi-supervised, multi-instance learning, and transfer learning in medical image analysis, covering both deep learning and traditional segmentation methods. {\brown The surveys by \cite{hesamian2019deep,taghanaki2019deep} reviewed deep learning techniques suggested for medical image segmentation but with a particular focus on architectural advancements and training schemes. The most relevant surveys to our work are \cite{zhang2019survey}, which reviewed the solutions that tackle the small sample size problem for the broad medical image analysis, and \cite{karimi2019deep} where the authors surveyed the methods suggested for handling label noise in both natural and medical image datasets.}

In contrast, the current survey has focused on problems of scarce and weak annotations with respect to medical image segmentation. Our focus is motivated by the fact that image segmentation requires the strongest supervision among other vision tasks such as classification and detection; and thus, most vulnerable to the quality and quantity of annotations. The specific scope and deep review of this survey distinguish it from \cite{yi2018generative,cheplygina2019not} that broadly cover deep learning for general medical image analysis, from \cite{hesamian2019deep} that focuses on architectural advancements for medical image segmentation, and from \cite{zhang2019survey} that investigates only the small sample size problem in medical image segmentation, {\brown and from the work of \cite{karimi2019deep} that primarily considers label noise in the medical image datasets.}

 %In contrast, the current survey has focused on medical image segmentation rather than the broad medical image analysis, covering strategies for handling both scarce and weak annotations. The specific scope and deep review of this survey distinguish it from \cite{yi2018generative,cheplygina2019not} that broadly cover deep learning for general medical image analysis, from \cite{hesamian2019deep} that focuses on architectural advancements for medical image segmentation, and from \cite{zhang2019survey} that investigates only the small sample size problem in medical image segmentation.

%  similar reviews in the literature, which either broadly cover deep learning for general medical image analysis \cite{yi2018generative,cheplygina2019not} or discuss on partially discuss the techniques for handling limitations with medical image segmentation datasets.

%% file: chapters/traditional_augmentation.tex
Traditional data augmentation has proved effective in reducing over-fitting and improving test performance for both natural and medical images \cite{zhang2016understanding}. The data augmentation methods used in medical imaging can be grouped by the image property they intend to manipulate \cite{zhang2019unseen}. These common image properties consists of image quality, image appearance, and image layout.

%To improve the model performance of medical imaging segmentation task, transfer learning is widely used to fine-tune a pre-trained model obtained from a known domain data set. To further improve the model’s generalization, data augmentation has been proved to be able to reduce over-fitting and improve test performance \cite{zhang2016understanding}. Different from natural 2D images, even within the same modality, 3D images could be visually different from because of different device vendors or scanning protocols. In the field of medical imaging, traditional data augmentation method could be grouped into three main categories \cite{zhang2019unseen}: 1) Image Quality. 2) Image Appearance 3) Spatial Transforms. 

\vspace{4pt}
{\noindent \emph{By image quality:}} Similar to the data augmentation for 2D natural images, image quality can be affected by sharpness, blurriness and noise. \cite{christ2016automatic} apply Gaussian noise to CT scans as part of data augmentation. \cite{sirinukunwattana2017gland} employ Gaussian blur on colon histology images for the task of gland segmentation. \cite{zhang2019unseen} show that data augmentation by adjusting image quality enables the largest performance gain in MR images, with largest improvement coming from image sharpening through the application of unsharp masking. 

\vspace{4pt}
{\noindent \emph{By image appearance:}} Data augmentation by adjusting image appearance consists in manipulating  the statistical characteristics of the image intensities such as brightness, saturation and contrast. \cite{liskowski2016segmenting} apply gamma correction of saturation and value of the HSV color space prior to segmenting retinal blood vessels. \cite{dong2017automatic} employ random enhancement of brightness in 3D MR volumes to enrich the training set for brain tumor segmentation. Contrast augmentation is usually helpful when images exhibit inhomogeneous intensities. For instance, \cite{fu2017nuclei} apply a contrast transformation function on fluorescence microscopy images to enrich the dataset for the task of nuclei segmentation. \cite{alex2017semisupervised} use histogram matching as a form of pre-processing where the 3D MR images are matched against an arbitrarily chosen reference image from the training data.

\vspace{4pt}
{\noindent \emph{By image layout:}} Data augmentation by changing image layout consists of spatial transformations such as rotation, scaling and deformation. \cite{ronneberger_u-net:_2015} show that augmenting the training set with random elastic deformations is key to training a segmentation network with very few annotated images. \cite{milletari2016v} also apply a dense deformation field through a 2x2x2 grid of control-points and B-spline interpolation on the training images. \cite{cciccek20163d} first sample random vectors from a normal distribution in a grid with a spacing of 32 voxels in each direction and then apply a B-spline interpolation.                  

%% file: chapters/mixing.tex
{\brown

 Mixup is a data augmentation method wherein new training images and the corresponding labels are generated through a convex combination of pairs of training images and their labels.  Mixup was originally proposed for the task of image classification; however, its extension to image segmentation is straightforward. Given two images $x_i,x_j$ and their corresponding masks $y_i,y_j$, the new image and mask are computed as follows:
 
 $$\Tilde{x} = \lambda x_i+ (1-\lambda)x_j$$
 $$\Tilde{y} = \lambda y_i+ (1-\lambda)y_j$$
 
 \noindent where $\lambda$ is sampled from a beta distribution. Despite its simplicity, mixup has been highly effective for both natural and medical images. \cite{panfilov2019improving} report improved generalization of a knee segmentation model when mixup is used for data augmentation. Using a linear combination of existing labels, mixup typically generates soft labels. \cite{li2019overfitting} propose asymmetric mixup that turn soft labels generated by mixup into hard labels, which, according to their experiments improve the segmentation of brain tumors in various data regimes.  The success of mixup at the input data space has further inspired its use in the latent feature space, a technique called manifold mixup \cite{verma2019manifold}. Manifold mixup has recently proved effective for prostate  cancer segmentation on MR image \cite{jung2019prostate}, improving Dice by two to four points depending on the neural architecture used for segmentation.
 
 %Another variant of mixup is mixmatch by \cite{} wherein the two labeled images $x_i,x_j$ are not selected randomly; rather, they are sampled to equally represent various classes present in the dataset. 

 }

%% file: chapters/data_synthesis.tex
% \usepackage{graphicx}
\begin{table*}[t]
\caption{Comparison between image synthesis methods suggested for medical image segmentation.}
\label{tab:image_synthesis}
\resizebox{\textwidth}{!}{%
\begin{tabular}{llll}
\hline
Publication                     & Synthesis Type         & Domains                 & Description                                                                                                                                                                             \\ \hline
\cite{chartsias2017adversarial} & Cross-domain synthesis & CT $\rightarrow$ MRI    & \begin{tabular}[c]{@{}l@{}}CycleGAN is used to generate pairs of synthesized MR images from  labeled CT slices \end{tabular}                  \\ \hline
\cite{zhang2018translating}     & Cross-domain synthesis & CT $\leftrightarrow$MRI & \begin{tabular}[c]{@{}l@{}}CycleGAN with shape consistency loss is used to translate between  MR and CT scans %Segmentation and synthesis networks are trained jointly
\end{tabular} 
\\ \hline

\cite{fu2018three} & Same-domain synthesis  & 3D Microscopy & CycleGAN with spatially constraints is used to synthesize 3D microscopy images\\ \hline

\cite{guibas2017synthetic}      & Same-domain synthesis  & Fundus                  & \begin{tabular}[c]{@{}l@{}}Conditional GAN and Vanilla are used to generate a vessel mask and the corresponding fundus image\end{tabular}                               \\ \hline

\cite{shin2018medical}          & Same-domain synthesis  & MRI                     & \begin{tabular}[c]{@{}l@{}}Conditional GAN to generate synthetic MR images given a lesion mask  and a brain segmentation mask\end{tabular}                                            \\ \hline

\cite{jin2018ct}                & Same-domain synthesis  & CT                      & \begin{tabular}[c]{@{}l@{}}Conditional GAN is used to synthesize pleural nodules in the nodule-free CT slices\end{tabular}                                                           \\ \hline

\cite{tang2018ct}                & Same-domain synthesis  & CT                      & Conditional GAN is used to synthesize higher contrast preprocessed images \\ \hline

\cite{tang2019ct}               & Same-domain synthesis  & CT                      & Conditional GAN is used to synthesize CT lymph node images given lymph node mask \\ \hline

\cite{tang2019xlsor}         & Same-domain synthesis  & X-ray                   & Conditional GAN is used to synthesize X-ray images with desired abnormalities \\ \hline

\cite{mahapatra2018efficient}   & Same-domain synthesis  & X-ray                   & Conditional GAN is used to synthesize X-ray images with desired abnormalities \\ \hline

\cite{Kumar2019Mask}     & Same-domain synthesis  & Skin images                   & Conditional GAN is used to synthesize skin images from user-defined lesion masks                                                                        \\ \hline

\cite{zhao2019data}             & Same-domain synthesis  & MRI                     & \begin{tabular}[c]{@{}l@{}}Hybrid spatial-intensity transformation network is used to synthesize MR images \end{tabular}                                       \\ \hline

\cite{chaitanya2019semi}             & Same-domain synthesis  & MRI                     & \begin{tabular}[c]{@{}l@{}}Hybrid spatial-intensity transformation network is used to synthesize task-driven MR images \end{tabular}                                       \\ \hline

\cite{xu2019deepatlas}             & Same-domain synthesis  & MRI                     & \begin{tabular}[c]{@{}l@{}} Spatial transformation network is used to synthesize task-driven MR images \end{tabular}                                       \\ \hline

\end{tabular}%
}
\end{table*}

Synthetic data augmentation methods for medical image segmentation can be broadly grouped into same-domain and cross-domain image synthesis. The former consists of synthesizing labeled data directly in the target domain. The latter, on the other hand, consists of projecting labeled data from another domain to the target domain, which is closely related to the subject of domain adaptation. We therefore postpone a detailed review of cross-domain image synthesis methods until Section~\ref{sec:domain_ada}, where we present a detailed study of domain adaptation techniques.  

{\brown We have summarized the representative approaches of same- and cross-domain image synthesis methods in Table~\ref{tab:image_synthesis}. As seen, cross-domain synthesis is based primarily on CycleGAN whereas same-domain synthesis uses various methodologies including CycleGANs, conditional GANs, and transformation networks. In the following, we review the methods suggested for same-domain image synthesis.}

% \vspace{4pt}
% \noindent\emph{cross-domain synthesis:}
% \cite{chartsias2017adversarial} use cycle GAN to generate pairs of synthesized MR images and the corresponding myocardium masks from pairs of CT slices and their myocardium segmentation masks. The authors base the image synthesis module on CycleGAN, because it does not require the CT and MR images to be registered nor do they have to belong to the same patient. Once the synthetic data generated, the authors train a myocardium segmentation model using both synthetic MR and real MR images, demonstrating 15\% improvement over the myocardium segmentation model trained using only the real MR images. However, \cite{zhang2018translating} demonstrate that the above offline data augmentation may only be partially effective and in some cases can even deteriorate the performance. Instead, they propose a framework wherein both data synthesis model and segmentation model are trained jointly. The authors show that the proposed joint framework enables segmentation models that are superior to those trained using offline augmentation or those with only real data.

% \vspace{4pt}
% \noindent\emph{same-domain synthesis:}
\vspace{4pt}
\noindent\textit{\textbf{Via CycleGANs:}}
{\brown \cite{fu2018three} propose a spatially constrained CycleGAN to generate synthetic 3D microscopy images. The spatial constraints guide the CycleGAN so that the nuclei appear in desired locations and orientations in the synthetic images. The results show that the synthetic images generated by spatially constrained CycleGAN are more effective than CycleGAN in improving the performance of the base segmentation model.}

\vspace{4pt}
\noindent\textit{\textbf{Via conditional GANs:}}
\cite{guibas2017synthetic} propose a framework consisting of a GAN and a conditional GAN to generate pairs of synthetic fundus images and the corresponding vessel masks. Specifically, the GAN takes as input a random vector and then generates a synthesized vessel mask, which is then sent to the conditional GAN to generate the corresponding photo-realistic fundus image. The authors verify the fidelity of the synthesized images by examining whether a classifier can distinguish the synthetic images from the real images, but do not demonstrate whether the synthesized examples enable training a more accurate segmentation model.

{\brown \cite{tang2018ct} train a stacked GAN (SGAN) to pre-process CT images, where the first GAN generates a denoised image and the second GAN generates a high resolution image. The SGAN was trained on a large external dataset (DIV2K- 1000 images \cite{Agustsson_2017_CVPR_Workshops}). Pre-processing using this method resulted in significantly improved segmentation performance on both deep learning (HNN) and non deep learning (GrabCut) approaches on the DeepLesion dataset
\cite{yan2018deeplesion}. \cite{tang2019xlsor} propose a 2-stage framework for lung segmentation in chest X-ray (CXR) where a segmentation model is first trained on 280 real images, and then fine-tuned using 5000 synthetic CXR. The authors use 
a pix2pix network \cite{huang2018multimodal} for image synthesis, which transforms an image of a healthy CXR into one with pathology. They observe that across different segmentation models, this augmentation significantly increases precision, recall, and Dice score.}

\cite{shin2018medical} use a conditional GAN to generate synthetic MR images given a lesion mask and a brain segmentation mask. Once  trained, the synthesis network can generate synthesized MR images given a user-defined tumor mask. The elegance of this approach is in how the user can rescale or relocate a tumor in the mask and then the synthesis network can generate the MR image in accordance to the new size and location of the tumor. Without typical data augmentation, the tumor segmentation model trained using both synthetic and real MR images achieves a significant performance gain over the model trained using only real MR images. However, the performance gap is largely bridged in the presence of typical data augmentation. 

{\brown \cite{tang2019ct} use a mask-guided GAN to augment their lymph node segmentation dataset. For this purpose, the authors use  pairs of lymph node images and segmentation masks from 124 patients. The trained GAN then generates 5000 lymph node images, each generated based on a user-provided mask. Augmenting the dataset with 5000 synthesized images significantly improves all performance metrics.}
In a similar spirit, \cite{mahapatra2018efficient} use a conditional GAN to synthesize X-ray images with desired abnormalities. The model takes as inputs an X-ray with an abnormality and a lung segmentation mask, and then it generates a synthesized X-ray that has the same diseases as the input X-ray while taking the image appearance that matches the provided segmentation mask. This approach has the capability of generating many synthesized diseased images from one real diseased image. {\brown A similar approach is also adopted by \cite{Kumar2019Mask} where conditional GAN is trained to generate synthesized skin images from user-defined lesion masks. The authors show that synthesized images, when combined with traditional augmentation, achieve 4 points increase in Dice over the same model trained using only traditional data augmentation.}

Lung segmentation is challenging in the presence of large pleural nodules,  which are  often under-represented in the training sets. To overcome this limitation, \cite{jin2018ct} train an image in-painting model based on a conditional GAN that can synthesize pleural nodules in the nodule-free CT slices.  %Receiving as the input an image  with the nodule region being masked, the CGAN, trained using reconstruction and adversarial losses, attempts to infer the nodular content. Having trained the CGAN, the authors feed to the network the images with masked regions along the pleura, and then the network injects a nodule in each masked pleural region that blends well with the image context. 
{\brown The authors test the lung segmentation model using 34 images with peripheral nodules from the LIDC dataset, demonstrating that the model trained with the synthetic data achieves 2 points increase in Dice over the model trained using only real images.}

\vspace{4pt}
\noindent\textit{\textbf{Via transformation networks:}} \cite{zhao2019data} propose a data synthesis method to generate pairs of brain MR images and the segmentation masks from only one  labeled MR image. For this purpose, the authors suggest a hybrid spatial-intensity transformation model. The spatial transformation network deforms the labeled image so it takes the spatial layout of a given unlabeled image. Once the layout is taken care of, the intensity transformation network changes the intensity at each pixel so the labeled image takes the appearance of a given unlabeled image. Together, the two transformation networks enable the generation of new labeled examples from a reference labeled image and a number of unlabeled images. For the task of brain structure segmentation, the suggested data augmentation method enables four points increase in Dice over a model trained using traditional data augmentation and 3 points increase in Dice over atlas-based data augmentation. {\brown Noteworthy, the suggested method is tested in a 1-shot medical image segmentation setting, where only one labeled  example is available for training. It is not clear whether the performance gain holds up in the presence of larger labeled training sets. 

\lref{nt:task_based_aug} Concurrent to the work above,  \cite{chaitanya2019semi} propose a few-shot image segmentation model based on a task-driven data augmentation method, wherein an intensity and  a deformation network generate synthetic pairs of image-mask to enrich the training set. The two transformation networks are conditional generators, which are trained in an adversarial manner so that the transformed images resemble the appearance of labeled and unlabeled images in the dataset. Also, to ensure that the synthetic images are relevant to the target task (segmentation), the transformation networks are trained jointly with the segmentation network by feeding the synthetic images to the segmentation network. The authors test the model for cardiac segmentation in MR images from 20 subjects, demonstrating marked improvement in Dice when only 1 or 3 labeled images are used for training. The authors report even larger improvement when they combine their synthetic  images with mixup~\cite{zhang2018mixup}.}

%% file: chapters/multiple_datasets.tex
%%% LEVERAGING EXTERNAL LABELED DATASETS %%%

% {\brown 
% The problem of scarce annotations can be mitigated through model pre-training or joint training based on external labeled datasets. The former is commonly known as transfer learning, which consists of pre-training the target model on an external labeled dataset followed by fine-tuning the model on the target scarce dataset. The latter, on the other hand, consists of training the target model using both the scarce and  external labeled datasets. The three common forms of this paradigm are domain adaptation with target labels, domain adaptation without target labels, and dataset fusion. In the following, we describe the representative works of the above four methodologies.}

\begin{figure*}
    \includegraphics[width=0.99\linewidth]{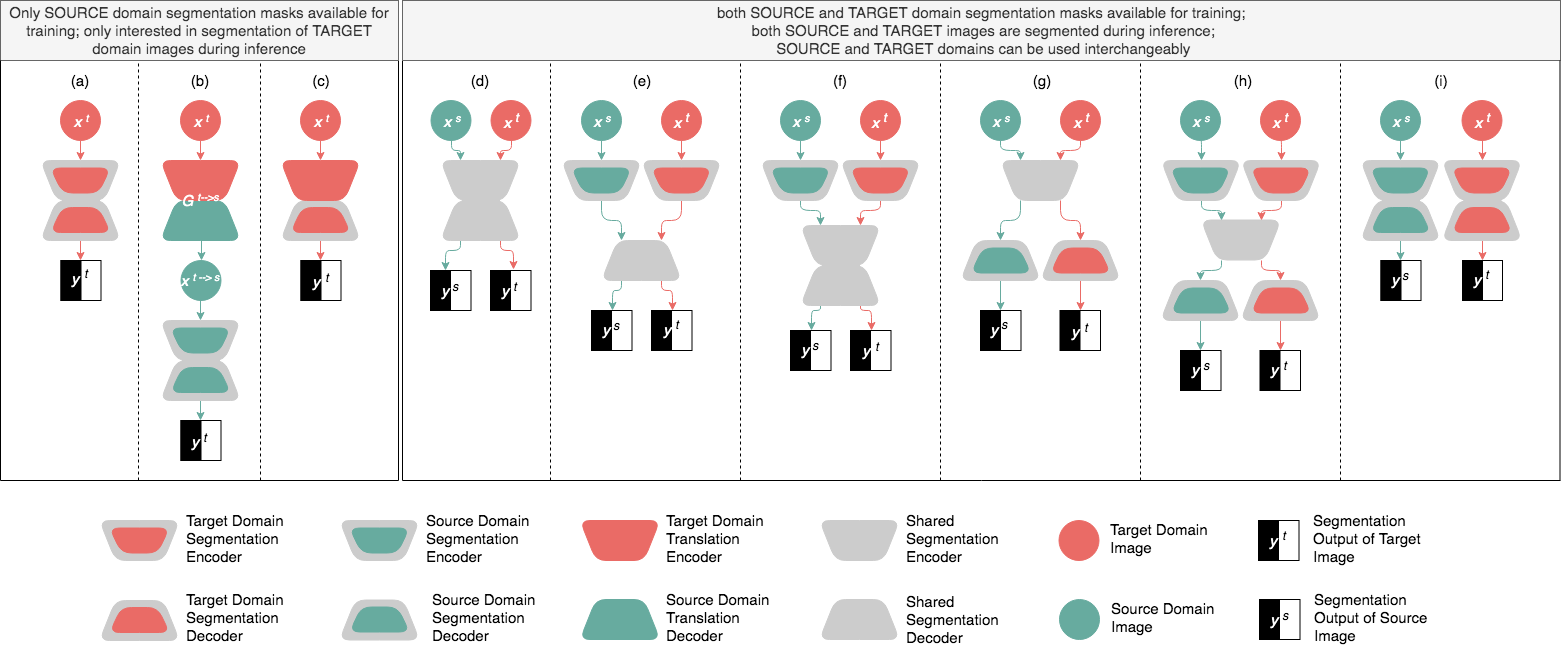}
    \caption{Leveraging external labeled datasets is effective for the problem of scarce annotations. This figure compares the  data flow during inference for the related solutions: (a)-(c) cover approaches that only use source domain labels during training while (d)-(i) cover approaches that make use of both source and target domain labels, in which case the terms 'source' and 'target' are no longer meaningful and can be used interchangeably. (a) Target domain images are directly passed through the segmentation network trained on images from the target domain and images translated to the target domain. (b) Target domain images are converted to the source domain and then sent to the segmentation network trained on the source domain and source-like images. (c) Target domain images are sent through the target domain encoder (belonging to the domain translation network) and then sent to the segmentation decoder trained on target domain and target-like images. (d) Images from either domain can be passed through the segmentation network trained jointly on both domains. (e) Both domain images are passed through their own domain specific encoders and then through the segmentation decoder trained on both domains. (f) Similar to (e) but now the domain specific encoded feature maps are sent through a jointly trained segmentation encoder and decoder, (g) Images from both domains pass through a jointly trained segmentation encoder and then pass through domain specific decoders, (h) Each domain has its own specific encoder and segmentation decoder, but pass through a shared segmentation encoder in between. (i) Each domain has its own segmentation network during inference which is trained using data from its own domain augmented using domain translation.}
    \label{fig:dom_ada}
\end{figure*}

{\brown The problem of scarce annotations can be alleviated by employing external labeled datasets via transfer learning, domain adaptation or dataset fusion techniques. Transfer learning typically involves model pre-training, wherein a large external labeled dataset is used to train an initial model, which can then be fine-tuned using the target dataset. Domain adaptation techniques attempt to bridge the distribution gap between the different datasets by either learning a common latent representation or by learning to translate images from one domain to
the other. Dataset fusion, on the other hand, simply utilizes data from one or multiple external datasets to train a general segmentation model having superior performance to those trained on each individual dataset. We have compared the inference stage of the aforementioned methodologies in \figurename~\ref{fig:dom_ada} where \lref{lj:fig_skip_connections} architecture details such as skip connections and dense blocks are not shown for the sake of grouping the overarching ideas together. We have further listed the representative works of the above methodologies in Table~\ref{tab:domada_methods}.}

%Note that this figure does not show the building blocks of the architectures used. Hence, skip connections and dense blocks which are now commonly used in segmentation networks may be used in these approaches but are not explicitly shown for the sake of grouping the overarching ideas together.

%In the following, we describe the representative works of each of these three methodologies.

% A bird's eye view of the papers discussed in this subsection is shown in Table~\ref{tab:domada_methods}.

%\lref{lj:fig_skip_connections}{\brown \figurename~\ref{fig:dom_ada} shows a simple schematic of the data flow at inference for the papers reviewed in this section. Note that this figure does not show the building blocks of the architectures used. Hence, skip connections and dense blocks which are now commonly used in segmentation networks may be used in these approaches but are not explicitly shown for the sake of grouping the overarching ideas together.}

%This subsection covers papers that use domain adaptation when the annotations for any one domain are limited or even absent and also discusses approaches that leverage multiple datasets for building universal models in lieu of individual models tailored to each dataset.

% FIGURE
% \begin{center}

\begin{table*}[t]
\caption{Overview of the papers leveraging external labeled datasets. The suggested method, among other factors, differ in terms of presence of target domain labels and the domain in which segmentation is performed. The Figure column on the right shows the matching data flow from \figurename~\ref{fig:dom_ada}.}
\label{tab:domada_methods}
\begin{tabular}{lcccc}
\hline
\multicolumn{1}{c}{Publication} & \begin{tabular}[c]{@{}c@{}}Availability of Target\\ Domain Segmentation Masks\end{tabular} & Segmentation Domain & Modality & Figure \\ \hline
\multicolumn{5}{l}{Transfer Learning} \\

\cite{ma2019neural} & \cmark & Target &  2D$\rightarrow$2D& (a) \\ 
\cite{qin2019transfer} & \cmark & Target & 2D$\rightarrow$2D & (a) \\ 
\cite{liu20183d} & \cmark & Target & 2D$\rightarrow$3D  & (a) \\ 
\cite{yu2018recurrent} & \cmark & Target & 2D$\rightarrow$3D   & (a) \\ 

\hline
% \multicolumn{5}{l}{Domain Adaptation without} \\ Target Labels \\
\multicolumn{5}{l}{Domain Adaptation} \\
\cite{huo2018adversarial} & \xmark & Target & MRI, CT & (a) \\
\cite{huo2018synseg} & \xmark & Target & MRI, CT & (a) \\
\cite{chen2019unsupervised} & \xmark & Target & bSSFP, LGE & (a) \\
\cite{chen2018semantic} & \xmark & Source & X-ray & (b) \\
\cite{zhang2018task} & \xmark & Source & DRR, X-ray & (b) \\
\cite{chen2019synergistic} & \xmark & Target & MRI, CT & (c) \\
\cite{Giger2018} & \xmark & Source & MRI, CT & (b) \\ 
% \hline
% \multicolumn{5}{l}{Domain Adaptation with} \\ Target Labels \\ 
\cite{chartsias2017adversarial} & \cmark & Both & MRI, CT & (i) \\ 
\cite{zhang2018translating} & \cmark & Both & MRI, CT & (i) \\ 
\cite{dou2018pnp} & \cmark & Both & MRI, CT & (e) \\
\cite{Valindria2018} & \cmark & Both & MRI, CT & (d),(e),(f),(g),(h) \\
\hline
\multicolumn{5}{l}{Dataset Fusion} \\ 
\cite{Harouni2018} & \cmark & All domains & MRI,CT,US,X-ray & (d) \\
\cite{dmitriev2019learning} & \cmark & All domains & CT & (d) \\ \hline
\end{tabular}
\end{table*}

% % TABLE
% \begin{table*}[]
% \caption{Overview of the papers leveraging multiple datasets. The suggested method, among other factors, differ in terms of presence of target domain labels and the domain in which segmentation is performed. The Figure column on the right shows the matching data flow from \figurename~\ref{fig:dom_ada}.}
% \label{tab:domada_methods}
% \begin{tabular}{|l|c|c|c|c|}
% \hline
% \multicolumn{1}{|c|}{Publication} & \begin{tabular}[c]{@{}c@{}}Availability of Target\\ Domain Segmentation Masks\end{tabular} & \textbf{Segmentation Domain} & \textbf{Modality} & \textbf{Figure} \\ \hline
% \multicolumn{5}{|l|}{\textbf{Domain Adaptation without Target Labels}} \\ \hline
% \cite{huo2018synseg} & \xmark & Target & MRI, CT & (a) \\
% \cite{chen2018semantic} & \xmark & Source & X-ray & (b) \\
% \cite{zhang2018task} & \xmark & Source & DRR, X-ray & (b) \\
% \cite{chen2019synergistic} & \xmark & Target & MRI, CT & (c) \\
% \cite{Giger2018} & \xmark & Source & MRI, CT & (b) \\ \hline
% \multicolumn{5}{|l|}{\textbf{Domain Adaptation with Target Labels}} \\ \hline
% \cite{dou2018pnp} & \cmark & Both & MRI, CT & (e) \\
% \cite{Valindria2018} & \cmark & Both & MRI, CT & (d),(e),(f),(g),(h) \\
% \cite{zhang2018translating} & \cmark & Both & MRI, CT & (i) \\ \hline
% \multicolumn{5}{|l|}{\textbf{Dataset Fusion}} \\ \hline
% \cite{Harouni2018} & \cmark & All domains & MRI,CT,US,X-ray & (d) \\
% \cite{dmitriev2019learning} & \cmark & All domains & CT & (d) \\ \hline
% \end{tabular}
% \end{table*}

\lref{lj:new_sec_tl}
\subsubsection{Transfer Learning}
\label{sec:transfer_learning}
\lref{lj:fine-tuning}{\brown When dealing with small medical image datasets it is possible to leverage the power of non-medical image data as well. Transfer learning from natural images has been widely adopted for medical image classification \cite{tajbakhsh16,shin2016deep}; however the application to medical image segmentation has been scarce. This trend is in part due to the 3D nature of medical images, which hampers transfer learning from 2D models trained on natural images, and also partially due to the promising performance of shallower segmentation networks in medical imaging, which unlike deep models may not benefit from fine-tuning. Nevertheless, we briefly describe the two common scenarios for transfer learning in medical imaging and then introduce the works that enable transfer learning from 2D models to 3D medical applications.

\vspace{4pt}
\noindent\textit{\textbf{$2D\rightarrow 2D$:}} There are two main approaches to transfer knowledge from natural images to 2D medical image segmentation models. The first approach  (e.g., \cite{ma2019neural}) is to fine-tune an autoencoder that is pre-trained for the task of image segmentation in natural images. The advantage of this approach is that both encoder and decoder are pre-trained, but the disadvantage is that natural image segmentation datasets are not massive. The second approach (e.g., \cite{qin2019transfer}) is to append a randomly initialized decoder to an encoder pre-trained for the task of image classification in natural images, followed by fine-tuning the entire network. This approach has the advantage of knowledge transfer from a massive natural image classification dataset, but the disadvantage is that the decoder needs to be initialized from scratch.

%The knowledge learned from natural images can be leveraged for medical image segmentation by 1) fine-tuning segmentation models whose encoders and decoders are pre-trained the task of image segmentation in natural images on datasets such as Microsoft COCO (e.g., \cite{ma2019neural}), or 2) training segmentation models whose encoders are pre-trained for the task of image classification in natural images on massive datasets such as ImageNet.

\vspace{4pt}
\noindent\textit{\textbf{$2D\rightarrow 3D$:}} The aforementioned approaches, while effective, are applicable to only 2D medical image segmentation. Knowledge transfer from 2D models pre-trained on natural images to the models targeted at 3D medical applications has been a little explored topic. \cite{yu2018recurrent} transfer models based on natural scene video by treating the third dimension of medical scans as a temporal axis. This approach however may fail to capture the 3D context of medical scans.   \cite{liu20183d} propose to turn a 2D model into a 3D network by extending 2D convolution filters into 3D separable anisotropic filters. With this approach, one can use 2D models to initialize 3D models for target medical image segmentation applications.}

%For the task of segmentation, however, there do not exist such large datasets at the same scale as ImageNet, so network pre-training is not always a viable solution. It is still possible to use these pre-trained weights in the encoder of the segmentation network. In transfer learning the external dataset is only needed for the first stage of pre-training and is no longer used, hence publicly available pre-trained models can ease the burden of having to perform this additional training step.}

% What are the sub-problems and how does each solution address them
% Annotations from external labeled datasets can be leveraged either through domain adaptation techniques or through dataset fusion. 

\subsubsection{Domain Adaptation}
\label{sec:domain_ada}

% Why is this a `Scarce Annotations' Solution
A frequently encountered obstacle in medical imaging is that of a distribution shift between the data available for training and the data encountered in clinical practice. This shift could be caused by using different scanners and image acquisition protocols or due to imaging different patient populations and ethnicities. As individual datasets tend to be small and typically originate from a single institution, they are inherently biased and the resulting models tend to perform poorly in the real world. Given the limitations of individual datasets, a natural workaround is to incorporate multiple datasets for training. %However, as discussed in the previous subsections, access to high quality labeled annotations is hard to come by even for single datasets, much less for several datasets. 
Domain adaptation techniques attempt to bridge the gap between multiple domains by either learning a latent representation that is common to these various domains or by learning to translate images from one domain to the other. These domains may consist of different imaging modalities or different image distributions within the same modality.

\lref{lj:gan_intro}{\brown A recurring theme in many of the domain adaptation papers discussed in this section is the use of GANs, CycleGANs, or some sort of adversarial loss for the purpose of image reconstruction. Therefore, we first briefly explain these methods and then cover their applications to medical image segmentation. GANs by \cite{goodfellow2014generative} make use of dual networks: a generator and discriminator, which are trained to compete against each other. The discriminator is trained to distinguish between real and synthetic images and the generator is trained to synthesize realistic images that the discriminator cannot distinguish from the real images. When this is used in the context of domain translation, the generator learns a mapping from one domain to another. CycleGANs by \cite{zhu2017unpaired} achieve this by using two pairs of generators, each with its own discriminator, one to map from the source to the target domain and the other for the inverse mapping. In addition to the adversarial loss a cycle consistency loss ensures that the result of the mapping followed by the inverse mapping is identical to the input.} \lref{lj:munit}{\brown Multimodal unsupervised image-to-image translation (MUNIT) \cite{huang2018multimodal} differs from CycleGAN in that the generators, that translate between the domains, are each composed of an encoder that explicitly disentangles the domain-invariant structure of the images from their domain-specific style before passing both to the decoders. While CycleGAN provides a one-is-to-one mapping between the two domains, MUNIT allows a one-is-to-many mapping by sampling from the style encoding distribution.}

\vspace{4pt}
\noindent\textit{\textbf{Domain Adaptation without Target Labels:}}
% \subsubsection{Domain Adaptation without Target Labels}
% \label{sec:domain_wo}
When the test domain (a.k.a the target domain) labels are unavailable, but we only have access to labels from a different domain (a.k.a the source domain), the popular approach is to convert one domain to the other.

\vspace{4pt}
\noindent\textit{Source $\,\to\,$ Target:}
In the absence of target labels, one approach is to convert the source domain images to have the style of the target domain while retaining the anatomical structure and thereby the segmentation masks of the source domain. Then, a segmentation network trained on the target-styled images and source masks can be used to make predictions on the target images.
\cite{huo2018adversarial} suggest a joint image synthesis and segmentation framework that enables image segmentation for the target domain using  unlabeled target images and labeled images from a source domain. The intuition behind this joint optimization is that the training process can benefit from the complementary information between the synthesis and segmentation networks. In this framework, the main job is done by the image synthesis network, a CycleGAN, that converts the labeled source images to synthesized target images. The synthesized target images are used to train the segmentation network. At test time, the real images from the target domain are directly submitted to the segmentation network to obtain the desired segmentation masks. The authors evaluate this framework for the task of spleen segmentation in CT scans where the {\brown 19 target abdominal} CT scans do not have the segmentation masks, but the {\brown 60 source abdominal} MR images come with spleen masks. Experimental results show that the model trained {\brown (leave-one-out cross validation using 19 CT scans)} using synthesized CT scans can achieve a performance level at par to the model trained using real CT scans with labels. The authors further extend their work in
\cite{huo2018synseg} for the task of splenomegaly and total intracranial volume segmentation, reporting 2\% improvement in Dice over the existing state of the art---a 2-stage CycleGAN followed by a separate segmentation network. For the task of total intracranial volume segmentation, the Dice coefficient using domain adaptation is only 1\% lower than the Dice coefficient of the model trained with the target labels (upper bound).

\cite{chen2019synergistic} perform domain translation from the MR to CT domain for the task of heart CT segmentation using only MR image masks. They propose the use of a CycleGAN for conversion from MR to CT and vice-versa {\brown (20 volumes each with a 80-20 training-testing percentage split)} with a segmentation network trained on the real and generated CT (target) images. The novelty of their approach lies in the use of a shared encoder common to both the CT segmentation network and the CT to MR generator network, which makes use of this multitask setting to prevent the segmentation encoder from over-fitting. The authors report a 9\% improvement in Dice over the existing state of the art domain adaptation techniques. 

{\brown \cite{chen2019unsupervised} makes use of MUNIT to translate between balanced steady-state free precession (bSSFP) images having masks for 3 cardiac structures and late-gadolinium enhanced (LGE) images that don’t have any masks. The framework is trained and evaluated on the multi-sequence cardiac MR segmentation challenge (MS-CMRSeg 2019) dataset. The translation network is trained with 40 images from each domain and this is used to create a synthetic dataset of 150 target domain LGE images from 30 bSSFP images (sampled 5 times from the style encoding distribution). The synthetic LGE images with the original bSSFP masks are then used to train the segmentation network. The authors evaluate their approach on the five validation images provided by the challenge and show a 10\% increase in Dice compared to a registration-based approach. \lref{lj:style_xfer}}

\vspace{4pt}
\noindent\textit{Target $\,\to\,$ Source:}
Alternatively, one can convert the target domain images into the source domain followed by training the segmentation model using the source images. During inference, the target images are first converted to the source domain and then fed to the segmentation network to generate the segmentation maps.
 
 \cite{Giger2018} propose converting the CT (target) domain to the MR (source) domain and then using an existing atlas-based algorithm (MALP-EM) to perform the segmentation on the converted MR images. The motivation is that it is easier to obtain segmentation annotations for brain MRI  than brain CT scans. They use a modified U-Net for the domain conversion, which requires the {\brown 10 pairs of} CT and MR images to be registered beforehand {\brown and then perform multi-modal registration using 15 atlases}. %The ground truth for each CT image is its paired registered MR image. 
 On average, they improve the Dice score by 9\% over a baseline that performs segmentation in the CT domain.

 \cite{chen2018semantic} use the CycleGAN with an additional semantic adversarial loss, which is used to distinguish between source segmentation masks and segmentation predictions of the converted target to source images. The authors evaluate their proposed method on 2 different X-ray datasets, which vary in disease type, intensity, and contrast {\brown (source: Montgomery set with 138 cases; target: JSRT set with 247 cases; 70-10-20\% training-validation-testing split)}. They achieve 2\% improvement in Dice  over the baseline CycleGAN performance.

 Given a set of annotated CT scans, \cite{zhang2018task} aim to segment X-ray images without having any X-ray segmentation annotations. For this purpose, the authors first convert annotated CT scans to digitally reconstructed radiographs (DRRs) via a 3D to 2D projection, and then learn a mapping between {\brown 815} DRRs and {\brown 73 training} X-ray images. The mapping is performed by a task-driven GAN, which is a CycleGAN with an additional segmentation loss to generate segmentation masks for the DRR-style images. 
%The discriminator is also conditioned on both the image and its segmentation prediction. 
With these new constraints, {\brown tested on 60 X-ray images,} the suggested method improves the segmentation Dice by two or three points over using either one of them alone and over the vanilla CycleGAN.

%\cite{gsaxner2019exploit} use domain adaptation on the labels rather than the images by thresholding PET images to act as the label masks for their corresponding paired CT images. They are then able to train a segmentation network using the CT images and synthetic ground truth. They are able to bring up the Dice score by 3 points using this form of label augmentation.

\vspace{4pt}
\noindent\textit{\textbf{Domain Adaptation with Target Labels:}}
% \subsubsection{Domain Adaptation with Target Labels}
% \label{sec:domain_w}
If the segmentation masks are available for both domains, there is no longer a distinction between the choice of source and target domains. In this scenario, domain adaptation is achieved by learning a shared feature encoding, allowing the segmentation network to predict meaningful masks regardless of the input domain.

\cite{chartsias2017adversarial} use CycleGAN to generate pairs of synthesized MR images and the corresponding myocardium masks from pairs of CT slices and their myocardium segmentation masks. The authors base the image synthesis module on CycleGAN, because it does not require the {\brown 15 training} CT and {\brown 15 training }MR images to be registered nor do they have to belong to the same patient. Once the synthetic data is generated, the authors train a myocardium segmentation model using both synthetic MR and real MR images, demonstrating 10\% improvement over the myocardium segmentation model trained using only the real MR images{\brown , when tested on 5 MR images}. 
% However, \cite{zhang2018translating} demonstrate that the above offline data augmentation may only be partially effective and in some cases can even deteriorate the performance. Instead, they propose a framework wherein both data synthesis model and segmentation model are trained jointly. The authors show that the proposed joint framework enables segmentation models that are superior to those trained using offline augmentation or those with only real data.

However, \cite{zhang2018translating} demonstrate that the above offline data augmentation may only be partially effective and in some cases can even deteriorate the performance. Instead, they propose a framework wherein both data synthesis model and segmentation model are trained jointly. They develop a segmentation network that can segment heart chambers in both CT and MR images by learning a translation between the two domains. They use a CycleGAN as their backbone and further add a shape consistency loss to ensure anatomical structure invariance during translation. {\brown Their dataset makes use of 142 CT volumes to match the number of MRI volumes. For both modalities, 50\% data is used as training and validation, and the remaining  50\% as testing data.} They improve the Dice score on CT images by eight points and MR images by two points over other methods that use both real and synthetic data for training.

\cite{dou2018pnp} train a cardiac segmentation network, consisting of two parallel domain-specific encoders and a shared decoder. During training, the decoder takes its input from a single encoder depending on the domain of the input image. The network is trained so that the decoder yields similar high-level semantic embedding for images of both domains. This is achieved by a discriminator that is trained to distinguish between the two domains. {\brown The authors use the challenge dataset by \cite{zhuang2019evaluation}, which contains 20 subjects with MR images and masks, and an additional 20 non-overlapping subjects with CT images and masks. Using 16 subjects from each modality for training and 4 for testing,} they achieve substantial performance boost over single domain training and 2\% improvement in Dice over other domain adaptation techniques. 

For the case where both source and target labels are available, domain adaptation is achieved using shared latent representations between the two domains, but the location of the shared features is a network design choice. \cite{Valindria2018} evaluate the performance of four different locations for the shared latent representations: 1) separate encoders with a shared decoder (see \figurename~\ref{fig:dom_ada}(e)), 2) separate initial streams, followed by a shared encoder and decoder (see \figurename~\ref{fig:dom_ada}(f)), 3) shared encoder and separate decoder streams (see \figurename~\ref{fig:dom_ada}(g)) and finally, 4) separate encoder and decoder streams with a shared latent representation in-between (see \figurename~\ref{fig:dom_ada}(h)). They compared these variants with a baseline consisting of a single-stream encoder-decoder segmentation network, which is trained with data from both domains (see \figurename~\ref{fig:dom_ada}(d)).  {\brown The authors perform 2-fold cross-validation based on 34 subjects for MRI and 30 subjects for CT.} Their results showed that the baseline was actually at par with or in some cases outperformed most variants, the only exception being the fourth variant, which consistently outperformed the baseline and other dual stream variants.

\subsubsection{Dataset Fusion}
\label{sec:fusion}

% Dataset fusion, on the other hand, simply utilizes data from different sources to train a general segmentation model having superior performance to those trained on each individual dataset.
 
Dataset fusion techniques leverage multiple datasets to train a universal segmentation model based on heterogeneous, disjoint datasets, offering two advantages: 1) more efficient training, as multiple models are consolidated into a single model, and 2) enhanced regularization, as data from multiple sources can provide further supervision. Domain adaptation and Dataset fusion both aim to leverage multiple datasets; however, they take different approaches: the former does this by minimizing the domain shift, whereas the latter does so by learning to discriminate between domains.

%When trying to leverage data from multiple datasets, apart from trying to minimize the distribution shift, there is also work on learning the distribution shift, wherein the network learns to segment images from different datasets by learning to discriminate between them.
It is inefficient to have modality-specific models to segment the same organs across different modalities. 
\cite{Harouni2018} propose a modality independent model that is jointly trained using data from all modalities. The network architecture is a modified U-Net with the base U-Net performing multi-organ segmentation and  a classification head added to the bottleneck layer, which performs the modality/viewpoint classification (7 classes: X-ray, short axis MRI, 2-chamber MRI, 4-chamber MRI, CT, ultrasound 4-chamber B-mode, Doppler ultrasound). The authors compared their jointly trained universal network against individually trained U-Nets for each task, {\brown using data from multiple sources split at the patient-level such that 65\% (2781 2d images) was used for training and the remaining 35\% (1016 2d images) for validation.} The results show that  the universal network usually performed at par with or outperformed the specialized networks. The exception to this performance was seen for left ventricle segmentation, where a dedicated MRI model showed significantly higher performance.

\cite{dmitriev2019learning} train a multi-organ segmentation model using data from multiple single organ datasets. For this purpose, the authors add an additional channel, which is filled with a class-specific hash value, to each layer of the decoder network, conditioning the segmentation predictions on the class labels. The drawback, however, is that the test image with an unknown organ label must be fed in `m` times sequentially to condition on all the possible classes.  {\brown The authors use 20 volumes with liver masks from the publicly available Sliver07 dataset, 82 volumes with pancreas masks from the publicly available NIH pancreas dataset, and 74 volumes of their own additional dataset of liver and spleen segmentation wherein each dataset was divided into training and testing sets with an 80/20 ratio.} The multi-dataset training scheme achieves 1.5\% improvement in Dice over the state of the art single dataset approaches.

%at inference time to get the complete multi-class segmentation result (with m-classes), the test image needs to be fed in `m` times sequentially to condition on all the possible classes. Their experiments show that this form of conditioning performs best when only the layers of the decoder are conditioned. On average their Dice score improves by 1.5\% over the state of the art single dataset approaches.

\subsubsection{Summary}
% In this section, we reviewed papers that have an edge over the existing state of the art through the assistance of additional labeled datasets. The most obvious way to accomplish this is to train a joint segmentation model with data collected from multiple sources. However, as discussed in Section \ref{sec:fusion}, the novelty of these approaches lies in discriminating between the different sources in addition to learning a joint segmentation model. This still requires annotated data from multiple datasets. Alternatively, it has been seen that domain adaptation lends itself to bridging the divide between different image distributions (Section \ref{sec:domain_w}) and helps improve the segmentation performance over single distribution training by increasing the input variability. However, this comes with the added complication of adversarial training, particularly when training in the absence of target domain labels (Section \ref{sec:domain_wo}).

In this section, we reviewed techniques that utilize additional labeled datasets to enhance the segmentation performance over the counterpart models trained using data from a single domain. {\brown 
These methods fell into three distinct categories: (1) transfer learning (Section~\ref{sec:transfer_learning}); 2) domain adaptation with and without target annotations (Section ~\ref{sec:domain_ada}), where the former is used to translate in the absence of target domain annotations whereas the latter learns a shared feature representation between the two domains, and (3) dataset fusion (Section~\ref{sec:fusion}), which learns to discriminate between the domains in order to condition the segmentation based on the domain of the input image. In general, the models trained with target domain annotations are bound to generalize better to the target domain; however, the appealing feature of domain adaptation methods without target annotations is independence from the target domain annotations, which makes these methods the only viable solution to deal with unlabeled target domain images.
The majority of domain adaptation techniques require some form of adversarial training, making them tricky to train. On the other hand, transfer learning and dataset fusion require minimal changes to the network architecture, allowing for simple and successful joint dataset training.}

%% file: chapters/active_learning.tex
% \begin{figure}
%     \centering
%     \includegraphics[width=0.9\linewidth]{pics/al_code.png}
%     \caption{The pseudocode for active learning. Given a base segmentation model and an unlabeled dataset, active learning algorithm judiciously selects the next batch of samples for annotation by experts.}
%     \label{fig:al_code}
% \end{figure}

\begin{table*}[ht]

\resizebox{0.99\textwidth}{!}{%
% \small
\begin{tabular}{llcccl}
%  \begin{tabular}{c  c c  c  c  c  c } 
 \hline
 Publication          & Query mode \quad \quad& \multicolumn{3}{c}{Sample selection strategy}    &  Annotation unit  \\ 
 %\cline{2-4} 
                      &  & Informativeness &  Diversity & Annotation cost   \quad \quad    &                                              \\
\hline
\cite{gorriz2017cost} &  Iterative& \checkmark &                   &              &  Whole 2D image                                 \\
\cite{yang2017suggestive} &  Iterative& \checkmark & \checkmark          &         &  Whole 2D image                               \\
\cite{ozdemir2018active}&  Iterative& \checkmark &  \checkmark      &         &  Whole 2D image                                \\
\cite{kuo2018cost}    &  Iterative&  \checkmark & \checkmark          & \checkmark  &  Whole 3D image                               \\
\cite{sourati2018active} &  Iterative& \checkmark &  \checkmark      &         &  2D image patch                               \\
 ~\cite{mahapatra2018efficient} & One-shot & \checkmark   &           &   &  Whole 2D image\\
\cite{sourati2019intelligent}  &  Iterative& \checkmark &  \checkmark      &         &  2D Image patch                               \\
\cite{zheng2019biomedical} & One-shot&   &    \checkmark               &              &  2D Image patch\\
\hline

\end{tabular}
}
% \end{center}
\caption{Comparison between active learning methods for medical image segmentation. The suggested methods differ in terms of the definition of the annotation unit and the criteria by which these units are selected for the next round of annotation.}
\label{tab:al_methods}
\end{table*}

Active learning is a cost-effective approach to enlarge the training datasets; and thus, it is highly amenable to the problem of limited annotation budget in medical image segmentation where clinical experts have limited availability,  annotation cost is high, and the amount of unlabelled data is usually non-trivial. Active learning, in its general form, requires the availability of a base segmentation model; thus, a minimal set of base annotations is necessary. Therefore, datasets with no segmentation masks or those with only weak annotations may not directly benefit from active learning unless a pre-trained model from a similar domain is available to serve as the base segmentation model. In what follows, we present a high-level overview of the active learning paradigm and then review the active learning methods for medical image segmentation.

Active learning is an iterative paradigm wherein the unlabeled samples for each round of annotation are selected judiciously to maximally improve the performance of the current model.
Algorithm \ref{algo:al} shows the pseudocode of active learning. In each iteration, the segmentation model is run against the unlabeled images, and then a set of selection criteria, which are defined on model outputs, are used to select the next batch of samples for annotation. Once annotated, the new batch is added to the training set and the segmentation model is fine-tuned using the augmented training set. This process is repeated until the performance on a validation set plateaus. Active learning methods differ in  their sample selection criteria and their definition of annotation unit (the whole or only a part of the image is to be annotated). Table~\ref{tab:al_methods} compares the active learning methods suggested for medical image segmentation.% in terms of selection criteria and annotation unit.

\begin{algorithm}[t]
    \SetKwInOut{Input}{Input}
    \SetKwInOut{Output}{Output}

    % \underline{function Euclid} $(a,b)$\;
    \Input{Initial model $\mathcal{M}_0$, unlabeled dataset $\mathcal{U}_0$, size of query batch $k$, iteration times $\mathcal{T}$, active learning algorithm $\mathcal{A}$}
    \Output{Labeled dataset $\mathcal{L}_{\mathcal{T}}$, updated model $\mathcal{M}_{\mathcal{T}}$}
    $\mathcal{L}_0\leftarrow \varnothing $\;
    \For {$i\leftarrow 1$ \KwTo $\mathcal{T}$}{
        \tcc{phase 1: query batch selection}
        $\mathcal{Q}_t\leftarrow\mathcal{A}(\mathcal{U}_{t-1},\mathcal{M}_{t-1}, k)$\;annotate samples in $\mathcal{Q}_t$\;
        \tcc{phase 2: update model}
        $\mathcal{L}_t\leftarrow\mathcal{L}_{t-1}\cup \{(\mathbf{x},y)|\mathbf{x}\in\mathcal{Q}_t,y\in\mathcal{Y}_t\}$\;
        $\mathcal{M}_t\leftarrow$ fine-tuning $\mathcal{M}_{t-1}$ using $\mathcal{L}_t$\;
        $\mathcal{U}_t\leftarrow\mathcal{U}_{t-1}\backslash\mathcal{Q}_t$\;
    }
    \Return $\mathcal{L}_{\mathcal{T}}, \mathcal{M}_{\mathcal{T}}$
    \caption{Active learning}
    \label{algo:al}
\end{algorithm}
%, bearing in mind that the underlying datasets must have a reasonably-sized segmentation masks available for training.

% In this review, we mainly focus on active learning algorithms suggested for medical image segmentation, referring the readers to \cite{} for more related work.

% Active learning has successfully been applied to many computer vision vision problems including image classification, object detection, and image segmentation, but in this paper, we focus on its applications to medical image segmentation.

%The key to active learning for medical image segmentation is to determine segmentation uncertainty.
\cite{yang2017suggestive} propose a framework called suggestive annotation where the candidate samples for each round of annotation are selected through a 2-stage screening process. First, uncertain samples are identified through the application of an ensemble of segmentation models. The uncertainty at pixel-level is computed as the variance of predictions generated by individual models in the ensemble. Pixel level uncertainty is then averaged to form one uncertainty value for the entire image. Second, the uncertain images are further refined by removing the samples that have high visual similarity. {\brown The authors evaluate suggestive annotation on a histopathology dataset for gland segmentation (85 training and 80 test images) and a CT dataset for lymph node segmentation (37 training and 37 test images), achieving the full-dataset performance with only 50\% of training data}. 

\cite{kuo2018cost} propose an active learning framework based on sample uncertainty and annotation cost. In fact, this work is the first of its kind in the context of medical image segmentation to account for annotation cost when selecting the samples for the next round of annotation. Without considering annotation cost, active learning frameworks treat the images equally, ignoring the fact that some images in practice incur substantially higher annotation cost due to the larger size or quantity of contained target structures (organs and abnormalities). Concretely, they formulate active learning as a knapsack 0-1 problem where the objective is to select a batch of samples for annotation so as to maximize the model uncertainty while keeping the annotation cost below a given threshold. To measure sample uncertainty, they propose to train FCNs at the patch-level rather than the image-level because a patchFCN is less likely to overfit to the global image context. To estimate annotation cost for each unlabeled image, they use a regression model where the predictor variables are the total perimeter and number of connected components in the segmentation mask. {\brown The authors evaluate the suggested active learning method for intracranial hemorrhage  segmentation  on  1247 head  CT scans (934 training/313 test), achieving the performance of a full-dataset model with 50\% of the training set and 20\% of annotation cost. These results are comparable to \cite{yang2017suggestive}, but are obtained using datasets that are two orders of magnitude larger in size.}

The methods suggested by \cite{kuo2018cost,yang2017suggestive}, despite their differences, both employ
an ensemble of FCNs to estimate sample uncertainty, which is slow and computationally expensive to train, as one needs to iteratively train an ensemble of segmentation models after each round of annotation. A more computationally efficient approach to quantifying model uncertainty is to run a given sample through the model several times with the dropout layers on \cite{gal2016dropout}. Pixel uncertainty is then estimated as the entropy of averaged probabilities over different classes. 
%A more computationally efficient approach is the work of \cite{devries2018leveraging} where dropout layers are retained during the test stage to quantify model uncertainty for a given sample. The segmentation result for each image is the average of 20 runs of the image through the model and pixel uncertainty is estimated as the entropy of averaged probabilities over different classes. 
This efficient sample uncertainty estimation  is used  by \cite{gorriz2017cost} to realize  a cost-effective active learning framework. Specifically, they compute an uncertainty value for a given unlabelled image by first obtaining an uncertainty map using the aforementioned dropout-based method
followed by reducing the map to a single value through a weighted averaging scheme where the weights come from a distance transform map over the segmentation result. The idea is to assign higher importance to the uncertain pixels that are located farther away from the object boundaries. Once uncertainty values are computed for all unlabelled images, at each round of active learning, they select samples with high uncertainties as well as a batch of random samples for annotation. In each round, they also directly add samples that have the lowest levels of uncertainty along with their predicted masks to the training set. The rationale is that if the sample uncertainty is low, then the model has probably created a high-quality segmentation mask, which can be used for training without any further corrections. {\brown The authors evaluate this approach for melanoma segmentation in ISIC 2017 challenge dataset for Skin Lesion Analysis (1600 training and 400 test images) , demonstrating a 55\% reduction in the annotation cost.}

Similar to \cite{yang2017suggestive}, \cite{ozdemir2018active} propose a 2-stage active learning framework where stage 1 identifies uncertain examples whereas stage 2 selects the representative examples among the uncertain examples. The suggested method is however different in how uncertainty and representativeness are measured. The authors use the dropout-based approach \cite{gal2016dropout} to estimate an uncertainty map for each unlabelled image. To identify representative examples, they use the latent space learned by the segmentation network; however, to increase the discrimination power of the latent space, they train the segmentation network using an entropy-based regularization technique, which encourages diversity among the features of the latent space. The farther an uncertain sample is located from other examples in the latent space, the more representative the example. The uncertainty and representativeness metrics are further fused using Borda count.
{\brown The authors evaluate their method for muscle and bone segmentation in MR images of 36 patients diagnosed with rotator cuff tear (25 training and 11 test)}. By ranking the examples using the fused metric metrics, the authors achieve similar performance to the model trained with the full dataset while using only 27\% of the entire training set.

\cite{sourati2018active} propose a probabilistic active learning framework where the probability of an unlabeled sample being queried in the next round of annotation is estimated based on its Fisher information. A sample has higher Fisher information if it generates larger gradients with respect to the model parameters. To incorporate Fisher information in the sample selection process, the authors formulate active learning as an optimization problem where the unknowns are the probabilities by which unlabelled samples are queried for the next round of annotation; the constraints are that the querying probabilities should add up to one and that they should change disproportionately to their Fisher information; and the objective is to assign the querying probabilities so as to maximize the overall Fisher information. The optimization problem above is solved for a batch of samples, as such, the sample inter-dependency is already taken into consideration, eliminating the need for a secondary stage that further selects the representative samples from the informative samples. This one-shot behaviour sets this approach apart from the previous works where informativeness and representativeness are accounted for sequentially (e.g., \cite{yang2017suggestive}). One limitation of this work, however, is that computational complexity is super quadratic with respect to the number of parameters, because the Fisher matrix has as many rows and columns as the number of parameters in the network. This limitation has been addressed in a follow-up work from the authors (\cite{sourati2019intelligent}) where the number of rows and columns of the Fisher matrix reduces to the number of layers in the network. {\brown This scheme was tested for brain extraction from MR images of 25 normal newborns and 26 subjects with tuberous
sclerosis complex aged younger than 2.5 years old.} When trained with only 0.5\% of the training voxels, this scheme achieves the same performance as the model that is trained using the entire training set. 

\cite{mahapatra2018efficient} use a Bayesian neural network for active learning where the informative samples are selected using a combined metric based on aleatoric uncertainty (noise in the data) and epistemic uncertainty (uncertainty  over  the  CNN parameters). Thus, this sample selection strategy  differs from the previous approaches, where user-defined heuristics such as standard deviation of predictions are used to identify informative samples for annotation. {\brown This scheme was evaluated for the segmentation of the clavicles, lungs and heart on chest X-ray images. For this purpose, the authors use training and test sets consisting of 247 and 400 images, respectively.} Combined with an image synthesis network, the suggested method achieves the full-data performance with only 30-35\% of annotated pixels.

%\cite{mahapatra2018efficient} propose an active learning framework empowered by a data synthesis module and a Bayesian neural network. The data synthesis module is a conditional GAN, which takes as inputs an image and a segmentation mask, and then generates a synthesized image that resembles the appearance of the input image but also matches the input segmentation mask. Note that one can obtain many synthetic images by feeding the data generator with the same input image while using different variants of the corresponding segmentation mask. Conditioning image synthesis on the segmentation mask is indeed attractive, because the synthetic images already have their corresponding segmentation masks, eliminating the need for an expert to provide annotations for the synthetic images. A Bayesian neural network then governs active learning where the informative synthetic images are selected according to a combined metric based on aleotaric uncertainty (noise in the data) and epistemic uncertainty (uncertainty  over  the  CNN parameters). This sample selection strategy thus differs from the previous approaches where user-defined heuristics such as standard deviation of predictions are used to identify informative samples for annotation.

 Different from the previous works, \cite{zheng2019biomedical} propose a 1-shot active learning method, which eliminates the need for iterative sample selection and annotation. The suggested method consists of a feature extraction network, which projects each image patch to a latent space, and a clustering algorithm, which discovers representative images for image annotation in the latent space. Being a 1-shot active learning approach, the feature extraction network must be trained using unlabeled data. For this purpose, the authors use various unsupervised models such as auto-encoders and variational auto-encoders. For clustering, the authors use a hybrid method based on  K-means and max-cover algorithms. {\brown This method was evaluated using a fungus dataset (4 training and 80 test images), a histopathology dataset (84/80), and an MR dataset (10/10).} The results for both 2D and 3D datasets suggest that the 1-shot active learning method performs comparably to an iterative alternative by \cite{yang2017suggestive}.

%% file: chapters/interactive_segmentation.tex
Creating segmentation masks is not only tedious and time-consuming  for expert annotators, but also incurs substantial annotation cost particularly for volumetric medical images where the same lesion or organ must be delineated across multiple slices. Interactive segmentation can accelerate the annotation process by allowing the expert annotators to interactively correct an initial segmentation mask generated by a model. Interactive segmentation complements active learning in achieving cost-effective annotation: the latter identifies which images to be annotated whereas the former reduces the time required to complete the annotation of a selected image.

 Algorithm~\ref{algo:is} shows the pseudocode for interactive segmentation. As seen, interactive segmentation may require an initial segmentation model, whose output is reviewed by human experts to provide feedback on possible segmentation error. The user feedback, as the core part of interactive segmentation methodologies, can take varying forms of interactions such as mouse clicks, bounding boxes, and scribbles. The user interactions then translate to foreground or background annotations, which the initial segmentation model can use to improve itself. The updated model re-generates the segmentation mask for the users' feedback, and this process repeats until the desired segmentation mask is obtained. Interactive segmentation is highly effective to cope with a model's inevitable segmentation mistakes, which are typically caused by domain shifts or unrepresentative training sets. 
%  Besides, it achieves greater efficiency by minimizing manual operations from human experts, therefore, reduced annotation time may translate to annotation cost savings overall.
%The effectiveness of an interactive segmentation method is measured by total annotation time saving relative to manual segmentation from scratch, therefore, an effective interactive method should reduce the overall annotation cost.
%  {\red An effective interactive segmentation method should require as short  user  time  as  possible  to  lessen  the  burden  on  users and reduce the annotation cost.}  
 In what follows, we summarize the recent interactive segmentation methods that are  suggested for medical image segmentation.

%\figurename~\ref{fig:interactive_segmentation} further% shows the schematic overview of interactive segmentation.
 \begin{algorithm}[t]
    \SetKwInOut{Input}{Input}
    \SetKwInOut{Output}{Output}

    % \underline{function Euclid} $(a,b)$\;
    \Input{Initial model $\mathcal{M}_0$, unlabeled image $\mathcal{I}$, number of iterations $\mathcal{N}$, feedback operation $\mathcal{R}$, conversion operation $\mathcal{C}$}
    \Output{Updated model $\mathcal{M}_{\mathcal{N}}$}
    \For {$i\leftarrow 1$ \KwTo $\mathcal{N}$}{
            \tcc{generate segmentation map}
        $\mathcal{S}_i\leftarrow\mathcal{M}_{i-1}(\mathcal{I})$\;
        \tcc{get feedback from an expert}
        $\mathcal{F}_i\leftarrow\mathcal{R}(\mathcal{S}_i, \mathcal{I})$\;
        \tcc{convert to a new annotation}
        $\mathcal{A}_i\leftarrow\mathcal{C}(\mathcal{F}_i)$\;
        $\mathcal{M}_i\leftarrow$ fine-tuning $\mathcal{M}_{i-1}$ with $\mathcal{A}_i$\;
    }
    \Return $\mathcal{S}_{\mathcal{N}}$
    \caption{Interactive segmentation}
    \label{algo:is}
\end{algorithm}

\cite{sun2018interactive} propose an interactive  method for segmenting fuzzy boundaries, wherein the user first places a point roughly at the center of the object, and then the model performs object delineation for the user-specified structure. For accurate boundary segmentation, the authors suggest a segmentation model that delineates the structures by comparing the appearances of image patches from inside and outside of the structure, imitating inside-outside comparison that physicians perform in order to precisely localize boundaries. The authors model the inside-outside comparison with a bidirectional convolutional recurrent neural network, which is trained using the image patch and ground truth mask sequences bidirectionally, allowing the network to learn appearance changes from foreground to background and vice versa. This method, however, only allows users to specify a seed point at the onset of segmentation, that is, the resulting segmentation masks are not responsive to the subsequent user interaction.

\cite{wang2018deepigeos} proposes a framework consisting of a proposal network and a refinement network where the former generates a base segmentation mask whereas the latter refines the base mask according to the suggestions provided by the user. However, the suggested framework lacks  adaptability to unseen image contexts. The authors have overcome this limitation in their followup work
\cite{wang2018interactive}. Given a test image and a pre-trained segmentation model, the suggested framework alternates between 2 steps: 1) refining the current segmentation mask through the application of Graph Cut~\cite{boykov2001interactive},  and 2) minimizing segmentation loss for the test image by creating a pseudo ground truth segmentation mask. This approach can be viewed as a self-learning method with the difference being the pseudo ground truth depends on both model predictions and user-provided scribbles. Specifically, the pseudo ground truth mask is the predicted segmentation mask wherein the labels of the scribble pixels are overwritten by the labels provided by the user. The segmentation loss is then a weighted cross entropy function, which receives large contributions from the scribble pixels and zero contributions from uncertain pixels.    The authors treat a pixel as uncertain if it is located near a scribble but has a predicted label other than that of the scribble or if the posterior distribution predicted by the model has high entropy (low confidence predictions). {\brown The suggested framework  is evaluated in two applications: brain tumor segmentation using the BraTS dataset~\cite{menze2014multimodal} (249 training and 25 test MR images) and multi-organ segmentation in fetal brain MR images (10/8).} The results show that the suggested interactive segmentation method outperforms traditional interactive segmentation methods in both accuracy  and  speed of annotation.

\cite{sakinis2019interactive}  propose a semi-automated image segmentation method that enables a high quality segmentation with only a few user clicks.  The authors choose a mouse click as the means of user interaction because it enables quick feedback and ease of simulation. The segmentation model is a U-Net that receives as input the image stacked with the foreground and background attention maps, where attention maps are constructed by placing a Gaussian blob at each foreground and background user click. The U-Net is then trained by minimizing the Dice loss between the predicted segmentation and ground truth. Since it is unfeasible to have true user interaction during training and large-scale testing, the authors propose a simulation scheme that has the effect of a hypothetical user clicking on regions with larger and more noticeable segmentation error. This method proves effective in segmenting both structures that exist in the training set and the structures that the model has never seen during training. To put this in perspective, with only 1 user click, this semi-automated method can achieve a Dice of 0.64 {\brown for segmenting an unseen structure, colon cancer, in 126 Abdomen CT scans}, outperforming the best automated model with a Dice of 0.56.

%% file: chapters/self_supervised.tex
% \usepackage{multirow}
% \usepackage{graphicx}
\begin{table*}[t]
\caption{Comparison between self-supervised training methods that can directly or indirectly aid medical image segmentation. }
\label{tab:self_sup}
\resizebox{\textwidth}{!}{%
\begin{tabular}{lllll}
\hline
\multicolumn{1}{c}{\multirow{2}{*}{Publication}} & \multicolumn{1}{c}{\multirow{2}{*}{Network}} & \multicolumn{3}{c}{Surrogate task} \\ \cline{3-5} 
\multicolumn{1}{c}{} & \multicolumn{1}{c}{} & \multicolumn{1}{c}{Type} & \multicolumn{1}{c}{Description} & Annotation \\ \hline
\cite{jamaludin17} & Encoder & Image-to-scalar & Predict if two longitudinal studies belong to the same patient & 1(same)/0(different) \\ \hline
\cite{zhang2017self} & Encoder & Image-to-scalar & Predict the order of two slices random selected from the same CT scan & 0(top)/1(bottom) \\ \hline
\cite{tajbakhsh2019surrogate} & Encoder & Image-to-scalar & Predict the degree of rotation applied to a chest CT scan & $\frac{\theta}{90^{\circ}}$ ($\theta \in\{0,90,180,270\}$) \\ \hline
\cite{spitzer2018improving} & Siamese & Image-to-scalar & Predict the distance between  two patches  sampled  from  the  same  MR image & Float distance \\ \hline
\cite{gildenblat2019self} & Siamese & Image-to-scalar & Predict if  two patches  sampled  from  the  same  MR image are spatially near & 1(near)/0(far) \\ \hline
\cite{alex2017semisupervised} & Encoder-decoder & Image-to-image & Learn how to remove noise from MR image patches & Original patch before injecting noise \\ \hline
\cite{ross18} & Encoder-decoder & Image-to-image & Learn how to colorize gray-scale colonoscopy frames & Original frame before removing color \\ \hline
\cite{tajbakhsh2019surrogate} & Encoder-decoder & Image-to-image & Learn how to colorize gray-scale tele-med skin images & Original image before removing color \\ \hline
\cite{zhou2019models} & Encoder-decoder & Image-to-image & Learn how to restore the image from various degradation transformations & Original image before degradation \\ \hline
\cite{bai2019self} & Encoder-decoder & Image-to-image & Learn how to weakly localize anatomical landmarks in MR images & Approximate landmark positions \\ \hline
\end{tabular}%
}
\end{table*}

%Transfer learning has commonly been used to tackle the limited training sample size problem in medical imaging, where models pre-trained on ImageNet are fine-tuned for target medical image analysis tasks. Despite promising results~\cite{tajbakhsh16,hoo16}, this approach has two major limitations. First, it may limit the designer to architectures that have been pre-trained on ImageNet, which are often needlessly deep for medical imaging, thus retarding training and inference. Second, transfer learning is barely applicable to 3D medical image analysis applications, because the 2D and 3D kernels are not shape  compatible. Therefore, transfer learning from natural images is only a partial solution to the common problem of insufficient labeled data in medical imaging.

Self-supervised model pre-training has recently been studied as an alternative to transfer learning from natural images. The key idea consists of pre-training the model using unlabeled medical data, which is easier to obtain, and then fine-tune the pre-trained model for the target medical vision task using the limited labeled data available for training. Specifically, self-supervised pre-training consists of assigning surrogate or proxy labels to the unlabeled data and then training a randomly initialized network using the resulting surrogate supervision signal. The advantage of model pre-training using unlabeled medical data is that the learned knowledge is related to the target medical task; and thus, can be more effective than transfer learning from a foregin domain (e.g., \cite{tajbakhsh2019surrogate} and \cite{ross18}). 

Self-supervised learning methods differ in the composition of the surrogate task. Reviewing the literature, we have identified two types of surrogate tasks: 1) image-to-scalar where an encoder network is pre-trained for a surrogate image classification or regression task; 2) image-to-image where an encoder-decoder network is pre-trained for a surrogate image regression task such as image colorization or image denoising. While the former approach seems particularly suitable for a downstream image classification task, it can still be used to initialize the encoder of a segmentation network, in which case the decoder should be initialized with random weights. We have summarized the representative examples of both categories in Table~\ref{tab:self_sup}, and further review them as follows.  %\figurename~\ref{} shows the major surrogate tasks used for pre-training models in medical imaging.

\vspace{4pt}
\noindent {\bf \textit{Image-to-scalar:}} \cite{jamaludin17} propose longitudinal relationships between medical images as the surrogate task to pre-train model weights. To generate surrogate supervision, they assign a label of 1 if two longitudinal studies belong to the same patient and 0 otherwise.  \cite{zhang2017self} propose a surrogate task wherein two slices are randomly selected from a CT volume and then the encoder is to predict if one slice is above or below the reference slice. The pre-trained model is then fine-tuned for the task of body part recognition in CT and MR images. \cite{tajbakhsh2019surrogate} use prediction of image orientation as the surrogate task where the input image is rotated or flipped and the network is trained to predict such a transformation. The authors show that this surrogate task is highly effective for diabetic retinopathy classification in fundus images and lung lobe segmentation in chest CT scans. \cite{spitzer2018improving} propose a new surrogate task that can be used to pre-train a Siamese network by predicting the  3D  distance  between  two patches  sampled  from  the  same  brain regions. The pre-trained model is then fine-tuned for cytoarchitectonic segmentation of the human brain. {\brown The authors use different sections of one MR scan for  training and testing. The segmentation model trained from the self-supervised model achieves a Dice score of 0.8, outperforming the model trained from scratch with a Dice of 0.72.}  Similarly, \cite{gildenblat2019self} suggest a surrogate scheme to pre-train a Siamese network by learning similarity between image patches. Specifically, the network is trained to distinguish between similar patches (nearby patches) and dissimilar patches (spatially distant patches). The pre-trained Siamese network is then fine-tuned for tumor tile retrieval in histopathology images.

\vspace{4pt}
\noindent {\bf \textit{Image-to-image:}}
\cite{alex2017semisupervised} use noise removal in small image patches as the surrogate task, wherein the surrogate supervision was created by mapping the patches with user-injected noise to the original clean image patches. \cite{ross18} use image colorization as the surrogate task, wherein color colonoscopy images are converted to gray-scale and then recovered using a conditional GAN. The pre-trained models are then fine-tuned for the task of instrument segmentation in colonoscopy videos {\brown with varying fractions of the training set. When only 25 images are used for fine-tuning, the instrument segmentation model pre-trained via self-supervised learning achieves a Dice score of 0.61, which outperforms the counterpart model pre-trained using Microsoft COCO dataset and the model trained from scratch, both with a Dice score of 0.57. However, weights pre-trained via self-supervised learning tend to lose their edge over randomly initialized weights when the size of the training set changes from 25 to 400 images.} A similar study is also done by \cite{tajbakhsh2019surrogate} wherein colorization is used as a surrogate task for skin segmentation in tele-medicine images. While image colorization proved more effective than random initialization {\brown  in training with both small ($\approx$140 images) and large ($\approx$1400 images) training sets}, it was outperformed by transfer learning from ImageNet, presumably because the distance between tele-medicine skin images and ImageNet is small. \cite{bai2019self} propose anatomical position prediction as a self-supervised scheme. The landmarks are however obtained through an annotation-free process based on the relative views of image planes. {\brown The authors evaluate the effectiveness of the pre-trained models for the task of cardiac MR segmentation. In a low data regime where the training set has fewer than 50 MR images, self-supervised pre-training achieves a significant gain over random initialization. However, the performance gain becomes only marginal when the segmentation model is trained with 100 MR images.}

The self-supervised learning methods described above are limited to a specific surrogate scheme. Models Genesis suggested by \cite{zhou2019models} is a significant shift from this paradigm where a library of diverse self-supervised schemes, all formulated as an image restoration task, is used to generate self-supervision signal. The suggested framework is scalable to a large library of surrogate tasks, because all tasks in the library can share the same encoder and decoder during training, eliminating the need for task-specific decoders. The authors have evaluated Models Genesis for both  image classification and segmentation in seven 2D and 3D medical datasets, demonstrating three  and five points increase in IoU over 3D models trained from scratch for lung nodule segmentation and liver segmentation in CT scans, respectively. {\brown These results are significant because, unlike the previous works, the performance gains hold up when the models are trained using the full-size training datasets. One caveat, however, is that the models trained from scratch and those trained from pre-trained weights are all constructed without data augmentation. It is not clear how the performance gains change in the presence of data augmentation.}% transferability of pre-trained weights across various diseases, organs, and modalities. 
%The first comprehensive study of self-supervised learning in medical imaging was conducted by \cite{tajbakhsh2019surrogate} wherein the authors evaluated the effectiveness of three surrogate supervision schemes, namely rotation, reconstruction, and colorization, in training more preferment 2D and 3D classification and segmentation models. Their results show that pre-training with surrogate supervision is effective for small training sets, enabling the models to achieve a higher level of performance than their counterparts trained from scratch. They further show that, when the medical domain is distant from the natural images, pre-training models using unlabeled medical images with surrogate supervision is more effective than transfer learning from an unrelated domain (e.g., natural images).

%% file: chapters/self_learning.tex
% Limitation of self-training:  if  initial  class  priors  given by the network are inaccurate, segmentation errors can occur and be propagated back to the network which then re-amplifies these  errors. 
% \begin{figure*}
%     \centering
%     \subfigure[]{\includegraphics[width=0.35\linewidth]{pics/self_learning_p1.png}}\hspace{15pt}
%     \subfigure[]{\includegraphics[width=0.5\linewidth]{pics/self_learning_p2.png}}
%     \caption{Self-learning paradigm. (a) A common scenario in self-learning where an unlabeled dataset and a small labeled training set are available for training. (b) An extreme scenario where only a large unlabeled dataset is available for training.}
%     \label{fig:self_learning_paradigm}
% \end{figure*}
{\brown Semi-supervised learning with pseudo annotations consists of assigning pseudo annotations to unlabeled data and then training the segmentation model using both the labeled and pseudo labeled data. Pseudo labeling, serving as the backbone of this paradigm, is commonly done in an iterative manner wherein a model iteratively improves the quality of pseudo annotations by learning from its own predictions on unlabeled data. Semi-supervised learning with pseudo annotations has shown promising performance, producing models that outperform the counterparts trained using only labeled data.}

%A common assumption behind self-learning is that a labeled training set, even though small, is available for training. A more extreme yet less common scenario is where no initial labeled data is available for training.

\begin{table*}[ht]
\caption{{\brown Comparison between the  methods based on semi-supervised learning with pseudo annotations for medical image segmentation.} The suggested methods differ in how the initial labeled dataset is constructed, how pseudo annotations for unlabeled data are generated, and whether or not any special treatment is applied to the unreliable regions in pseudo  annotation masks.
}
\resizebox{\textwidth}{!}{%
\begin{tabular}{llll}
 \hline
 Publication & Initial annotations by & Pseudo masks generated by & Label noise handled by\\ [0.5ex] 
  \hline
 \cite{zhang2018self} & K-means & Single segmentation model & N/A\\
  \cite{bai2017semi} & Expert & Single segmentation model + CRF & N/A\\
  \cite{zhou2018semi} & Expert & Ensemble segmentation model & N/A\\
            \cite{zhao2019multi} & Expert & Single segmentation model & N/A\\
          \cite{nie2018asdnet} & Expert & Single segmentation model & Discriminator network\\
    \cite{min2018two} & Expert & Ensemble segmentation model & Consensus by two parallel networks\\
    
        \cite{xia20183d} & Expert & Ensemble segmentation model & Consensus by multi-view networks\\
 \hline

\end{tabular}
}

\label{tab:self_learning}
\end{table*}

\begin{algorithm}[t]
    \SetKwInOut{Input}{Input}
    \SetKwInOut{Output}{Output}

    % \underline{function Euclid} $(a,b)$\;
    \Input{Small labeled dataset $\mathcal{L}$, unlabeled dataset $\mathcal{U}$, iteration times $\mathcal{T}$, masks generation function $\mathcal{F}$}
    \Output{Updated model $\mathcal{M}_{\mathcal{T}}$}
    $\mathcal{M}_0\leftarrow$ training base model with $\mathcal{L}$\;
    \For {$i\leftarrow 1$ \KwTo $\mathcal{T}$}{
        \tcc{generate pseudo segmentation masks}
        $\mathcal{S}_i\leftarrow \mathcal{F}(\mathcal{M}_{i-1}, \mathcal{U})$\;
        $\mathcal{D}_i\leftarrow \mathcal{L}\cup\{(\mathbf{x},s)|\mathbf{x}\in\mathcal{U},s\in\mathcal{S}_i\}$\;
        $\mathcal{M}_{i}\leftarrow$ fine-tuning $\mathcal{M}_{i-1}$ using $\mathcal{D}_i$\;
    }
    \Return $\mathcal{M}_{\mathcal{T}}$
    \caption{Semi-supervised learning with pseudo annotations}
    \label{algo:sl}
\end{algorithm}

Algorithm~\ref{algo:sl} shows the pseudo code for semi-supervised learning with pseudo annotations where one has access to a small labeled dataset and a fairly large unlabeled dataset. First, a base model is trained using limited labeled data. The base model is then applied to unlabeled data to generate pseudo segmentation masks. The limited labeled data is then merged with pseudo-labeled data to update the base model. This training paradigm alternates between the two steps above until a desired level of performance on the validation set is achieved. In a less common scenario, no initial labeled dataset is available for training, in which case, an unsupervised segmentation method such as K-means is used to generate pseudo masks for unlabeled data. While the semi-supervised learning methods based on pseudo annotations commonly follow the iterative process stated above, they differ in how they initialize the base model, how they generate pseudo masks, and whether or not they use a mechanism to handle label noise in pseudo segmentation masks. We have compared the semi-supervised learning methods that use pseudo annotations  from these perspectives in Table~\ref{tab:self_learning}, and further review them as follows.

\vspace{4pt}
\noindent {\bf \textit{Without initially labeled dataset}}:  \cite{zhang2018self} train a cyst segmentation model using unlabeled chest CT scans. Since the dataset is completely unlabeled, the authors generate the initial ground truth using K-means clustering followed by a refinement stage through graph cuts. The segmentation model is trained using the pseudo masks and then the model is applied back to the data to generate refined pseudo masks. The training process  alternates between updating the segmentation model and refining pseudo masks. {\brown The authors train the segmentation model on 166 CT scans and use 17 CT scans  including 5 mild, 6 moderate, and 6 severe cases for testing.} In 3 iterations, the suggested method achieves 12-point increase in Dice over a model trained using the initial pseudo mask generated by K-means. 

% To generate the initial seed ground truth, \cite{can2018learning} use a random walk image segmentation approach with a high threshold, following which the segmentation network with a CRF-RNN layer is trained iteratively. Test time dropout is used to make multiple predictions for each image and assess the uncertainty of each pixel-level prediction in the image. The new ground truth in each iteration is made up of only the certain predictions. They achieve a 3\% average Dice improvement over the baseline.

\vspace{4pt}
\noindent {\bf \textit{With initially labeled dataset:}}
 \cite{bai2017semi} propose a two-step framework to segment the heart chambers in MR images. The training process alternates between two steps: 1) estimating the ground truth for unlabeled data using the current segmentation model followed by a refinement stage through the application of CRF, 2) updating the current model using both the labeled data with expert annotations and the unlabeled data with pseudo annotations. This approach is simple to implement; however, hindered by the quality of pseudo annotations, the resulting model achieves only a moderate level of improvement over the model trained using only labeled data. Similarly, \cite{zhou2018semi} propose an iterative framework, but, at each iteration, the authors train three segmentation models for the axial, sagittal, and  coronal planes. Once trained, the three models scan each unlabeled 3D image slice-by-slice, generating three segmentation volumes, which are further combined through a majority voting scheme to form the final segmentation mask. The unlabeled images with their estimated segmentation masks are added to the labeled set to train three new segmentation models in the next iteration. {\brown The authors test their semi-supervised paradigm in segmenting 16 structures in 80 contrast-enhanced
abdominal clinical CT images  in the portal venous phase. The training set consists of 50 labeled CT and 100 unlabeled CT scans. Under this data split, the suggested method achieves on average four points increase in Dice over the model trained using only the labeled data. The performance gain however drops to 1 point in Dice when the authors increase the size of labeled training set to 100 CT scans.}

{\brown In the semi-supervised work by \cite{zhao2019multi} pseudo annotations are generated through data distillation where the segmentation masks obtained for different transformations of the same image are averaged to produce the final pseudo mask. The suggested model is trained for the task of brain region segmentation using 12 labeled and 749 unlabeled MR images. On a tests set with 20 MR scans, the suggested model achieves 1.6 points increase in Dice over the base model trained using only labeled data.} 

%show that using three 2D segmentation models is more effective than using one 3D model. They further show that the segmentation models trained with artificially labeled dataset surpass the models trained using only labeled data.

A limitation with the previous approaches is that the images that have expert annotation and images with pseudo segmentation masks are treated equally during training. As such, errors in the pseudo labels can degrade the quality of the resulting models. {\brown Recent methodologies overcome this limitation by estimating the reliability of pseudo annotations during training, by means of model consensus \cite{min2018two,xia20183d} or through the use of an evaluation network \cite{nie2018asdnet}. We cover these methods as follows:}

\cite{min2018two} propose a two stream network where each stream has its own independent weights. During training, if a training sample receives the same class prediction from both streams, then the sample is deemed as easy or hard, in which case it will not contribute to the overall loss, where the training samples refer to each individual pixel. The rationale is that the easy examples do not add much value to the model and the hard examples may have label noise; therefore, it is safe to exclude them from backpropagation. To obtain the pseudo segmentation masks, the authors propose a hybrid method based on model distillation \cite{gupta2016cross} and data distillation \cite{radosavovic2018data}, which essentially consists of model-ensembling and test-time data augmentation. %Specifically, the authors first train three segmentation networks using labeled data. Then, each unlabeled image and its transformed variations (rotation and flipping) are fed to each of the three trained models. The final pseudo mask is obtained by averaging the resulting prediction after performing the corresponding inverse transforms. 
{\brown The authors use two public datasets for evaluation: BraTS 2015~\cite{menze2014multimodal} with 244 MR images for training and 30 images for testing, and HVSMR 2016 with 10 MR images for training and 10 images for testing. The results show that the suggested semi-supervised learning framework based on pseudo-labeling and the two-stream network is noise-resilient, significantly outperforming  fully supervised models under varying levels of label noise present in labeled images.}

\cite{nie2018asdnet} use only the reliable regions of the pseudo segmentation masks during training. Specifically, they propose a framework consisting of a segmentation network (generator) and a confidence network (discriminator), which are trained through an adversarial game. The discriminator, a fully convolutional network, serves two purposes: 1)  distinguishing between ground truth and predicted masks at the pixel level, 2) providing a confidence (reliability) value for each pixel in the predicted mask. The former functionality is used during adversarial training whereas the latter functionality is used to identify reliable regions in the pseudo masks of unlabeled data. During training, the pseudo mask for an unlabeled image is first masked by the binarized confidence map and is then used as ground truth to compute the segmentation loss. The authors show the effectiveness of the suggested framework over a pure supervised approach across multiple datasets.

\lref{nt:cotr} {\brown \cite{xia20183d} have suggested a semi-supervised framework based on co-training of multiple networks wherein pseudo annotations are generated as the consensus of network predictions in the ensemble. Specifically, $N$ views of the same input are fed to $N$ parallel convolutional networks, which each is trained using labeled data and the corresponding ground truth, as well as, unlabeled data with the pseudo annotations generated through the combination of the predictions from the other $N-1$ networks. To mitigate the unreliable regions in pseudo annotations,  network predictions are combined through an uncertainty-aware scheme where each prediction is weighted by its uncertainty. The authors use the dropout scheme to estimate uncertainty for predictions---a common scheme in active learning, see Section~\ref{sec:active}. 
For pancreas  segmentation in the NIH dataset (62 training and 20 test CT scans), the authors report seven points and five points increase in Dice over a full supervised model, when 10\% and 20\% of the training set is used as the initial labeled training set.}

%% file: chapters/semi_supervised_learning.tex
{\brown Semi-supervised learning without pseudo annotations consists of training a model with both labeled and unlabeled data,
where the unlabeled data generate a supervision signal through an unsupervised loss function. In this paradigm, the domain of unlabeled data may be from a different domain or from the same domain as the labeled dataset. The former is commonly known as unsupervised domain adaptation, which aims to mitigate the domain shift problem by adapting the model to a target domain. We discuss unsupervised domain adaptation methods in Section \ref{sec:domain_ada}. The latter enables training the model in the source domain using both labeled and unlabeled data, increasing the effective size of the training set, mitigating the problem of limited annotations in the source domain. In this section, we focus on this form of semi-supervised learning.}

% Unlike self-learning, majority of semi-supervised methods do not attempt to generate psuedo annotations for unlabeled data, rather, they use unlabeled data as is during training. The domain of unlabeled data may be the same as the labeled dataset, in which case semi-supervised learning can serve as an effective solution for the problem of limited annotations. This is because the model is presented with a larger number of samples during training; and thus, it may generalize better to the unseen test set. On the other hand, the unlabeled dataset may be from a domain other than that of the labeled dataset, in which case semi-supervised learning mitigates the domain shift problem by adapting the model to a target domain. 

Algorithm~\ref{algo:ssl} shows the pseudocode for semi-supervised learning without pseudo annotations in its most general form. As seen, the semi-supervised framework consists of two loss functions: a supervised loss function to which only labeled data contribute; and an unsupervised loss function or a regularization term, which is computed for both labeled and unlabeled data. The total loss is the summation of the two terms, which is minimized for batches of labeled and unlabeled data. 

{\brown We have listed the representative semi-supervised methods for medical image segmentation in Table~\ref{tab:ssl_methods}. As seen, the suggested methods differ in terms of the unsupervised task, which can be as simple as image reconstruction or transformation consistency or representation similarity based on adversarial loss. For better readability, we have grouped the related semi-supervised methods by  the underlying unsupervised task and explain them as follows:} 

\begin{algorithm}[t]
    \SetKwInOut{Input}{Input}
    \SetKwInOut{Output}{Output}

    % \underline{function Euclid} $(a,b)$\;
    \Input{Limited labeled dataset $\mathcal{L}$, unlabeled dataset $\mathcal{U}$, shared backbone $\mathcal{M}_c$, branch model and loss function for labeled data $\mathcal{M}_l, \ell_l$, branch model and loss function for unlabeled data $\mathcal{M}_u, \ell_u$}
    \Output{Fine-tuned model $\mathcal{M}$}
    $\zeta_l\leftarrow\ell_l(\mathcal{M}_l(\mathcal{M}_c(\mathcal{L})))$\;
    $\zeta_u\leftarrow\ell_u(\mathcal{M}_u(\mathcal{M}_c(\mathcal{U}))+\ell_u(\mathcal{M}_u(\mathcal{M}_c(\mathcal{L}))$\;
    $\text{minimize}(\zeta_l+\zeta_u)$\;
    \Return $\mathcal{M}$
    \caption{Semi-supervised learning without pseudo annotations}
    \label{algo:ssl}
\end{algorithm}

\iffalse
\begin{enumerate}
    \item The framework of
semi-supervised learning provides the means to use both labeled data
and arbitrary amounts of unlabeled data for training.

    \item can be effective for both limited annotation and domain-shift, which later cannot be helped even with ample labeled data from the source domain
    
    \item  semi-supervised  learning  framework  is a form of  domain  adaptation where we  try  to  improve generalization  of  a  baseline  model  by  fine-tuning  it  with  unlabeled  data  from the target domain.
    
    \item semi-supervised loss $L= L_s(X^l;Y^l)+L_ss(X^l;X^u)$ where the second term is some regularization term that can be applied to unlabeled data or both labeled and unlabeled data.
 \end{enumerate}
 \fi

\begin{table*}[t]
\caption{{\brown Semi-supervised  methods suggested for medical image segmentation that do not use pseudo annotations}. The suggested methods combine the segmentation task with an unsupervised task, allowing the model to use both labeled and unlabeled images during training.}
\resizebox{\textwidth}{!}{%
\begin{tabular}{lll}
 \hline
 Publication &  Unsupervised task & Description\\
\hline

\cite{li2019transformation} & Transformation consistency &   Segmentation model is trained to achieve equivariance to image rotation or flipping\\ 

\cite{bortsova2019semi} & Transformation consistency & Segmentation model is trained to achieve equivariance to elastic image deformation\\

\cite{yu2019uncertainty} & Transformation consistency& Segmentation model is mentored by a mean teacher network to achieve equivariance to image perturbations\\

\cite{cui2019semi} & Transformation consistency& Segmentation model is mentored by a mean teacher network to achieve equivariance to image perturbations\\

\cite{chen2019multi} &  Image reconstruction& Segmentation model is trained along with a class-specific image reconstruction network\\

\cite{chartsias2018factorised}  &   Image reconstruction& Segmentation model decomposes an image into a mask and a vector and then feeds them to image reconstruction\\

% \cite{chaitanya2019semi} &  Image synthesis\\
% \cite{zhao2019data}  &   Image synthesis \\
\cite{sedai2017semi}  &  Representation similarity& Segmentation model is trained to mimic the latent space of an autoencoder that is trained by unlabeled data\\

\cite{Zhang2017deep}  & Representation similarity& Segmentation model is trained adversarially to similarize the segmentation of labeled and unlabeled images \\

\cite{baur2017semi}  &    Representation similarity& Segmentation model is trained to bring  the  feature  embedding  of labeled and unlabeled images close\\

\cite{mondal2018few} & Representation similarity& Segmentation model is trained to maximize the similarity between the logits of labeled and unlabeled data\\

% the following refs have moved to other sections:
%\cite{zhou2018semi}  &   \checkmark & segmentation & N/A & multi-organ seg. (CT)\\
%\cite{huo2018adversarial}  & Image synthesis \\ 
%\cite{bai2017semi} & Embedding consistency \\
\hline

\end{tabular}
}

\label{tab:ssl_methods}
\end{table*}

\vspace{4pt}
\noindent {\bf \textit{Via image reconstruction:}}
\cite{chartsias2018factorised} propose a solution to the problem of domain shift based on a disentangled image representation where the idea is to separate information related to segmenting the structure of interest from the other image features that readily change from one domain to another. By doing so, the segmentation network focuses on the intrinsic features of the target structure rather than variations related to imaging scanners or artifacts. The authors show that the suggested framework is highly effective for myocardial segmentation in low-data regime, but the performance gap closes as the size of training set increases.

\lref{nt:mlt_att} {\brown \cite{chen2019multi} suggest a semi-supervised learning framework with one encoder and two decoders. The first decoder is used for the segmentation task and is trained with the labeled data. The second decoder is trained using unlabeled data for the task of class-specific image reconstruction. Specifically, given a $k$-class segmentation problem, the reconstruction decoder generates $k$ maps, which each is compared against the input image multiplied by the corresponding segmentation mask  obtained from the segmentation decoder. For tumor segmentation in the BraTS dataset~\cite{menze2014multimodal} (120 training and 50 test MR scans), the authors report six and seven points increase in Dice over supervised model when 20 and 50 labeled MR images are used for training. The semi-supervised model, however, continues to outperform the supervised model by four points in Dice even when the entire large set of the labeled dataset is used for training---a significant finding rarely reported before.}

\vspace{4pt}
\noindent {\bf \textit{Via transformation consistency:}}
\lref{nt:transf_consistency}{\brown It is known that convolutional networks are  inherently not rotation equivariant\footnote{{\brown Researchers have attempted to instill this property through various approaches such as group equivariant convolutional neural networks \cite{cohen2016group}, which has also proved effective in segmenting histopathology images  \cite{veeling2018rotation}.}}. The recent semi-supervised learning frameworks have turned this weakness of regular convolutional networks into an opportunity to leverage unlabeled data. Specifically, they attempt to achieve transformation equivariance by minimizing a transformation consistency loss, to which both labeled and unlabeled data can contribute. Let $x$, $T$, $S$   denote the input image, an image transformation, and the segmentation network, respectively. The transformation-consistent regularization is formulated as $ \|S(T(x))-T(S(x))\|$, which is essentially the mean squared error loss between the segmentation map of the transformed image, $S(T(x))$, and the transformed segmentation map of the original image, $T(S(x))$. By imposing this additional regularization term, not only does the model behave more predictably, but also it can utilize the unlabeled data available for training. In  what follows, we review how transformation consistency has been used in modern semi-supervised learning methods.

\cite{li2019transformation} propose a semi-supervised learning framework consisting of a segmentation loss whose optimization requires labeled data, and a rotation equivariant loss  where both labeled and unlabeled data can contribute to. Simply, if an image, being labeled or unlabeled, is rotated by 90 degrees, then the resulting segmentation map should also appear with 90 degrees of rotations with respect to the segmentation mask of the original image. Transformation equivariance is also adopted by \cite{bortsova2019semi} where the segmentation model is trained to achieve equivariance to elastic deformation. Performance evaluation on multiple datasets show that both methods above are effective in boosting segmentation performance in low data regime; however, the performance gains tend to be limited in the presence of large labeled training sets. 

The works above enforce consistency by imposing constraints on the outputs of the segmentation network for the original and the transformed images. A different approach to ensure transformation consistency is through the mean teacher paradigm. The student network is the segmentation model, which receives the original image. The teacher network has the same architecture as the student network and receives a transformed version of the image sent to the teacher network. The unsupervised loss minimizes the dissimilarity between the segmentation masks of the original and transformed images, which are generated by the teacher and student networks, respectively. In this paradigm, the student network is trained as usual by minimizing the segmentation and consistency loss; however, the teacher network is not trained through back-propagation---it merely tracks the weights of the student network through an exponential moving average rule. Mean teacher paradigm, a.k.a self-ensembling, is expected to outperform the previous paradigm where the segmentation network serves as both teacher and student. This is because the teacher network typically outperforms the student network; and thus, it can generate more reliable targets for the consistency loss. In the following, we review the self-ensembling methods suggested for medical image segmentation.

\cite{cui2019semi} adopts the mean teacher paradigm for the task of stroke lesion segmentation in MR images. In this method,  the segmentation consistency is measured under additive and multiplicative Gaussian noise. For training, the authors use a dataset with 20 labeled and 196 unlabeled training subjects. Test results on MR images from 30 subjects demonstrate that the semi-supervised model achieves three points increase in Dice over a supervised model, further outperforming semi-supervised methods by \cite{Zhang2017deep}. This mean teacher model is further extended by  
\cite{yu2019uncertainty} where  the consistency loss is masked through an uncertainty map. The idea is that regions with low uncertainty should contribute to the segmentation consistency loss. The authors estimate segmentation uncertainty by using  Monte Carlo dropout \cite{kendall2017uncertainties} in the teacher network. For the task of atrial segmentation in a dataset with 80 training and 20 test MR images, the authors report four points increase in Dice over a supervised model trained using 16 labeled images. This improvement is however similar to that of other competing semi-supervised methods used for comparison (e.g., \cite{li2019transformation}).}

\vspace{4pt}
\noindent {\bf \textit{Via representation similarity:}}
{\brown 
 Semi-supervised learning method often use adversarial loss functions to bridge the dissimilarities between the representations of labeled and unlabeled images, thereby leveraging unlabeled data during the course of training. In this paradigm, the term representation may refer to the feature maps at the latent space of an encoder \cite{baur2017semi,sedai2017semi}, or to the final segmentation map of a decoder \cite{Zhang2017deep,mondal2018few}. We present a detailed description of these methodologies as follows.}

The semi-supervised framework suggested by \cite{baur2017semi} consist of a U-Net with two loss functions: a Dice-based segmentation loss that is computed based on the labeled data; and an embedding loss, which, given a batch of labeled and unlabeled data, brings the feature embedding of the same-class pixels as close as possible while pushing apart the feature embedding  of the pixels from different classes. To identify same-class pixels between labeled and unlabeled images, the authors assume the availability of a noisy label prior for unlabeled images. Also, to reduce the number of pair-wise comparisons between feature embedding of all pixels within the batch, they employ a pixel sampling scheme. The suggested semi-supervised framework proves promising in improving the segmentation of multiple sclerosis  in the presence of limited data and domain shift. {\brown However, one caveat is that the suggested method is evaluated only in a low data regime where the labeled and unlabeled training sets have images from 3 and 9 subjects, respectively.}

\cite{sedai2017semi} propose a semi-supervised learning framework consisting of a segmentation network and an auto-encoder. The training process begins with training a variational auto-encoder, which stores the knowledge learned from the unlabeled images in its latent space. Next, the segmentation network, which is also a variational auto-encoder, is trained using the labeled data. To leverage the knowledge learned from the unlabeled data, in addition to the segmentation loss, the segmentation network benefits from an l2-loss between its latent feature vector and the one generated by the reconstruction network for a given labeled image. {\brown The authors evaluate this framework in segmenting optic cup segmentation in fundus images using an unlabeled training set of 11400 images and a labeled training set of 400 images. On a test set with 200 images, the semi-supervised model achieves four points and one point increase in Dice over the supervised model, when 12.5\% and 100\% of the labeled set is used for training, respectively.}  

\cite{Zhang2017deep} propose a semi-supervised learning framework according to an adversarial game between a segmentation network (U-Net) and an evaluation network (encoder). Given an input image, the segmentation network generates a segmentation map, which is then stacked with the original image and fed to the evaluation network, resulting in a quality score. During training, the segmentation network is updated with two objectives: 1) minimizing the segmentation loss for the labeled images and 2) making the evaluation network assign a high quality score to the unlabeled images. On the other hand, the evaluation network is updated so as to assign a low quality score to unlabeled images but a high quality score to labeled images. Owing to this adversarial learning, the segmentation network enjoys a supervision signal from both labeled and unlabeled images. {\brown The authors evaluate this semi-supervised learning framework in two applications, one of which being gland segmentation in histopathology images, where the labeled and unlabeled training sets consist of 85 and 100 images. The suggested model however yields mixed results, yielding comparable Dice to a supervised model on the larger test set (60 images) while  achieving 1 point increase in Dice on the smaller test set (20 images).}

\cite{mondal2018few} propose a semi-supervised framework that integrates labeled and unlabeled  images for the task of brain tissue segmentation in MR images. For this purpose, the authors train the segmentation network by introducing an additional fake class, resulting in a $k+1$ class segmentation problem where $k$ is the number of classes present in the dataset. During training, the segmentation network seeks to maximize the probability of the correct class for each pixel in the labeled images. For unlabeled images, the segmentation network minimizes the probability of each pixel belonging to the fake class, which has the effect of maximizing the probability by which an unlabeled pixel belongs to one of the $k$ classes. The suggested method proves effective in segmenting brain tissues from MR images when the training set contains only a few annotated images.

%% file: chapters/altered_img_rep.tex
Altered image representations consist of projecting or transforming the images into a more informative or compact representation, which present deep models with an easier problem to solve, thereby reducing the need for large training sets. Informative representations can be particularly effective for 2D medical image segmentation whereas compact representations can benefit 3D applications where the curse of dimensionality requires large annotated datasets. This section reviews altered representations for both 2D and 3D images.

%3D models are the natural choice for 3D medical scans, but they are parameter rich requiring many training samples. On the other hand, 2D models require fewer training samples but are unable to capture the 3D context. To overcome this, some frameworks attempt to alter the image representation to make the best use of both worlds. This section looks at different input representations, besides the typical 2D and 3D image inputs, that are used to improve the segmentation performance. Altered image representations can also be used to aid with 2D image segmentation problems by providing representations that are more informative for the task being solved - \cite{fu2018joint}. This section begins with altered representations of 3D images and then reviews altered representations of 2D images.

\vspace{4pt}
\noindent {\bf \textit{Altered 3D image representation:}}
 Training segmentation models with altered 3D image representations include training 2D models with multi-scale and multi-view patches \cite{wang2017multi}, fusing 2D models trained for the three clinical views \cite{xia2018bridging}, training a 2D model with a 2.5D image representation \cite{angermann2019projection}, and finally training a 3D model with a 3D representation augmented with handcrafted features \cite{ghafoorian2017location}. We explain these methods in more detail as follows.

\cite{wang2017multi} make use of multi scale 2D patch-based pixel predictions for the task of lung nodule segmentation. The network has three shallow branches, one for each of the three orthogonal clinical views that share the same central pixel. Each branch is fed a 2D 2-channel input that captures the nodule at two different scales. The three branches are then fused to provide a binary prediction for the central pixel of the patch. Thus, there are six new 2D patches used for every voxel in the 3D volume being segmented. The segmentation results for all voxels are finally put together to obtain the 3D segmentation mask.   {\brown Testing on 393 nodules from the LIDC dataset, they achieve 7\% increase in Dice over GrabCut;} however, they have not compared their method against a single view CNN approach.

\cite{xia2018bridging} propose a 2-stage approach where the first stage uses a set of 2D segmentation networks whose outputs are further fused in the second stage through a 3D volumetric fusion network. The 2D networks generate slice-by-slice predictions along each of the 3 orthogonal views---axial, sagittal and coronal. The stacks of predicted segmentation masks for the 3 views are then concatenated with the original image, creating a 4-channel input to the second network, which learns to fuse the predictions to produce the final 3D prediction. {\brown For the task of pancreas segmentation, the authors experiment with 4-fold cross validation on the NIH pancreas segmentation dataset, which contains 82 abdominal CT volumes. For their experiments on multi-organ segmentation, they curated an in-house dataset of high resolution abdominal CT scans and tested their approach on 50 scans. In both scenarios,} the suggested method improves Dice score by 1\% over baselines that use majority voting for volume fusion. 

\cite{angermann2019projection} make use of intensity projections, specifically maximum intensity projections (MIP) at multiple angles, which are then fused to create a 2.5D representation of magnetic resonance angiography images. {\brown The authors use their approach for the task of volumetric blood vessel segmentation with 18 test volumes.} However, the authors have not demonstrated a significant gain over the 2D and 3D performance baselines.
% Performance is worse than 2D and 3D approaches

% 2D model - hand crafted location features used
\cite{ghafoorian2017location} use an image representation based on registered T1 and FLAIR MR images augmented with dense handcrafted features to segment white matter hyperintensities. {\brown On a test set of 46 cases,} their multi-scale architecture equipped with the hand-crafted features achieves 6\% increase in Dice over their single scale baseline that does not incorporate the handcrafted features. Noteworthy, the improved segmentation performance is mainly due to the contribution of handcrafted features rather than the multi-scale paradigm.

%authors experiment with different multi-scale fusion techniques by changing where the different scaled inputs are fused, which only marginally improves the Dice scores. However, the main performance boost is due to the inclusion of hand-crafted features, which achieves a 6\% Dice score improvement over their single scale baseline that does not incorporate hand-crafted features.

\vspace{4pt}
\noindent {\bf \textit{Altered 2D image representation:}} The aforementioned methods offered altered representations for 3D images; however, even problems that are inherently 2-dimensional can benefit from using a different representation.
\cite{fu2018joint} utilize an image representation based on the polar coordinate system for the purpose of joint cup and disc segmentation in fundus images. Specifically, the authors use a circular image crop around the cup and disc region, which is then converted to a rectangular image through a transformation from Cartesian to polar coordinates. Scale-based data augmentation is  performed by varying the radius of the circle prior to coordinate conversion. The authors have used the area under the ROC curve for evaluating the binary glaucoma classification performance using the cup to disc ratio. {\brown They test their model on 325 images from the ORIGA dataset and the entire SCES dataset containing 1676 fundus images.} The suggested image representation enables a 4\% gain in AUC compared with the existing state of the art trained using the standard Cartesian plane image representation.

%% file: chapters/multi_task_learning.tex
%\iffalse Without explicitly specifying a loss function tailored to the shape and border of the target [(see loss functions section)], \fi 

Multi-task learning \cite{zhang_survey_2017} refers to the paradigm in which multiple tasks are derived from a single learned representation. In modern applications, this can be realized by a single feature extractor (encoder) on which multiple tasks (e.g. classification, detection, segmentation) are performed. Intuitively, this paradigm encourages the encoder network to learn a latent representation that generalizes across the required tasks, with each task serving as a regularizer for the others. The studies outlined below demonstrate that adding a parallel task generally results in improved segmentation performance {\brown at little or no cost in terms of additional data}. These tasks may be supervised (e.g. classification, detection), {\brown requiring additional annotations}), or unsupervised (e.g. image reconstruction) {\brown with no additional labeling required}.

% In an early application of multi-task learning to gland segmentation in histology images, \cite{chen_dcan:_2016} designed their model to explicitly segment the contour of glands, in addition to the glands themselves. By implicitly encouraging the shared encoder network to capture information about contours in the image, they were able to win the 2015 MICCAI Gland segmentation challenge \cite{sirinukunwattana_gland_2017}.  % J.C. I think this is covered in more detail in the shape prior section

Most multi-task learning applications to medical image segmentation involve a variant of the U-net. The upsampling (segmentation) branch can be understood as just one of multiple output ``heads'' connected to a feature extractor, which allows for a natural extension to other tasks from the abstract feature space. \cite{mehta_y-net:_2018} apply such a multi-task U-net to segment different tissue types in breast biopsy histopathology images, where an additional classification head  is trained to classify whether the image is malignant or not. Simply adding this additional branch, which encouraged the model to learn a feature space relevant to diagnosis in addition to tissue segmentation, significantly improves the IoU of the vanilla network by 7\%. 

% ok this paper is more complicated than i thought
Similarly, \cite{jaeger_retina_2018, huang_3d_2018} propose a joint segmentation and detection framework. \cite{huang_3d_2018} use their method for colorectal tumor segmentation in MRI volumes. Their proposed model resembles Mask-RCNN \cite{he_mask_2017}, in which a global image encoder network detects regions of interest (ROIs), and a local decoder performs tumor segmentation on the proposed regions. The feature pyramid representing each ROI is passed to the local decoder, which, unlike Mask-RCNN, only segments the region of interest. The authors show that the local (rather than global) decoding approach preserves spatial details as well as decreases GPU footprint. This approach outperforms other U-net variants by several points and the popular Mask R-CNN algorithm by 20 points when the Dice score is used for performance evaluation. 

% in this case, the image reconstruction was also denoising so they had a GT image target... so technically it's not unsupervised
\cite{sun_joint_2018} apply multi-task learning to brain tissue segmentation, combining the segmentation task with image denoising. The base architecture of their model can still be described as a modified U-net, in which a shared encoder generates a latent feature representation for two separate decoders (for image reconstruction and segmentation). However, instead of training the network from scratch, Sun and colleagues pre-train the reconstruction head before training the full model end-to-end. By jointly performing image reconstruction with segmentation, they observed improved segmentation performance in all types of brain tissue across different metrics. % Their proposed system was a cascade of these networks, and 

{\brown The objective function in the multi-task paradigm is generally presented as the summation of the loss functions for each respective task, implicitly giving each an equal weight. In a recent work, \cite{li_recurrent_2019} present a network for echocardiographic sequence segmentation comprised of a feature pyramid based CNN for segmentation and classification. They introduce an aggregation loss which frames the final loss as a weighted combination of the segmentation and classification loss functions, allowing these weights to be empirically calculated during training. Their ablation studies show that multi task learning (segmentation and image-level classification) significantly increase segmentation when compared to the same network without the additional classification task.}

% is there not a paper for just image reconstruction using a regular autoencoder?!
One drawback of the above approaches is that the companion task (classification or detection) requires additional labels for each image. Image reconstruction can be considered an unsupervised task that still provides the regularization benefits of multi-task learning. \cite{myronenko_3d_2018} propose a framework combining image reconstruction and segmentation for the task of brain lesion segmentation.  The proposed model is a variational auto-encoder with an asymmetrically large encoder backbone and two decoders: the first decoder is trained to reconstruct the input MR image whereas the second decoder generates the segmentation maps. The suggested method outperforms other modifications such as CRF post-processing as well as sophisticated test-time data augmentation, winning the first place in the 2018 BraTS challenge~\cite{menze2014multimodal}.

% Advances in image generation can also be incorporated into the reconstruction branch. 

% could also frame as a future direction
In addition to providing robustness to the learned feature representation via regularization, the multi-task paradigm provides a feasible framework for consolidated biomedical image segmentation. As Harouni and colleagues \cite{harouni_universal_2018} note, given the sheer number of potential conditions, a single model per condition is not a scalable approach for clinical use. As a proof of concept, Harouni and colleagues use an architecture similar to Y-net \cite{mehta_y-net:_2018} to segment different organs in several different imaging modalities using a single network. A single U-net is trained to segment nine different targets encountered in thoracic imaging, which is complemented by a classification branch trained to determine the input domain (e.g., CT, MRI, Ultrasound). While this model does not exceed the state of the art performance for any one given modality or target, performance is competitive across all domains. The authors also report that the presence of the image classification branch results in slightly improved segmentation performance.

%% file: chapters/shape_prior.tex
\begin{figure*}[t]
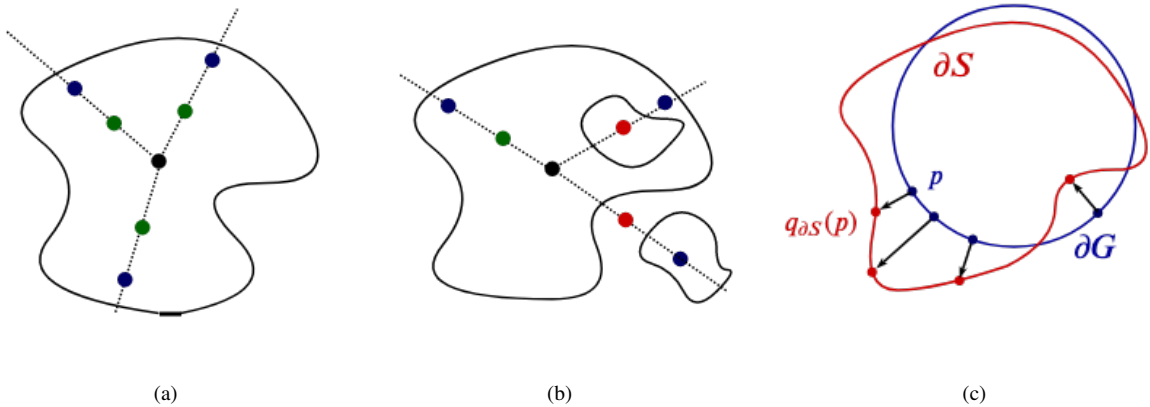

    \centering
    \subfigure[]{\includegraphics[width=0.27\linewidth]{pics/star_obj.png}}\hspace{4pt}
    \subfigure[]{\includegraphics[width=0.27\linewidth]{pics/non_star_obj.png}}\hspace{4pt}
    \subfigure[]{\includegraphics[width=0.3\linewidth]{pics/b_loss.png}}
    \caption{Shape regularization can combat the limited annotation problem by imposing additional constraints on predicted segmentation masks. Two common shape regularization methods are 1) star shape prior where any point in between the center and an interior point is constrained to be interior, and 2) boundary regularization. (a) An example of a segmented region that meets star shape prior. The black circle indicates the center of the segmented shape, the blue circles refer to interior points near the boundary, and the green circles are interior points that satisfy the star shape criteria.(b) An example of a segmented region with a hole and an isolated island, which does not meet requirements of star shape prior. The red circles indicate the pixels that lie exterior to the segmented region. (c) Boundary regularization improves segmentation accuracy around boundaries by minimizing point-wise deviation between the segmentation mask $S$ relative to the ground truth $G$.}
    \label{fig:star_and_shape}
\end{figure*}

Shape defines a region of interest (ROI) in segmentation problems under certain constraints, e.g. smooth and semantically sound. Such constraints can be effectively encoded as regularization towards more realistic appearance of the segmentation output, especially when well-annotated data is scarce. Specifically, shape regularization consists of imposing a prior, highlighting certain geometric and structural characteristics, on the segmented ROIs, by operating either at pixel-level with an emphasis on shape and boundary explicitly, or in depth to capture high level features related to semantic meanings. In this section, we refer to the methods serving the former and latter objective as \emph{Shallow regularization} and \emph{Deep regularization}, respectively.

%which typically induce extra term on boundary pixels in addition to regional loss like pixel-wise cross entropy; whereas \emph{Deep regularization} directly regularizes high level feature maps.

\vspace{4pt}
\noindent {\bf \textit{Shallow shape regularization:}}
Shallow shape prior may regularize boundary pixels towards a certain class of shapes. \cite{mirikharaji2018star} leverage a star shape prior via an extra loss term on top of a binary cross entropy loss. To regularize a segmented ROI towards a star shape, any point on the linear path in between the ROI center and an interior point is expected to be interior as well (Fig~\ref{fig:star_and_shape} a \& b), ensuring a smooth segmentation mask without holes. This definition comprises a broad class of objects even including convex shapes as a special case. {\brown With this additional term, authors evaluate the effectiveness of the star prior for skin lesion segmentation on the ISIC 2017 dataset (2000 training 600 test images) and report a $3.0\%$ gain in Dice using U-net as segmentation network over the same network without star shape regularized loss.}%, ranking them top in the ISBI 2017 competition.

Another class of shallow priors operate on boundary pixels to further improve segmentation accuracy around the boundaries. Boundary points, the first order derivative of a region, better capture the proximity of two shapes by inducing an extra penalty term between the estimated and expected pixels along the boundary. Inspired by the optimization technique for computing gradient flows of curve evolution, \cite{kervadec2019boundary} introduce a non-symmetric $L_2$ loss to regularize boundary deviation of the segmentation mask $S$ relative to the ground truth $G$,

%Another class of shallow prior operates on boundary points to compliment the regular cross entropy loss. Essentially, cross entropy loss is based on the regional overlap between prediction and ground truth masks, treating each pixel equivalent. Boundary points, regarded as the first order derivative to region, better capture proximity of two shapes by inducing extra penalty between an estimated and expected point along the boundary. Inspired by the optimization technique for computing gradient flows of curve evolution, \cite{kervadec2019boundary} introduces a non-symmetric $L_2$ loss to regularize boundary deviation of the segmentation mask $S$ relative to the ground truth $G$,

\begin{equation}
    \text{Dist}(\partial G, \partial S) = \int_{\partial G} ||q_{\partial S}(p) - p ||^2 \mathrm{d}p,
\end{equation}
where a boundary point $p$ on $\partial G$ (boundary of GT) is aligned against its counterpart $q$ on $\partial S$ (boundary of prediction), which is written as $q_{\partial S}(p)$, such that $p\rightarrow q$ is norm to GT boundary at point $p$ (see Fig~\ref{fig:star_and_shape} c). {\brown By using the boundary loss for brain lesion segmentation, authors evaluate on two MRI datasets (ISLES with 74 training plus 20 testing and WMH of 50 training plus 10 testing) and report an $8\%$ gain in Dice and a $10\%$ gain in Hausdorff score over a baseline that uses generalized Dice as the loss function on ISLES dataset compared to marginal improvement on the WMH dataset. \lref{ql:karimi} \cite{karimi2019reducing} explore three approximations to Hausdorff distance such that it can be directly minimized. The authors report performance on multiple applications including 2D prostate ultrasound (450/225), 3D prostate MRI (80/30), 3D Liver CT (100/31) and 3D Pancreas CT (200/82), resulting in 18\% to 45\% reduction in Hausdorff distance without degrading other performance metrics such as Dice similarity coefficient.

More recently, \cite{duan2019automatic} propose a refinement approach following a segmentation network that utilizes shape information by registering a cohort of atlas masks to the target segmentation mask. In their work, the first step consists of a 2.5D FCN trained to segment ROIs and localize 6 landmark points, e.g. center of the mitral valve. This step is conducted on both high-resolution (HR) and low-resolution (LR) cardiac MR images whereas HR volumes will later on be used as atlases. In the refinement stage, a sequence of landmark alignment, atlas selection and deformable registration are used to map a cohort of HR volumes to the target LR volumes, then labels from this cohort of atlases are fused to derive a shape-refined mask of LR images in the HR mode. The authors evaluate on MR scans of both healthy and pathological cases. For healthy cases which only contain HR images, LR images are simulated by downsampling and adding artefacts and ground truth masks in HR are used for evaluation. With a 1000/600/231 data split, the 2.5D FCN trained on the LR cases with refinement achieves comparable performance relative to the 2.5D FCN trained on the HR cases, trailing by 2.0\% for most regions in Dice and 0.5mm worse in Hausdorff distance. As for pathological cases, 629 volumes are in the LR mode only and another 20 volumes have both HR and LR images. For the task of myocardium segmentation, the authors report performance superior to 3D-ACNN~\cite{oktay2017anatomically}, achieving  4.3\% increase in Dice and 4.15mm (40\% less) improvement in Hausdorff distance . However, the refinement stage is computationally expensive, taking 15-20 minutes per case with their multiple CPU implementation.

Shape priors are also leveraged in training with partially labeled data. \cite{zhou2019prior} tackle the problem of multi-organ segmentation when more datasets are available as single organ annotated. Specifically, the proposed method assumes that the average organ size distribution approximates what is learned from fully labeled data, therefore, it can be utilized as a organ-specific prior guiding the training on single organ datasets. In their experiments, an FCN is initialized on 30 fully labeled abdominal CT scans  to segment 13 anatomical structures, followed by partially labeled datasets on spleen, pancreas and liver, each of which containing 40 scans. Their approach outperforms a naive partial supervision implementation, which does not benefit from shape prior regularization, by 1.69\% in Dice with ResNet-101 as backbone for 2D task and by 0.45\% in Dice using 3D U-Net as the backbone models, both based on 5-fold cross-validation.}

% \begin{figure*}
%     \centering
%     \includegraphics[width=0.75\linewidth]{pics/star_and_boundary.png}
%     \caption{a.star shape object; b. non-star shape object; c. boundary loss}
%     \label{fig:star_and_shape}
% \end{figure*}

\vspace{4pt}
\noindent {\bf \textit{Deep shape regularization:}}
Shape regularization can also be applied to high level semantic features. Compared to shallow approaches, deep regularization, a.k.a deep supervision or deep priors, is less prone to image noise and more semantically and structurally aware. \cite{ravishankar2017learning} incorporate deep prior within segmentation framework where a segmentation FCN is followed by a shape regularization FCN, which functions as a \emph{convolutional denoising autoencoder} (CDAE), consisting of an encoder that projects the segmentation mask to the shape space and a decoder that samples a segmentation mask from the shape space. In addition to the reconstruction loss, the regularization FCN has a projection loss that constraints ground truth and predicted segmentation to have similar encodings in the shape space.
%and does not have skip connection between encoder and decoder blocks. 
%The total loss is the summation of 1) projection loss between the predicted and de-noised segmentation; 2) the encoded error between CADE's encoded ground truth and predicted segmentation; and 3) the original segmentation loss between the ground truth and predicted segmentation. To effectively train the shape regularization network, the authors feed incomplete shapes to CDAE as data augmentation, e.g. corrupted shapes and predictions from U-Nets that have not converged. 
{\brown Combining both data augmentation and deep regularization, \cite{ravishankar2017learning} report a $4.66\%$ gain in Dice relative to a vanilla U-Net segmenting kidney on 2D ultrasound B-mode images (100 training and 131 testing)}. 

 \cite{oktay2017anatomically} adopts a similar cascaded architecture with a major difference: the regularization FCN is first pre-trained as an auto-encoder with ground truth masks, and then only its frozen encoder is used as a regularizer during training the segmentation network. Therefore, the objective function reduces to a regular segmentation loss and a shape projection loss. {\brown The suggested model achieves  $1.2\%$ and $2.0\%$ improvement respectively in Dice over \cite{ravishankar2017learning} for 3D endocardium and myocardium segmentation on a cine-MR dataset (1000 training and 200 testing)}. The authors attribute the inferior performance by \cite{ravishankar2017learning} to over-regularization, which they have overcome by replacing CDAE with a frozen encoder during training. 

\cite{dalca2018anatomical} suggest a segmentation VAE that leverages shape prior in order to learn from unpaired images and segmentation masks. {\brown The VAE consists of an image encoder, which is initialized from scratch, and a frozen decoder, which is selected from an auto-encoder that has previously been trained for the task of mask reconstruction on 5000 T1-weighted brain MRI scans.} Since the VAE uses a segmentation decoder, it generates a segmentation mask given an input MR image. However, the input MR images have no corresponding ground truth segmentation; therefore, the VAE is trained by minimizing the L2-loss between the input MR image and the predicted segmentation after being transformed through a 1x1 convolution block. During inference, the authors use the decoder output as the segmentation result. {\brown For the task of brain structure segmentation, the suggested method is evaluated on 9000 MRI scans, achieving Dice scores ranging between $0.50$ and $0.80$ without any comparison against supervised methods.

\cite{he2019dpa} propose a segmentation network that benefits from priori anatomical features. To capture anatomy priors, the authors first train 
a denoising autoencoder with unlabeled data. To embed the shape peior in the segmentation network, the authors concatenate the latent space of the segmentation network with priori anatomical features generated by the frozen encoder of the autoencoder. The authors use 26 labeled and 118 unlabeled abdominal CTA volumes for training and report 1.0\% gain in Dice over the same architecture without anatomy priori on a test set of 26 volumes.} % As one of the pioneering works on zero-shot deep learning for medical image segmentation, this model is worth considering when annotations are too few to train a supervised network.

%In the absence of paired images and segmentation masks, \cite{dalca2018anatomical} make use of 2 variational auto-encoders (VAE)--- one for reconstructing images and the other for reconstructing the unpaired segmentation masks, which can potentially be from a different domain. The segmentation VAE is first trained to learn the shape prior with the help of a segmentation atlas, following which the image VAE is trained. The image VAE consists of 3 parts: (i) its own encoder which produces the latent feature vector, (ii) the decoder of the mask VAE with its weights frozen, which now takes as input the encoded image and produces the predicted masks, and (iii) an additional generator layer that acts as the last part of the image VAE decoder to reconstruct the image from the mask. This training scheme, thus, makes use of unsupervised training via image reconstruction while learning to produce segmentation maps as an intermediate output using the encoded shape priors. The authors employ this approach for the task of brain structure segmentation using MR Images and report Dice scores that range between $0.50$ and $0.80$ on T1w images without any comparison against supervised methods. As one of the pioneering works on zero-shot deep learning for segmentation, this model is worth considering when annotations are too few to train a supervised network.

%% file: chapters/post-processing.tex
{\brown Variants of auto-encoders have recently been adopted for automatic error correction in medical images \cite{Larrazabal2019Anatomical,tajbakhsh2019errornet}, outperforming conditional random fields (CRFs) especially in the presence of domain shift and limited annotations. Nevertheless, CRFs are still} the most commonly adopted and recognized approach to refine segmentation masks of both natural images (\cite{schwing2015fully}, \cite{chen2017deeplab} and \cite{zheng2015conditional}) and medical images (\cite{roth2015deeporgan}, \cite{chen2018end} and \cite{wachinger2018deepnat}). In the following, we briefly explain CRFs and then cover the papers that have used  variants of CRF for medical image segmentation.

To obtain more realistic predicted masks, a CRF model incorporates two regularization terms: a smoothness term that removes small isolated regions, and an appearance term that ensures nearby pixels with similar color will more likely belong to the same class. The segmentation result, inferred as a maximum a posteriori (MAP) estimate from the CRF defined across all pixels, is expected to capture both local features and spatial dependency more holistically. As a consequence, CRF is able to refine a collection of inaccurate and coarse pixel-level predictions, producing sharp boundaries and fine-grained segmentation masks. %, thereby that is expected to complement segmentation networks with significantly less bias.

Concretely, a CRF models pixel-wise labels, $\mathrm{X}$, collectively as a random field that is conditioned upon image/volume intensities, $\mathrm{I}$. This CRF can be characterized by its potentials, consisting of unary and pairwise terms \cite{krahenbuhl2011efficient},

\begin{equation}
    E(\mathbf{x}\mid\mathbf{I}) = \sum_i \phi_u(x_i) + \sum_{i \ne j}\phi_p(x_i, x_j).
    \label{eqn:gibbs}
\end{equation}

\begin{figure*}[t]
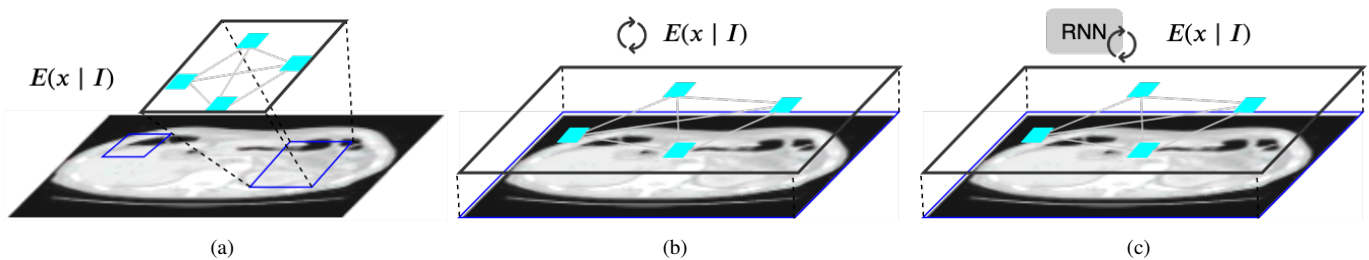

    \centering
    \subfigure[]{\includegraphics[width=0.32\linewidth]{pics/local_CRF.png}}
    \subfigure[]{\includegraphics[width=0.32\linewidth]{pics/FC_CRF.png}}\hspace{4pt}
    \subfigure[]{\includegraphics[width=0.32\linewidth]{pics/RNN_CRF.png}}
    \caption{Post segmentation refinement can, to some degree, correct the segmentation errors. (a) local CRF optimizes Gibbs energy over local patches weighing in pairwise pixel dependency. (b) Fully Connected CRF (FC-CRF) extends the local scope of the CRF  to the whole image in an efficient manner. (c) CRF as Recurrent Neural Networks (RNN-CRF) makes FC-CRF end-to-end trainable by replacing the iterative calculations with an RNN.
    %Illustrations of the common post segmentation refinement methods:
    }
    \label{fig:CRFS}
\end{figure*}

The unary potential $\phi_u(x_i)$ is computed independently per pixel, which incorporates shape, texture, location and color descriptors. The existing variants of CRF differ in terms of the definition of the pairwise potential term, $\phi_p(x_i, x_j)$, and the underlying optimization technique. These variants include \textit{Local CRF}, which considers neighboring pixels only, i.e. $j\in neighbor(i)$; \textit{Fully Connected CRF} (FC-CRF), which considers all pixel pairs with an iterative mean field approximation of Eq.~\ref{eqn:gibbs}; and  \textit{RNN-CRF}, which takes a similar approach to FC-CRF, but it is now end-to-end trainable using Recurrent Neural Networks (RNNs). 
% The seminal work of \cite{krahenbuhl2011efficient} presents a major breakthrough in optimizing the gibbs energy approximated by Eq.~\ref{eqn:gibbs}, where the authors reduce the computation complexity by an order, from $O(N^2)$ by a naive implementation with $N$ being the number of pixels in an image, making CRF a practically viable option. 
A visual comparison of the three types of CRF is provided in Fig~\ref{fig:CRFS}.

% \begin{figure*}
%     \centering
%     \includegraphics[width=0.99\linewidth]{pics/CRFs.png}
%     \caption{a. local-CRF; b. Fully-Connected CRF; c. RNN-CRF}
%     \label{fig:CRFS}
% \end{figure*}

\subsubsection{Locally Connected CRF}
\label{sec:local_crf}
Restricting the pairwise potentials to neighbouring pixels, the resulting CRF is designed to induce local smoothness. \cite{roth2015deeporgan} explore 2D CRF as well as a 3D Gaussian smoothing as post-processing for pancreas segmentation in CT images. The weights corresponding to pairwise and unary potentials are calibrated by a grid-search. {\brown In terms of performance, the authors evaluate on 82 contrast-enhanced abdominal CT volumes in a 4-fold cross-validation manner, and report an average of $3.3\%$ gain in Dice using CRF that falls short of a $6.9\%$ gain using Gaussian smoothing.} However, CRF does reduce the standard deviation of Dice in all experiments, demonstrating its regularization capability in reducing inference variance.
\cite{cai2016pancreas} use CRF to fuse mask and boundary predictions, which are separate branches off the same backbone during training, as a cascaded task that post-processes pancreas segmentation of MR Images. In this work, CRF still operates on neighbouring pixels in a feature space spanned by hand-crafted image features and the features learned by both segmentation branches. {\brown By using CRF for decision fusion, the authors report a  2.3\% gain in Dice (73.8\% $\pm$ 12.0\% $\rightarrow$ 76.1\% $\pm$ 8.7\%) over a baseline without CRF on a MRI dataset consisting of 78 scans (52 for training and 26 for testing).} %Comparatively against the previous work, Dice($\%$) is improved to $76.1\pm8.7$ in MRI whereas $71.8\pm10.7$ in CT by \cite{roth2015deeporgan}. 

\subsubsection{Fully Connected CRF (FC-CRF)}
\label{sec:fc_crf}
\cite{krahenbuhl2011efficient} provides an efficient inference approach to Eq.~\ref{eqn:gibbs} using mean field approximation. The resulting algorithm reduces the computational complexity from quadratic to linear in the number of pixels involved in the computation. FC-CRF has proven effective as a segmentation post-processing solution for both natural images \cite{chen2017deeplab} and 2D medical images, e.g. \cite{fu2016retinal} in retinal images, \cite{gao2016segmentation} on individual CT slices. The work by \cite{kamnitsas2017efficient} is the first to extend FC-CRF to 3D brain lesion segmentation in MR Images, leveraging intensity and spatial association under 3D context. {\brown However, their 3D generalization on BraTS dataset~\cite{menze2014multimodal} (274 training and 110 testing) leads to marginal performance gains in Dice, i.e. $3.7\%$ over Random Forests, $0.3\%$ over an ensemble method, and merely $0.7\%$ over their proposed architecture, which is a patch-based multi-scale 3D CNN network}\iffalse---with more marginal gains in sensitivity and Hausdorff distance\fi. In addition, the authors note that configuring 3D FC-CRF is a laborious task\iffalse, where they used a grid search, resulting in marginal benefits\fi. Other 3D FC-CRF endeavors include a U-net $+$ 3D FC-CRF by \cite{christ2016automatic} for liver and lesion segmentation in CT images  and a 3D FC-CRF with spectral coordinates characterization by \cite{wachinger2018deepnat} for neuroanatomy segmentation in MR Images. In the works above, FC-CRF refines segmentation masks that often exhibit small isolated regions and zigzag boundaries, but its effectiveness is greatly hindered by the extensive manual tweaking, or in other words, being not end-to-end trainable.

\subsubsection{CRF as Recurrent Neural Networks (RNN-CRF)}
\label{sec:e2e_crf}
RNN-CRF organically integrates CRF with CNNs, making it possible to train the whole network in an end-to-end manner. \cite{zheng2015conditional} reformulates the mean field approximation of FC-CRF as a stack of common CNN layers and the iterative optimization as hidden states in an RNN. {\brown \cite{fu2016deepvessel} combine a multi-scale and multi-level CNN that has auxiliary output layers with a RNN-CRF, and achieve the state-of-the-art performance of vessel segmentation on three public fundus datasets (DRIVE, STARE and CHASE-DB1)}. \cite{monteiro2018conditional} implement a 3D version of RNN-CRF for volumetric medical images on top of a V-net segmentation network. The authors evaluate their  3D RNN-CRF on multiple datasets. {\brown On the PROMISE 2012 dataset, which consists of 50 3D-MRI prostate images, 3D RNN-CRF improves the Dice from $76.7\% \pm 10.9\%$ to $78.0\% \pm 11.0\%$ evaluated in a 5-fold cross-validation manner. And on BraTS 2015~\cite{menze2014multimodal}, consisting of 220 multi-modal MR images of brain tumors, 3D RNN-CRF slightly improves the Dice for tumor segmentation from $73.5\% \pm 10.5\%$ to $73.8\% \pm 10.5\%$ based upon a split of 85\%/15\% (187 training and 33 testing)}. The authors acknowledge that the improvements by 3D RNN-CRF are inconclusive, and attribute that to the intrinsic differences between natural and medical images: 1) object segmentation in 2D RGB images is generally easier with greater contrast and better defined boundaries; 2) the relatively low resolution of 3D volumes causes a mosaic appearance, which poses further challenges on top of blurry edges; and 3) {\brown the local nature of ROIs in medical images downgrades the need to capture global image context beyond what is modeled by the segmentation network, leaving less room for improvement by the CRF.}
% ROIs in medical settings tend to be ``local'', rather than scattered, e.g. multiple birds in natural images. Hence, the segmentation network is already capable of capturing all of the relevant spatial and colour relations, leaving less room for improvement by the CRF.
More recently, \cite{chen2018end} report promising results with their 3D RNN-CRF implementation jointly trained with a 3D U-net for the task of brain lesion segmentation in MR Images. In their approach, CRF operates on high-level features learned by the CNN, which are less prone to image noise than raw intensity values directly taken from the input image. {\brown The authors evaluate on 60 MR scans from WMH 2017 Challenge (36 training, 12 validation and 12 testing) and report between six to seven points increase in Dice over a baseline U-Net and other implementations of CRF.}

%% file: chapters/imperfect_annotations.tex
\begin{figure*}
\centering
    \includegraphics[width=0.8\linewidth]{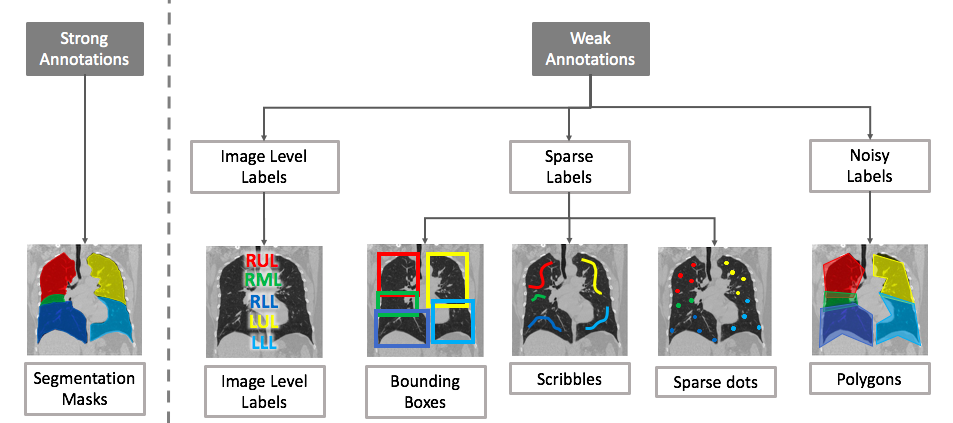}
    \caption{Comparing strong and weak annotations for lung lobes in the coronal view of a chest CT scan. (left) an example of strong annotations in the form of well contoured segmentation masks. (right) examples of different types of weak annotations discussed in this section, which can take the form of (i) bounding box or image level labels, (ii) sparse pixel annotations, or (iii) noisy annotations }
    \label{fig:weak_anno}
\end{figure*}

Creating manual segmentation masks, also known as strong annotations, is time consuming and tedious, particularly for 3D images. To combat the high cost associated with strong annotations, researchers have recently explored the use of weak annotations, which can be obtained at significantly lower annotation cost. Reviewing the literature, we have identified three types of weak annotations: 1) image level annotations; 2) sparse annotations, where only a fraction of the slices or pixels are annotated; and 3) model-generated annotations or noisy annotations, which tend to appear under- and over-segmented.  \figurename~\ref{fig:weak_anno} shows different types of weak annotations outlined above. While the absence of strong annotations may seem like an obvious handicap, recent research, as summarized below, has shown that it is possible to train fairly effective models with weak annotations.

\subsection{Learning with Image Level Labels}
\label{sec:weak_l}
Weakly supervised techniques can take advantage of  bounding boxes or image level labels. The common weakly supervised approaches suggested for medical image segmentation are based on class activation maps  or multiple instance learning. We review both approaches as follows. 

\subsubsection{Class Activation Maps (CAMs)}
\label{sec:activation}
A recurring idea in weakly supervised learning is the use of class activation maps \cite{zhou2016learning} and its variants (e.g., \cite{selvaraju2017grad}), where the idea is to  to combine the feature maps to generate class-specific saliency maps. In the following, we review how this technique can be used in conjunction with image-level annotations.

The trend for tackling the problem of having only image-level labels is to use some form of class activation maps (CAMs), {\brown which can be binarized to generate  a segmentation mask. For the task of diagnostic brain tumor segmentation in confocal laser endomicroscopy (CLE) glioma images, }\cite{izadyyazdanabadi2018weakly} use a multi-layer CAM in the form of a 3-stage inception network where the penultimate feature maps from each network are passed on to the next stage. In parallel, CAM followed by global average pooling is applied to these feature maps to obtain the image-level label prediction. The network performance is boosted further by upregulation of  confident predictions and downregulation uncertain predictions, wherein the regions activated in a single class map are determined to be confident and those activated in both class maps are uncertain. {\brown The dataset consists of 6287 CLE images from 20 patients with a 12-4-4 patient split for training, validation and testing.} Their average IOU improvement across different tests over the baseline of just using CAM is 20\%. \cite {feng2017discriminative} propose a 2-stage approach  consisting of a coarse image segmentation followed by a fine instance-level segmentation. The first stage makes use of CAM via an image classification model, which learns whether a slice has a nodule or not. In the second stage, a region of interest is selected around each localized instance in the class activation map and everything outside this region is masked out. Each masked image is then passed to the same classification network to obtain an instance-level segmentation mask, removing false positive regions produced in stage 1. {\brown The authors use the LIDC-IDRI dataset, but convert contour annotations to slice-level labels. From 1010 patients, 8345 slices are selected for each class: nodule and non-nodules, and the patients are divided in a 4:1:1 training, validation and testing split.} The suggested method achieves a 10\% Dice score improvement over the CAM baseline.

%wherein the first stage provides several regions of interest (ROIs) whereas the second stage performs segmentation within each ROI. The first stage consists of training an image classification model with CAM wherein the model learns whether a slice has a nodule or not. The Nodule Activation Map (NAM) produced is used to define the spatial scope for the second stage to generate a coarse segmentation map. To refine segmentation, each nodule candidate is masked out from the image in sequence. By feeding the masked image into the same network, a residual NAM (called R-NAM) is generated and used to select the true nodule. The suggested method achieves a 10\% dice score improvement over the CAM baseline.

\subsubsection{Multiple Instance Learning (MIL)}
\label{sec:mil}
% Explain MIL here in brief
% Why is MIL relevant to WSL
% How to get segmentation output
Multiple Instance Learning (MIL) refers to a classification scheme where the labels are provided for each bag of instances rather than each individual instance. If the label is negative, all instances in the bag are negative. However, if the label is positive, then at least one instance in the bag is positive. \lref{lj:noisy_labels}{\brown Since there is no information about which instances in the positive bag are actually positive instances, assigning the bag-level label to each instance in the bag would result in noisy and unreliable labels.} In the context of weakly supervised image segmentation, each image can be considered as a bag of instances where each instance can be a pixel or a tile in the image. By learning to classify the image as a whole, MIL learns to predict instance-level predictions, or equivalently the segmentation prediction for the input image.

\cite{jia2017constrained} use a multiple instance learning approach to generate pixel-level predictions given only image-level cancer labels for histopathology images. The suggested model consists of a VGG network that generates image level classification scores at multiple levels. For this purpose, the authors propose a soft aggregation layer that reduces the feature maps to a cancer classification score. By classifying each image into cancerous and healthy, the authors treat each histopathology image as a bag and each individual pixel in the image as an instance. To regularize training, the authors further impose area constraints (provided during annotation) on segmentation results by MIL. {\brown The authors use a histopathology image dataset of colon cancer, consisting of 330 cancer and 580 non-cancer images, of which 250 cancer and 500 non-cancer images are used for training; 80 cancer and 80 non-cancer images are used for testing.} The improvement in F-measure over the baseline (without area constraints and multiple-level predictions) is 6\% around the boundaries and 2\% overall.

%, by treating each pixel prediction as an instance that gets subjected to a max function to compute the resulting image classification score. To ensure non-zero gradients for even the non-max response pixels, the authors propose using a generalized mean function as their `softmax' function to get the image level prediction. In addition to the image-level labels, a rough estimate of the area of the cancerous region is provided during annotation. 
% They also propose fusing multiple side output predictions to account for multiple scales. Additionally, the authors define an area constant which measures the overall `positiveness’ of each image and constraint that with an L2 loss against the annotated area measurements. 
\cite{Campanella2019} train a weakly supervised segmentation model for whole slide histopathology images using only slide-level labels. The slides are divided into small tiles and the multiple instance training procedure includes a full inference pass of the dataset through a CNN, to rank the tiles according to their probability of being positive. The most suspicious tiles in each slide are sequentially passed to an RNN to predict the final slide-level classification, and  the heatmaps produced by the RNN are considered as segmentation predictions. {\brown The authors use a dataset consisting of 44,732 slides from 15,187 patients across three different tissue types: prostate, skin and axillary lymph nodes.} They achieve a 1\% increase in AUC when using the RNN classifier over just using the MIL approach directly. %{\brown When using an additional publicly available dataset with pixel-wise annotations, they show that weak supervision using large datasets leads to higher generalization performance than fully supervised learning on small curated datasets.}

\subsubsection{Summary}
The papers reviewed in this section address the problem of image segmentation with the weakest form of annotation: image-level labels. Section \ref{sec:activation} delved into the use of CAMs, which generate segmentation masks from intermediate outputs of a classification network. Section \ref{sec:mil}, on the other hand, treated this form of weak supervision as a label noise problem that can be solved via MIL. While both approaches improve the model performance over their respective baselines, it would be interesting to see how these two techniques can be combined together to achieve even higher gains. 
% courtesy: Xingjian

\subsection{Learning with Sparse Labels}
\label{sec:sparse_l}
Incomplete or sparse annotations refer to annotations where the masks are only provided for a fraction of the slices of a 3D volume, or for only a fraction of the pixels of a 2D image. There are various methods for dealing with these annotations as discussed below. Since the pixels (or voxels) are only partially labelled, the underlying theme in these methodologies is the use of a selective pixel loss wherein only the labeled pixels contribute to the loss. 

\subsubsection{Selective Loss with Mask Completion}
\label{sec:selective_w}
The papers discussed in this subsection all attempt to artificially reconstruct the incomplete regions of the ground truth masks and use the completed masks for training.

\cite{zhang2019sparse} propose a method for brain extraction and brain tissue segmentation in 3D MR images with a sparse set of annotations, where only a fraction of the slices are annotated at irregular intervals. To complete the sparse annotations, the authors use active learning. Specifically, the non-annotated slices are ranked by the Dice similarity coefficient between the output feature maps and attention maps, which are generated by a segmentation network equipped with channel-wise and spatial attention mechanisms. The slices with lowest Dice similarity are then presented to an expert for annotation. {\brown The authors evaluate their method on a brain MRI dataset of 40 neonates, but did not discuss their training-testing split, they have instead only shown the effect of incrementally utilizing the training set in an active learning framework.} For the task of brain extraction, the suggested method trained with just 15\% of the slices labeled yields results similar to that of a model trained with a 50\% annotation rate. For brain tissue segmentation,  the model requires 30\% of slices to be labeled in order to achieve a similar level of performance.  In addition to having spatially sparse annotations, when dealing with time sequence data, the annotations may be temporally sparse. \cite{bai2018recurrent} propose an image sequence segmentation algorithm by combining a fully convolutional network with a recurrent neural network, which incorporates both spatial and temporal information into the segmentation task. The missing annotations are then recovered through a non-rigid label propagation scheme. {\brown The authors use an aortic MR image set of 500 subjects, with 400 subjects used for training and the remaining 100 for testing.} The model trained using the additional masks recovered for unlabeled time frames achieves 1\% improvement in Dice over the U-Net baseline that was only trained on labeled time frames.
 
\cite{cai2018accurate} train a 3D segmentation model using only 2D annotations, which consist of diameter markings along the short and long axes of each lesion on the slice where the lesion appears the largest. The authors first refine the initial ground truth markings through the application of GrabCut. The suggested method then alternates between two steps. The first step consists of training the model using augmented ground truth masks. The second step expands the ground truth masks by running the trained model on the non-annotated slices adjacent to the annotated slices followed by applying GrabCut on the generated masks. This iterative process continues until all slices that contain a lesion are annotated. {\brown The authors make use of the DeepLesion dataset which is composed of 32,735 bookmarked CT lesion instances, of which 28,000 lesion volumes are used for training and the rest for testing.} This method achieves 10\% improvement in Dice over a model trained using the original  diameter markings. %the model morphs from a 2D segmentation network to a 3D network.

% initial masks are created using GrabCut with the aid of the RECIST markings. The model first learns to segment only the slices with markings. Next, the slices adjacent to the slices with markings are segmented using the trained model, forming the new initialization for the GrabCut. With new ground truth being created the number of slices used as input keeps iteratively increasing till it slowly builds up to the full 3D volume. Their approach shows a 10\% dice score improvement over just using the RECIST markings alone.

% This method alternates between training the model with ground truth segmented by GrabCut and running inference on the slices adjacent to the slices with annotations to create new GrabCut initialization. This iterative process  continues until the entire image segmentation is available and the model morphs from a 2D segmentation network to a 3D network
Scribbles have been recognized as a user-friendly alternative to overlapping bounding boxes. To generate the initial ground truth, \cite{can2018learning} use a random walk image segmentation approach with a high threshold to perform region growing around seed scribble annotations. Once the initial masks are generated, they suggest an iterative framework to incrementally refine the segmentation masks. Each iteration consists of two stages: 1) training the segmentation network with a CRF-RNN layer using the current annotations, 2) using test time dropout to make multiple predictions for each image and assess the uncertainty of each pixel-level prediction in the image. The new ground truth in each iteration is comprised of only the certain predictions. {\brown The authors make use of 2 datasets: 1) the ACDC cardiac segmentation challenge data, split into 160 training and 40 validation volumes, and an additional 100 images for evaluating the model using the challenge server, and 2) the NCI-ISBI 2013 prostate segmentation challenge containing 29 volumes split  into 12 training, 7 validation and 10 testing volumes.} They achieve a 3\% average Dice improvement over their non-iterative baseline.

{\brown \cite{matuszewski2018minimal} perform virus segmentation in electron microscopy images given minimal manual annotation in the form of points or lines. They artificially construct the ground truth masks by dilating the manual annotations with disc shaped structural elements, using a smaller disc for creating the foreground and a larger disc for creating the inverted background. Due to the different sized structural elements used, there exists unlabeled pixels at the boundary of the foreground and background that do not contribute to the segmentation loss while training. The authors use the Rift Valley virus dataset containing 143 images with 95 used for training and the remaining 48 for testing. They show a 7\% increase in Dice when using this scheme, over an approach where all pixels contribute to the loss and undefined boundary regions are penalized. However, no comparison was made against a fully-supervised oracle or other state-of-the-art weakly supervised approaches.}
% Unlike previous iterative approaches that attempt to generate the missing regions of the ground truth via Markov Random Fields (MRF) or Conditional Random Fields (CRF) and then use the completed masks for training, \cite{tang2018regularized} introduce regularization losses that incorporate a CRF in the loss function, thereby eliminating the need for a separate post-processing CRF. The authors also create a kernel cut loss to eliminate the need for a separate step of using graph cuts. These regularization losses are used in addition to the selective segmentation loss which compares the predicted output with the incomplete ground truth. They are able to achieve a 1\% improvement in Dice over the other weakly supervised methods that alternate between network training and proposal (training mask) updates. %For implementation details regarding the gradient calculation refer \cite{tang2018regularized}
 
\subsubsection{Selective Loss without Mask Completion}
Reconstructing the complete segmentation mask is not always a requirement. The papers reviewed below circumvent ground truth completion by modifying the objective function.
\label{sec:selective_wo}
\cite{silvestri2018stereology} recommend using a hexagonal grid for sparse annotations and show that a dense mask is not a requirement for training an effective pancreas segmentation model for abdominal CT scans. They compare using grids of different strides and the effect of padding the grid points to generate the training masks. The padding process consists of extending the ground truth masks around the sparse points to form discrete label blocks, but does not attempt to complete the segmentation masks.  {\brown The authors use a dataset containing 399 high-resolution abdominal CT scans, of which 79 scans are used for testing and the rest for training.} Their results show that using a grid stride of 9 pixels achieve a comparable performance to using a grid size of 3 pixels. The higher the stride, the fewer grid points that need to be annotated. %With a fixed stride and padding, the brush tool with the same size as the padding can be used for annotating the images, which would make it easier than marking individual pixels.
\cite{cciccek20163d} train a 3D model to segment kidney tubules in 3D confocal microscopy images using only sparse annotations. They propose using an additional class for unlabeled pixels. For annotated slices, the segmentation loss is class-balanced cross entropy where only the labeled pixels contributed to loss.  {\brown Using a dataset of three samples of Xenopus kidney embryos and a 3-fold cross-validation scheme}, the authors demonstrate a significant gain in IoU (~0.4$\rightarrow$0.86) when the annotation rate increases from 2.5\% to just 8.9\%. 

\cite{bokhorst2018learning} compare 2 different class-balancing methods that can be used to improve the segmentation performance given sparse annotations without trying to fill in the missing mask pixels. In the suggested method, only the labeled pixels contribute to a weighted segmentation loss. The loss-weighting to balance the classes is performed at the instance level or mini-batch level. {\brown Their dataset of whole slide images (WSI) contained 43 WSIs with sparse annotations and 8 with dense annotations in the training set, 11 WSIs with sparse annotations and 2 with dense annotations in the validation set and a test set containing 5 WSIs with only dense annotations.} The authors show that using instance-based balancing improves the Dice score by 1\% and mini-batch balancing improves it by 4\% when trained entirely on sparsely annotated images. %However, these balancing techniques did not improve the performance when they were used to train on densely annotated images. Adding a small subset of densely annotated images can drastically boost the segmentation performance and the authors have found that using a 1:4 ratio of densely annotated to sparsely annotated images with either of the class-balancing techniques results in a model performance comparable to that of using only dense annotations for training.
{\brown \cite{zhu2019pick} introduce a new quality awareness module (QAM), which is a CNN trained in parallel with the segmentation network to assess the quality of the masks. QAM uses the image with its mask as input and computes a new loss-weight for each sample in the mini-batch to ensure that the new segmentation loss penalizes different images differently based on the quality of the mask as measured by the secondary network. The authors use the JSRT dataset containing 247 X-ray images with segmentation masks for three types of organ structures: heart, clavicles, and lungs. They split the dataset with 165 training and 82 testing images. They artificially create noisy labels by eroding and dilating them with different sizes of structural elements. Their results show that as the level of label noise increases, the baseline segmentation model starts to perform significantly worse, but their model with QAM-weighted segmentation loss retains its high accuracy.}

\subsubsection{Summary}
The papers covered in this section deal with incomplete pixel labels and follow one of two distinct schools of thought: those that attempt to artificially complete the labels for the unlabeled pixels (Section \ref{sec:selective_w}) and those that do not (Section \ref{sec:selective_wo}). The former tends to be iterative, making these methods slow to train yet fairly reliable, because only confidently labeled pixels are added to the set of labeled pixels in each iteration. The latter, on the other hand, is more straightforward to implement, because it uses only the available labeled pixels with proper loss-weighting schemes. Although there is no direct comparison between the two approaches, we surmise that the approaches attempting to create pseudo ground truth masks will benefit from better gradient flow through larger portions of the image.

\subsection{Learning with Noisy Labels}
\label{sec:noisy_l}
%%%%%%%%% a subsection should start here
Noisy labels for the task of image segmentation  refer to ambiguities or inaccuracies in the boundaries of the segmentation masks. For medical images in particular, label noise could be induced by annotators unintentionally (random errors), or by inconsistencies between different readers due to human subjectivity concerning ambiguous lesions (expertise errors) \cite{Gu2018}. This type of annotation noise can be simulated by representing the labels as a polygon and then reducing the number of polygon vertices, which has the effect of creating segmentation error in the peripheral areas of segmentation masks. Label noise could also arise in  semi-supervised learning with pseudo annotations (Section \ref{sec:sl}) where the model learns from its own predictions on unlabeled data. Label noise, if left untreated, can degrade the performance of the segmentation model. It is therefore important to utilize strategies that mitigate the adverse effects of label noise during training.

%Annotation noise can be caused by human annotators when delineating the intricate boundaries of lesions and organs. This type of annotation noise can be simulated by representing the mask as a polygon and then reducing the number of vertices, which has the effect of creating segmentation error in the peripheral areas of segmentation masks. Annotation noise can also be caused by segmentation models when deployed in a self-learning paradigm where the model learns from its own predictions on unlabeled data. Both scenarios require proper treatment of label noise while training a segmentation model.

\subsubsection{Robust Loss without Mask Refinement}
\label{sec:robust}
\cite{mirikharaji2019learning} propose a learning algorithm resilient to the label noise in the segmentation masks. The suggested method consists of a weighted cross entropy loss function where the contribution of each pixel to the total loss is controlled by model's perception of the annotation quality for the pixels. During training, the weight matrices are updated based on the batches of images with clean annotations, and then used to scale the segmentation loss at the pixel-level for batches with noisy annotations. The authors simulate the annotation noise by replacing the segmentation masks with polygons of varying number of vertices. For the task of skin lesion segmentation, {\brown the authors use a skin image dataset consisting of 2000 training, 150 validation and 600 test images with their corresponding segmentation masks, and} show that a  model trained using the suggested loss and 3-vertex polygon masks performs comparably to the model trained using full annotation masks.

\subsubsection{Robust Loss with Iterative Mask Refinement}
\label{sec:iterative}
% this briefly summarizes methods to handle label noise in the context of self learning.
A proper handling of label noise is also studied in the context of semi-supervised learning with pseudo annotations where the model-generated annotations, commonly corrupted by noise, are used to fine-tune the model in an iterative manner. \lref{lj:iterative_refinement}{\brown The common solution in this context is to identify samples with noisy labels during training and then lower their impact on parameter updates by downplaying their gradients. In doing so, the model becomes capable of predicting more accurate labels for unlabeled data. The labeled data and pseudo-annotated unlabeled data are then used to improve the model and this process can be repeated as needed.} \cite{min2018two} propose a two stream network with independent weights whose concord determine the quality of segmentation mask. \cite{nie2018asdnet} propose a segmentation network with adversarial loss where the job of the discriminator network is to identify the reliable annotated regions from noisy annotations. Readers can refer to Section~\ref{sec:sl} for a detailed discussion of these approaches.

\subsubsection{Summary}
Similar to the methods for handling sparse labels (Section  \ref{sec:sparse_l}),
the papers reviewed in this section either attempt to refine the noisy masks (Section  \ref{sec:iterative}) or leave them as is (Section  \ref{sec:robust}). However, they differ from the sparse label techniques in that the pixels that contribute to the loss are not predefined but decided on the fly.
%Similar to Section  \ref{sec:sparse_l}, when dealing with label noise, the papers reviewed in this section either attempt to refine the mask (Section  \ref{sec:iterative}) or leave them as is (Section  \ref{sec:robust}).
The former approaches are iterative and slow to train, reducing the adverse effects of label noise on the segmentation model to a minimum. The latter approaches, on the other hand, are more straightforward to implement, because they use various loss-weighting schemes. 

%% file: chapters/discussion.tex
\begin{table*}[t]
\caption{Top-down overview of the methodologies suggested for the problems of scarce and weak annotations, where the methodologies are grouped by the underlying general and specific strategies. We have further used color encoding to show the required data resources of each methodology. Methodologies highlighted in green require no further data resources in addition to the original limited annotated dataset available for training; thus, they should be used wherever possible. Methodologies highlighted in orange require access to additional unlabeled data from the same domain or labeled data from a similar domain. Methodologies highlighted in red require experts in the loop; and thus, may not always be a viable option. }
\label{tab:discussion}
\resizebox{\textwidth}{!}{%
\begin{tabular}{llll}
\hline
{\color[HTML]{00009B} Problem I: Scarce Annotations} &  &  &  \\ \hline
General Strategy & Specific Strategy & Methodology & Description \\ \hline
\multicolumn{1}{l|}{} & \multicolumn{1}{c|}{} & \cellcolor[HTML]{009901}Same-domain data synthesis & \begin{tabular}[c]{@{}l@{}}Training a segmentation model with additional labeled data\\  generated by an image synthesis model\end{tabular} \\ \cline{3-4} 
\multicolumn{1}{l|}{} & \multicolumn{1}{c|}{} & \cellcolor[HTML]{009901}Data augmentation by mixing images & \begin{tabular}[c]{@{}l@{}}Training a segmentation model with additional labeled data\\  generated by blending the labeled images\end{tabular} \\ \cline{3-4} 
\multicolumn{1}{l|}{} & \multicolumn{1}{c|}{\multirow{-2}{*}{\begin{tabular}[c]{@{}c@{}}Augmenting the limited data with new\\  artificial examples\end{tabular}}} & \cellcolor[HTML]{009901}Traditional data augmentation & \begin{tabular}[c]{@{}l@{}}Training a segmentation model with additional labeled data\\  generated by spatial and intensity transformation\end{tabular} \\ \cline{2-4} 
\multicolumn{1}{l|}{} & \multicolumn{1}{c|}{} & \cellcolor[HTML]{F8A102}Semi-supervised learning with pseudo labels& \begin{tabular}[c]{@{}l@{}}Annotating unlabeled images using models' own predictions and\\  then using the augmented dataset for training a segmentation model\end{tabular} \\ \cline{3-4} 
\multicolumn{1}{l|}{} & \multicolumn{1}{c|}{} & \cellcolor[HTML]{F8A102}Semi-supervised learning without pseudo labels & \begin{tabular}[c]{@{}l@{}}Training a segmentation model with both labeled \\ and unlabeled data\end{tabular} \\ \cline{3-4} 
\multicolumn{1}{l|}{} & \multicolumn{1}{c|}{\multirow{-3}{*}{\begin{tabular}[c]{@{}c@{}}Leveraging additional unlabeled data\\  from the same domain\end{tabular}}} & \cellcolor[HTML]{F8A102}Self-supervised pre-training & \begin{tabular}[c]{@{}l@{}}Pre-training a model using unlabeled medical data and then \\ fine-tuning the model for the target segmentation task\end{tabular} \\ \cline{2-4} 
\multicolumn{1}{l|}{} & \multicolumn{1}{c|}{} & \cellcolor[HTML]{009901}Transfer learning & \begin{tabular}[c]{@{}l@{}}Training a segmentation model from the knowledge learned \\ from natural images (ImageNet or COCO)\end{tabular} \\ \cline{3-4} 
\multicolumn{1}{l|}{} & \multicolumn{1}{c|}{} & \cellcolor[HTML]{F8A102}Dataset fusion & \begin{tabular}[c]{@{}l@{}}Training a universal segmentation model from heterogeneous datasets \\ by learning to discriminate between the datasets\end{tabular} \\ \cline{3-4} 
\multicolumn{1}{l|}{} & \multicolumn{1}{c|}{} & \cellcolor[HTML]{F8A102}Domain adaptation w/ target labels & \begin{tabular}[c]{@{}l@{}}Training a segmentation model using shared feature representations \\ learned across multiple domains\end{tabular} \\ \cline{3-4} 
\multicolumn{1}{l|}{} & \multicolumn{1}{c|}{\multirow{-3}{*}{\begin{tabular}[c]{@{}c@{}}Leveraging external labeled data from\\  a similar domain\end{tabular}}} & \cellcolor[HTML]{F8A102}Domain adaptation w/o target labels & \begin{tabular}[c]{@{}l@{}}Training a segmentation model using only source domain labels \\ by translating from one domain to the other\end{tabular} \\ \cline{2-4} 
\multicolumn{1}{l|}{} & \multicolumn{1}{c|}{} & \cellcolor[HTML]{E13026}Active learning & \begin{tabular}[c]{@{}l@{}}Selecting unlabeled images for annotation judiciously \\ based on model predictions\end{tabular} \\ \cline{3-4} 
\multicolumn{1}{l|}{\multirow{-10}{*}{Expanding the dataset}} & \multicolumn{1}{c|}{\multirow{-2}{*}{\begin{tabular}[c]{@{}c@{}}Collecting additional annotations with\\  experts in the loop\end{tabular}}} & \cellcolor[HTML]{E13026}Interactive segmentation & \begin{tabular}[c]{@{}l@{}}Accelerating the annotation process by propagating the user\\ changes throughout the segmentation mask\end{tabular} \\ \hline
\multicolumn{1}{l|}{} & \multicolumn{1}{c|}{Leveraging additional tasks} & \cellcolor[HTML]{F8A102}Multi-task learning & \begin{tabular}[c]{@{}l@{}}Training a segmentation model with additional heads,\\ each for a separate classification task\end{tabular} \\ \cline{2-4} 
\multicolumn{1}{l|}{} & \multicolumn{1}{c|}{Imposing additional constraints} & \cellcolor[HTML]{009901}Shape regularization & \begin{tabular}[c]{@{}l@{}}Training a segmentation model by imposing shape constraints on\\  predicted segmentation masks\end{tabular} \\ \cline{2-4} 
\multicolumn{1}{l|}{\multirow{-3}{*}{Training w/ regularization}} & \multicolumn{1}{l|}{\begin{tabular}[c]{@{}l@{}}Leveraging more informative or\\  compressed input data\end{tabular}} & \cellcolor[HTML]{009901}Altered image representation & \begin{tabular}[c]{@{}l@{}}Training a segmentation model with a more compact or informative\\ image representation\end{tabular} \\ \hline
\multicolumn{1}{l|}{Post-training refinement} & \multicolumn{1}{l|}{\begin{tabular}[c]{@{}l@{}}Using post-processing methods to \\ refine segmentations\end{tabular}} & \cellcolor[HTML]{009901}CRF-based post segmentation & \begin{tabular}[c]{@{}l@{}}Using CRF as a post-processing or as a trainable module\\ in the segmentation network\end{tabular} \\ \hline
{\color[HTML]{00009B} Problem II: Weak Annotations} & \multicolumn{1}{l}{} &  &  \\ \hline
\multicolumn{1}{l|}{} & \multicolumn{1}{l|}{Learning with sparse annotations} & \cellcolor[HTML]{009901}\begin{tabular}[c]{@{}l@{}}Selective loss w/ and w/o mask\\  completion\end{tabular} & \begin{tabular}[c]{@{}l@{}}Training a segmentation model by excluding unannotated pixels\\  from backpropagation\end{tabular} \\ \cline{2-4} 
\multicolumn{1}{l|}{} & \multicolumn{1}{l|}{Learning with noisy annotations} & \cellcolor[HTML]{009901}\begin{tabular}[c]{@{}l@{}}Robust loss w/ and w/o iterative label\\  refinement\end{tabular} & \begin{tabular}[c]{@{}l@{}}Training a segmentation model with mechanisms that downgrade\\  unreliable annotations during training\end{tabular} \\ \cline{2-4} 
\multicolumn{1}{l|}{} & \multicolumn{1}{l|}{Learning with image-level annotations} & \cellcolor[HTML]{009901}Class activation maps & \begin{tabular}[c]{@{}l@{}}Training a classification model with global average pooling\\ and using activation maps as class-specific segmentation\end{tabular} \\ \cline{3-4} 
\multicolumn{1}{l|}{\multirow{-4}{*}{Leveraging weak annotations}} & \multicolumn{1}{l|}{} & \cellcolor[HTML]{009901}Multiple instance learning & \begin{tabular}[c]{@{}l@{}}Training a classification model with aggregation layers \\ and using activation maps as class-specific segmentation\end{tabular} \\ \hline
\end{tabular}%
}
\end{table*}

\lref{lj:scarce_weak_combo}{\brown In \figurename~\ref{fig:problem_tree}, we split the data limitations into scarce annotations and weak annotations, which allowed us to group similar methodologies with ease. However, when making use of multiple datasets, it is possible that the combined dataset now suffers from both scarce and weak annotations. A combination of each individual solution can be used in tackling such datasets. For example, if a small dataset of histopathology images had tumor segmentation masks available, but another much larger histopathology image dataset had only image level tumor classification labels available, then a multi-task framework incorporating a semi supervised approach like \cite{sedai2017semi} used in conjunction with a CAM-based approach like \cite {feng2017discriminative} could address the issue of both scarce annotations and weak annotations while utilizing the joint potential of this new larger dataset. When faced with the task of annotating a large dataset with a limited budget, it may also be helpful to get a small subset with dense segmentation mask annotations and only weak annotations for the remaining subset and use a combined strategy to train a robust model.}

Table~\ref{tab:discussion} presents a summary of the methodologies suggested for the problems of scarce and weak annotations. For clarity, the table is split into two sections, each focusing on one annotation problem. We have further grouped the methodologies in each section by the general and specific strategies the follow. Color encoding is also used to indicate the data requirements of each methodology.  We hope this table can serve as a strategy guideline, assisting the readers in choosing the right methodology according to the dataset problems they face and the data resources they have available. In what follows, we highlight the important messages of Table~\ref{tab:discussion}.   %The summary for each solution consists of a short description of how it works, the advantages it offers, and data requirements it demands. 

%Table~\ref{tab:discussion} compares the solutions reviewed in this survey at a high-level by highlighting the advantages and requirements of each solution, which can serve as guidelines assisting the readers in choosing the proper solutions according to their available resources. 

As indicated by the color encoding, the  methodologies suggested for the problem of scarce annotations can be placed in  three broad categories {\brown according to the data requirements:}

%1) the low-requirement group, where  the solutions require no further training data, 2) the mid-requirement group, where the solutions require access to additional labeled or unlabeled training data from the same or a similar domain, 3) the high-requirement group, where the solutions require access to medical experts. While the effectiveness of a solution is normally  proportional to its level of requirements, there are solutions in low-level and mid-level categories that have a high return compared to their relatively low data requirements.

\begin{enumerate}
    \item {\noindent \emph{Solutions with low {\brown data} requirements:}} This group of solutions rely solely on the available labeled segmentation dataset, requiring no additional labeled or unlabeled training data.  Therefore, they should be utilized wherever possible. Of the suggested methodologies, CRF-based post-processing has shown mixed results for 3D segmentation, and altered 3D image representations have achieved medium gains at the price of training several 2D models. Therefore, in addition to traditional data augmentation, which is the de facto solution to the scarce annotation problem, we recommend using shape regularization, data augmentation by mixing images,  and same-domain data synthesis for both 2D and 3D applications, and CRF-based post-processing for 2D applications.
    
    %Therefore, they should be utilized wherever possible. These solutions include 1) traditional data augmentation, which is a well-known strategy to combat data scarcity; 2) CRF-based post-processing, which is effective to some extent for 2D images, but its application to 3D images has yielded only mixed results; 3) shape regularization, which is effective particularly around boundaries, but often requires architectural modifications in the segmentation model; and 4) altered image representation, which is easy to implement but has mixed results across 2D and 3D applications. Considering the advantages, requirements and performance gains, we recommend using shape regularization and traditional data augmentation.
    
    \item \noindent \emph{Solutions with medium {\brown data} requirements:} This set of methodologies  requires access to additional labeled or unlabeled training data from the same or a similar domain. Therefore, depending on the application at hand and the availability of the corresponding auxiliary datasets, these solutions may or may not be applicable. Of the suggested methodologies, semi-supervised learning with pseudo annotations has shown mixed results with the exceptions being methods that adopt advanced architectures to handle annotation noise in model-generated annotations. Domain adaptation techniques are effective, but they can be difficult to adopt due to the
    instability of adversarial training, which lies at the core of these methodologies. Semi-supervised learning without pseudo annotations require only additional unlabeled data and are typically less demanding to implement compared to unsupervised domain adaptation methods. Multi-task learning and dataset fusions are both straightforward solutions with reasonable performance gains. In our opinion, self-supervised pre-training is one of the most promising approaches in this category, requiring only unlabeled data and typically only minor modifications to the architecture.

    %This group covers 1) synthetic data augmentation, which has proved effective particularly for few-shot image segmentation, but, admittedly, the suggested methods are heavy-weight and non-trivial to implement;  2) domain adaptation, which is effective in leveraging additional datasets, but often hard to train due to the adversarial networks at the core of these methods, 3) self-learning, which attempts to annotate unlabeled data, but its effectiveness can be limited by severe label noise in pseudo annotations; 4) semi-supervised learning, which leverages additional unlabeled data and is relatively less demanding to implement than domain adaptation methods; 5) self-supervised learning, which is easy to implement with impressive performance gains; and 6) multi-task learning, which is typically easy to implement, but the performance gain depends on the similarity between the auxiliary tasks and the target segmentation task. In summary, we recommend using self-supervised and semi-supervised learning due to their high return and the ease of implementation.}
    
    \item {\noindent \emph{Solutions with high {\brown data} requirements:}} These solutions require access to medical experts, but their elegance lies in the use of expert knowledge in a cost-effective manner. Two solutions in this category are active learning and interactive segmentation where the former determines which samples to be annotated by experts whereas the latter helps experts complete the annotation tasks quickly. If our hands are forced into annotating more data or if additional data annotation is deemed highly advantageous, then these two methodologies should be prioritized in practice.
\end{enumerate}

%From Table~\ref{tab:discussion}, it is however evident that the problem of weak annotations has received less attention compared to the problem of scarce annotations. Specifically, for each manifestation of weak annotations namely, sparse, noisy, and regional annotations, only few solutions with relatively similar methodologies are available. We therefore compare these solutions from the perspectives of performance gains and annotation cost, highlighting solutions with the best gain-cost trade-off.

%When faced with the task of annotating medical images, one must choose the type of annotation that best suits their segmentation task and budget. Weak annotations are an affordable way to to annotate a larger set of images with a relatively low cost, which in turn allows for a more representative training set. 

The methodologies suggested for handling weak annotations are closely related to the types of annotations that are readily available for training. For each type, we compare and recommend methodologies  that best suit the given limitation from the perspectives of performance gains and annotation cost. 

%The reviewed methodologies, color encoded in Table~\ref{tab:discussion}, come with low data requirements and aim to provide the optimal gain-cost trade-off under each scenario.

% Also leveraging weak annotations, the suggested methodologies all have low data requirements. We therefore compare the methodologies for each manifestation of weak annotations from the perspectives of performance gains and annotation cost, highlighting solutions with the best gain-cost trade-off. 
\begin{enumerate}
    \item\emph{Noisy annotations:} A common problem with medical datasets and in particular segmentation datasets is annotation noise where the annotated contours may not always follow the contours of the region of interest. Handling annotation noise is important, because not only does it reduce the adverse effects of inter-observer annotation variability on the trained model, but it also enables training with only rough annotations, which can be obtained in a cost-effective manner with significantly shorter annotation time than that of accurate annotations. For instance, the work by \cite{mirikharaji2019learning} shows that, with  a noise-resilient approach, a skin segmentation model trained with 3-vertex contours can achieve similar performance to a model trained using accurate segmentation masks. Handling annotation noise in medical segmentation datasets is still a fairly new topic and deserves further investigation.  
    
 %Due to human or machine error, annotation labels may not always follow the contours of the region being segmented. The weakly supervised approaches reviewed that tackle this obstacle, are useful in attempting to minimize the impact of this label noise without requiring an expert to rectify the annotations and are thus cost effective. However, the initial noisy annotations are still obtained by expert annotators, or in cases where they are machine generated, the methodologies for solving them attempt to iteratively correct the masks, slowing down the training process.
    
    \item\emph{Sparse annotations:} Of the weakly supervised approaches reviewed, the papers tackling sparse annotation have achieved the closest performance to their strongly supervised counterparts; however, the application of sparse annotations may not always be viable. For instance, while dot grids \cite{silvestri2018stereology} may be useful for larger organ segmentation, they would not be as effective for segmenting small lesions. Furthermore, even though sparse annotations are easier to obtain than strong segmentation masks, the annotation process is still not entirely user-friendly, and the training schemes tend to be iterative, leading to longer training periods.
    
    \item\emph{Image-level annotations:} Of the weak annotations reviewed, image-level annotation incur the least annotation cost. Comparing the suggested methodologies, we would recommend using the modified CAM-based approaches with image level-labels. Not only do they use the least expensive form of annotation, but they also show large improvement in Dice over the direct CAM approaches and only fall a couple of Dice points short of using full supervision with strong annotations \cite{feng2017discriminative}.

\end{enumerate}

% \lref{lj:scarce_weak_combo}{\brown The papers reviewed easily fit into the category of scarce annotations or weak annotations. However, when making use of multiple datasets, it is possible that the combined dataset now suffers from both scarce and weak annotations. A combination of each individual solution can be used in tackling such datasets. For example if a small dataset of histopathology images had tumor segmentation masks available, but another much larger histopathology image dataset had only image level tumor classification labels available, then a multi-task framework incorporating a semi supervised approach like \cite{sedai2017semi} used in conjunction with a CAM-based approach like \cite {feng2017discriminative} could address the issue of both scarce annotations and weak annotations while utilizing the joint potential of this new larger dataset. When faced with the task of annotating a large dataset with a limited budget, it may also be helpful to get a small subset with dense segmentation mask annotations and only weak annotations for the remaining subset and use a combined strategy to train a robust model.}